\documentclass[a4paper,useAMS,usenatbib]{mnras}
\usepackage{natbib}
\usepackage{graphicx}
\usepackage{amsmath}
\usepackage{float}
\usepackage{color}
\usepackage{changes}
\usepackage{verbatim} 

\newcommand{\myr}{\mbox {~$\rm M_{\odot}$/yr}}
\newcommand{\mdot}{\mbox {$\dot{M}$}}
\newcommand{\mch}{\mbox {$M_{Ch}$}}
\newcommand{\ms}{\mbox {$M_{\odot}$}}
\newcommand{\rs}{\mbox {$R_{\odot}$}}

\newcommand{\porb}{\mbox {$P_{\rm orb}$}}
\newcommand{\mamd}{\mbox {$M_{\rm acc} - M_{\rm don}$}}
\newcommand{\al}{\mbox {$\alpha_{ce}\lambda$}}
\newcommand{\ace}{\mbox {$\alpha_{ce}$}}
\newcommand{\sna}{\mbox {SN~Ia}}
\newcommand{\sne}{\mbox {SNe~Ia}}

\newcommand{\mt}{\mbox {$M_1+M_2$}}
\def\apgt{{\raise-.5ex\hbox{$\buildrel>\over\sim\,$}}}
\def\aplt{{\raise-.5ex\hbox{$\buildrel<\over\sim\,$}}}
\def\mn{MNRAS}
\def\acta{Acta.Astron.}
\def\aap{Astron.Astrophys.}
\def\za{{\bf A}}
\def\zb{{\bf B}}
\def\ze{{\bf E}}

\title[Merging white dwarfs and SN Ia]{Merging white dwarfs and SN~Ia}

\author[L.R. Yungelson and A.G. Kuranov]
{L.R. Yungelson$^1$\thanks{E-mail:lev.yungelson@gmail.com}, A. G. Kuranov$^2$\\
 $^{1}$Institute of Astronomy of the Russian Academy of Sciences, 48 Pyatnitskaya Str., Moscow, Russia\\
 $^{2}$Sternberg Asronomical Institute, Moscow M.V. Lomonosov State University, 2 Lomonosov prosp.,  Moscow, Russia}

\date{Accepted 2016 September 22, Received 2016 September 20; in original form 2016 July 25}

\pubyear{2016}
\pagerange{\pageref{firstpage}--\pageref{lastpage}}
\begin{document}
\label{firstpage}
\maketitle

\begin{abstract}
Using population synthesis, we study a double-degenerate (DD) scenario for SNe Ia, 
aiming to estimate the maximum possible contribution to the rate of SNe from this scenario
and the dependence of the delay-time distribution (DTD) on it. 
We make an extreme assumption that all mergers of super-Chandrasekhar pairs of CO white dwarfs (WDs) 
and mergers of CO WDs more massive than 0.47 $M_\odot$ with hybrid or helium WDs more massive than
0.37$M_\odot$ produce SNe Ia. 
The models are parametrized by the product of the common envelope efficiency and the parameter
of binding energy of stellar envelopes $\alpha_{ce}\lambda$, which we vary 
between 0.25 and 2. The best agreement with observations is obtained for 
$\alpha_{ce}\lambda$=2. A substantial contribution to the rate of SNe Ia is 
provided by the pairs with a hybrid WD. The estimated Galactic rate of SNe Ia is 
$\mathrm{6.5\times 10^{-3}\,yr^{-1}}$ (for the mass of the bulge and thin disk equal to
$7.2\times10^{10}\,M_\odot$), which is comparable to the observational estimate 
$\mathrm{(5.4\pm0.12)\times 10^{-3}\,yr^{-1}}$. 
The model DTD for $1\leq t \leq 8$\,Gyr satisfactorily fits DTD for SNe Ia in the field
galaxies \citep{2012MNRAS.426.3282M}. 
For this epoch, model DTD is  $\propto t^{-1.64}$. At earlier and later 
epochs our DTD  has a deficit of events, as in other studies. Marginal agreement
with observational DTD is achieved even if only CO+CO WD with $M_1\geq0.8\,M_\odot$
and $M_2\geq0.6\,M_\odot$ produce SNe Ia. A better agreement of observed 
and model DTD may be obtained if tidal effects are weaker than assumed and/or 
metallicity of population is much lower than solar.
\end{abstract}

\section{Introduction}
\label{sec:intro}

The exceptional role of SNe~Ia in 
exploration of the riddles of the Universe is a result primarily of the fact
that, with certain caveats,  \sne\ may be considered as 'standard candles' and serve as a measure of cosmic distance.
The nature of \sne\ is, however, still elusive. The exists only an agreement 
that they are related to the thermonuclear explosions of white dwarfs (WDs) \citep{hoyle_fowler60}. 

The evolution of close binaries leading to formation of potential precursors of \sne\ has been 
discussed  for more than 45 years,
starting from the papers of \citet{ty79a,web79,ty81,it84a,web84}. 
The current status of the problem is considered, for example, in the recent reviews by
\citet{2013FrPhy...8..116H,2014ARA&A..52..107M,2014LRR....17....3P,2014NewAR..62...15R}.

There are two main competing hypothetical scenarios of the evolution of close binaries to \sne: (a) the 
`single-degenerate' one (SD), implying the explosion of a CO WD that accumulated matter from a non-degenerate 
donor and (b) the `double-degenerate' scenario (DD), in which \sna\ is a result of the merger of a pair of WD. 
Other scenarios, such as  a  merger of WD with the core of a red giant (core-degenerate one in modern terms)
\citep{sparks-74,2012MNRAS.419.1695I}  or  direct
collisions of WD \citep{2009MNRAS.399L.156R,2009ApJ...705L.128R} are usually considered to be less important.
We do not consider \sne\ induced by tidal interactions
of WDs with black holes \citep{2009JPhCS.172a2036R} and 'resonant' ones   \citep{2016MNRAS.tmp.1194M}, 
as well as the possible enhancement of \sne\ rate owing to dynamical interactions in stellar clusters
\citep{2002ApJ...571..830S}. 
 
The topic of the present paper is the double-degenerate scenario.
Its main assumption is that the loss of angular momentum via gravitational wave radiation by a close binary 
with WD components reduces separation of thr WD and leads to the Roche lobe overflow (RLOF) by the lower
mass one (with the larger radius). Early simple analytical estimates using mass-radius relation 
\citep{ty79a} suggested that for mass-ratio of components
$\apgt 2/3$, mass-loss by the WD becomes unstable and the latter completely disrupts, presumingly       
on a time scale comparable to the orbital period of  the system \porb, forming 
a disk that gradually settles onto the WD.  Components of the binary thus 'merge'.
Later, \citet{npv+01,mns04} showed that the stability of mass loss depends also on the efficiency of the tidal 
interaction in the binary.
SPH-calculations taking into account the physical equation of state demonstrated  that WD disrupts  
in  \ensuremath{\sim100\porb} and merger occurs in the direct impact regime
\citep[e. g., ][]{souza_dynme06,2009JPhCS.172a2034D,2011ApJ...737...89D}. 
While the early versions of the scenario assumed as precursors of \sna\ only 
pairs of CO WD with $\mt \geq \mch$\ \citep{ty79,web79,ty81,web84,it84a}, currently, 
mergers of sub-\mch\ systems with CO accretors and CO, He and hybrid donors\footnote{M$\simeq$
(0.32 -- 0.60)\ms\ CO WD that descend 
from  helium stars and have mass abundances of He up to almost 90\% in the  
$\simeq$(0.20 -- 0.01)\,\ms\ envelopes; more massive post-He-stars WD have only traces of He 
in their envelopes.} are considered too.  
The evidence for the necessity of
consideration of sub-\mch\ \sne\ was summarized by \citet{2010ApJ...722L.157V}. 

However, an essential problem that is apparently unresolvable in the foreseeable future is that the disruption of 
the donor and formation of a quasi-steady configuration are modelled by SPH-methods, with resolution 
$\sim10^7$\,cm at best, while models of nuclear burning deal with scales down to  
$\sim1$\,cm. SPH-computations allow us only to determine whether the necessary condition 
of detonation  --  an energy release time-scale $\tau_{3\alpha}$ is shorter than the local dynamical time-scale 
$\tau_{dyn}$ -- is fulfilled, while the necessary and sufficient condition for 
self-sustained detonation is the supersonic speed of flame propagation.   
If it does not follow immediately from the computations of merger that the latter condition is fulfilled,  
one has to judge whether detonation is possible, using results of SPH-calculations as an input for
mesh-based codes of different degrees of sophistication. The aim is to find whether  
formation of 'hot spots' in the merger products is possible, that is, to calculate
the critical sizes of hot regions that ignite and yield propagating detonations.
The parameters of hot spots (density, temperature and its gradient, geometry) 
for He were found by 1-D calculations by \citet{2013ApJ...771...14H,2014ApJ...797...46S} 
and  for C/O mixture, most recently,  
by \citet{2009ApJ...696..515S,shen_bildsten_dd13}\footnote{Such a method may lead 
to the smoothing of temperature distribution and distortion of  the distribution of chemical species, potentially 
influencing results of calculations, see,for example,  the  detailed discussion by 
\citet{2014ApJ...797...46S,2016ApJ...819...94K}. Another important factor is sophistication of the nuclear reactions grid.}.  

The above circumstance makes the real role of the DD-channel in production of \sne\ uncertain.
The aim of the present study is to estimate possible contribution of the DD-scenario to
the rate of \sne\ and dependence of delay-time distribution for \sne\ on this 
scenario under an \textit{extreme assumption that all mergers of super-\mch\ CO+CO pairs of WD, as well as mergers of CO WD and massive He and hybrid WD result in \sne}.
      
In \S~\ref{sec:merger} we briefly review the extant results of merger computations and show that
they are still inconclusive. For this reason, we compute the total rate of mergers of WD pairs with He, hybrid- 
 and CO-donors and CO or ONe accretors for different parameters of population synthesis, aiming to
estimate the upper limit of contribution of mergers to \sne\ rate and, particularly, 
of different pairs to the latter. In \S~\ref{sec:model} we present
the assumptions we use, in \S~\ref{sec:res} our results are presented, followed by Discussion in \S~\ref{sec:anal} and 
Conclusions in \S~\ref{sec:concl}. Some additional information is provided in the Appendix. 

\section{Merger computations}
\label{sec:merger}

The merger of WD may proceed through several stages: 
(i) disruption of the less-massive WD and dynamical accretion;
(ii) relaxation of the resulting configuration to the  quasi-steady state;
(iii) evolution of the merger product to a SN or an accretion induced collapse (AIC).

Merger products have a similar structure \citep{guerrero04}: a cold virtually isothermal core, 
a pressure-supported envelope, a Keplerian disc, and a tidal tail. 
If explosion 
does not happen during dynamical stage of merger,  He or C/O-mixture may  explode in the envelope which
is the hottest part of the object.        
Detonation of He in the envelope may lead
to the detonation of carbon at the periphery of accretor  
('edge-lit detonation')  or in its central region which is compressed by
converging shocks ('double-detonation').
The explosion of a CO WD results in its complete disruption. 
While the mayor fraction of the accretor mass burns to radioactive Ni, which determines optical luminosity and
spectrum of the  \sna, the donor burns to intermediate mass elements which are responsible
for observational manifestations of  \sna\ at maximum brightness 
\citep{1992ApJ...386L..13S,2010ApJ...714L..52S}.
Thus, the main role of the donor explosion (or of its remnants after disruption) is to trigger the detonation 
of the accretor \citep{2009ApJ...705L.128R}.

Sub-\mch\ `double-degenerate' scenario is to a significant extent related to 
the also still hypothetical `double-detonation' scenario with  
a He-rich donor, either degenerate or non-degenerate, e. g., \citet{lg91,2014ApJ...785...61S}.  
\citet{2010ApJ...714L..52S}, based on a series of 1-D calculations of pure detonations of CO WD 
with post-processing nucleosynthesis and radiation transfer, concluded that sub-\mch\ 
explosions are a viable model for SNe Ia for \textit{any} evolutionary scenario leading to explosions 
in which the optical display is dominated by the material produced in the detonation of the primary WD. 

Systematic studies of mergers of WD pairs with He WD or CO/He hybrid WD donors and CO WD accretors 
with mass from  0.4 \ms\ to 1.2 \ms\ were performed by  \citet{2010ApJ...709L..64G} and 
\citet{2012MNRAS.422.2417D,2014MNRAS.438...14D}, using grid-based and SPH-codes, respectively.
They found that for He/hybrid+CO  pairs of WD the condition $\tau_{\rm 3\alpha}\leq\tau_{\rm dyn}$ is 
fulfilled prior to merger or at the surface contact provided 
$M_{CO} \leq 1.1$\,\ms\ and mass of He-rich donor $\apgt 0.4$\,\ms. In some cases detonation is due to 
instabilities in the accretion stream. It remains unclear, however, whether these surface detonations may
initiate detonation in the core. 
For CO+CO WD pairs dynamical burning conditions were met for $(\mt)\geq2.1$\,\ms\ only.   

\citet{2010Natur.463...61P,pakmor_violent12} studied 'violent' (see below) mergers of 
several pairs of equal and almost equal mass CO WD ($M_1=0.9$\,\ms)
in which 
they \textit{assumed}, based on computations of \citet{2009ApJ...696..515S} that detonation occurs if, during 
compression and heating of the matter encountering the accretor, the threshold 
$T \geq 2.5 \times 10^9$ K at the density
of about $2 \times 10^6 \mathrm{g\, cm^{−3}}$ is reached. As a critical mass ratio for 
detonation, Pakmor et al. inferred $M_2/M_1 =0.8$. Having in mind that the traces of He ($10^{-3} - 10^{-2}$)\,\ms\
should be present at the surface of all CO WD and that He is ignited more readily than carbon and, thus,
may facilitate nuclear ignition during the merger \citep{2010ApJ...709L..64G},
 and  based on the computation of merger of 1.1\,\ms\ and 0.9\,\ms\ WDs with 0.01\,\ms\ 
He-envelopes, \citet{2013ApJ...770L...8P} 
speculated that He-ignited (CO+CO) WD and (CO+He) WD mergers may present a unified model for normal and rapidly declining \sne. The model does not explain \sne\ with strongly mixed ejecta (SN~2002cx type ones).

\citet{2013MNRAS.429.1425R} used the relationship between accretor mass
and $\mathrm{M_{bol}}$ in maximum for detonating sub-\mch\ mass CO WD to construct
the brightness distribution expected for violent mergers of pairs of CO WD by population synthesis.
Under certain assumptions, the shape of brightness amplitude distribution  matches the observations well.
However, the model needs further elaboration in respect of nuclear display and spectral features. At the moment, it is worth to mention that 
the model encounters serious difficulties in reproducing spectrophotometric features of \sne\ because of 
asymmetries in the distribution of ejecta \citep{2016MNRAS.455.1060B}. 

\citet{2015MNRAS.454.4411D} analysed possibility of post-merger explosions for some 
previously found configurations, using them as an input for the grid-based 
2-D hydrodynamical calculations with account of rotation 
by the code  FLASH \citep{2000ApJS..131..273F,2009arXiv0903.4875D}.
It was found that \textit{immediately} after completion of merger detonation does not occur in either  
He+CO or CO+CO pairs, because of too low a density of hot regions (envelopes).
To the similar conclusion for CO+CO pairs came 
\citet{2013ApJ...767..164Z,2015ApJ...807...40T,2015ApJ...807..105S}. However, it was suggested 
that an insufficient resolution in  most computations, $\aplt 500k$  ($k=1024$ particles per \ms),  may be the 
reason for this \citep{0004-637X-821-1-67}.
In test computations with resolution up to $\approx 2000k$\ in which also $\mathrm{^{12}C+^{12}C}$ 
reaction was taken into
account, detonation conditions of \citet{2009ApJ...696..515S}, were met by pairs of WD with $M_1 \apgt 0.8\ms$ and 
\begin{equation}
M_2/M_1 \apgt 0.8\,(M_1/\ms)^{-0.84}. 
\label{eq:sato16}
\end{equation}
 
One may expect that a further increase of 
resolution, will allow to resolve smaller hot regions. \citet{0004-637X-821-1-67}
also noted that the study of possible
mergers at contact for He+CO pairs needs a higher resolution, since the occurrence
of helium detonation depends strongly on the mass of the helium layer. This may affect 'unified' 
model of \citet{pakmor_violent_13}.  

\citet{2015MNRAS.454.4411D} found that the criteria of spontaneous detonation discovered by 
Holcomb et al., \citet{2013ApJ...771...14H}, \citet{2014ApJ...797...46S},
\citet{2009ApJ...696..515S} may be met during  evolution of merger products in their 
most dense and hot regions. However, because of above-mentioned discontinuity in 
resolution of SPH and mesh-based computations,  hot spots were located 'manually', 
assuming  that the merger products evolved to conditions necessary for detonations.
It was found that He-detonation, may lead or not lead to the detonation in 
the centre, As well, it was found that the initial perturbation may not initiate 
detonation in the envelope, but converging shocks may cause detonation in the core. Some of
post-merger configurations do not lead to \sna\footnote{A caveat should be inserted that results of
Dan et al. may be influenced by assumption that their initial models of CO WD with mass 
0.45\,\ms\ to 0.6\,\ms\ have a 0.1\,\ms\ helium mantle; such an abundance  may be an overestimate
\citep{it85,2002MNRAS.335..948C}.}.

\citet{2012ApJ...746...62R}, who also computed the merger of WDs by SPH, have shown that in the models of 
merger of CO WD with He-envelopes 
($M_{1,2}=(0.64-1.06)$\,\ms,\,\,$M_{He} =(0.013-0.015)$\,\ms), He detonates in the merger process, but the 
released energy is not sufficient for initiation of detonation in the core of the merger product. 
In the continuation of this study, \citet{moll_prompt13}, have analysed possibility of detonation using grid-
based code with a higher  resolution and showed that secondary detonation is possible, if massive ($M_{1,2} 
\geq 1.06$\,\ms) WD merge. But we note that, like in the studies of
Pakmor et al., these computations rely on artificial ignition of detonation.

The long-term post-merger evolution of the merger product of  CO WDs which avoided detonation at 
the merger itself was simulated by 
\citet{2012ApJ...748...35S,2016MNRAS.463.3461S} using $\alpha$-viscosity prescription.
\citet{2012MNRAS.427..190S} performed a similar study for a  He+CO WD merger. 
It was found that merger products evolve on the viscous time scale towards spherical configurations.
A hot, slowly rotating, and radially extended envelope forms.
Certain amount of mass may be lost by the stellar wind. At the end of this stage, owing to
dynamical and viscous heating, the temperature at the base of the envelope may become high enough 
for off-centre burning. In the case of He/CO mergers, it is possible that He-detonation ensues and a \sna\ explodes. 
As argued by \citet{2016MNRAS.463.3461S} on the basis of 1-D computations, in the case of CO WDs, 
if the  mass of the object remains below 1.35\ms, 
inward propagation of the burning leads to formation of a massive ONe WD.
In merger remnants with higher mass, neon ignites off-centre. 
It is expected that a silicon WD forms. If the mass remains super-\mch, further nuclear evolution 
will result in formation of
an an iron core and collapse, producing a neutron star. The optical manifestation of an accompanying supernova 
is uncertain. In fact, the study  of \citeauthor{2016MNRAS.463.3461S} questions the role of 
mergers of CO WD with mass below $\approx 2\,\ms$ as progenitors of \sne\ in the DD-scenario. 

\citet{2013ApJ...773..136J} have shown (in 2-D) that the merger of a pair of (0.6+0.6)\,\ms\ 
CO WDs produces a rapidly-rotating WD
surrounded by a hot corona and a thick, differentially-rotating disk, which is strongly
susceptible to the magneto-rotational instability. Instability leads to the
rapid growth of the initially dynamically weak magnetic field in the disk, spin-down of the 
`new-born' WD, and to the central ignition of the latter. However, as the outcome of ignition depends on the 
temperature profile \citep{2009ApJ...696..515S}, this simulation also does not definitely tell whether \sna\ 
explodes. Consideration of magnetic field evolution of merger product of $\mt < \mch$\ WD
\citep{2015ApJ...806L...1Z} has shown that exponential amplification of the field strengths occurs. 
\citeauthor{2015ApJ...806L...1Z} speculated that  magnetic field 
provides a mechanism for angular momentum transfer and additional heating, facilitating carbon 
fusion. 

However it was pointed out earlier \citep{piers+03b} that,
if a Keplerian disk forms out of the disrupted component and persists,
as a result of the spin-up of
rotation of the WD by accretion, instabilities associated with
rotation, deformation of the WD, and angular momentum loss by a distorted configuration via 
gravitational waves, accretion rate onto WD  that is initially $\sim10^{-5}$\,\myr\ decreases
to $4\times10^{-7}$\,\myr\ and  close-to-centre ignition of carbon becomes possible. 
This self-regulated accretion mechanism is applicable for pairs with total mass (1.4--1.5)\,\ms\
at the onset of carbon ignition; its timescale is $\sim10^6$\,yr.

\citet{2015ApJ...800L...7K} performed a post-merger evolution simulation for a  (1.1+1.0)\,\ms\ CO+CO WD 
system and found that a spiral-mode instability developed in the accretion 
disk on the dynamical time-scale  and forced hot disk material to accrete onto the core of the  remnant. This  
process drives a thermonuclear outburst leading to \sna\
without the need for artificial ignition. This mechanism works on the time scale which is two-three orders of 
magnitude shorter then the magneto-rotational instability suggested by \citet{2013ApJ...773..136J}. Thus, 
a self-ignited detonation may be possible in post-merger stage at least for the most massive objects.  
However, follow-up modelling of light-curve and spectrum of the system 
\citep{2016ApJ...827..128V} led 
to the suggestion that a better agreement with observables of normal \sne\ 
will require lower masses of components, but whether similar mechanism will be effective for these systems remains 
an open question . Nevertheless,  as noted by \citet{2015ApJ...800L...7K},  a wide range
of He+CO WD mergers may ignite  unstable helium burning via the spiral-wave  mechanism. An attractive feature 
of the above-described mechanism is that detonation occurs within $\sim$100\,s after merger and ejectum 
will not interact with any significant amount of the matter lost from the disk and produce additional radiation in 
the early light-curve which may be misinterpreted in favour of the presence of a non-compact progenitor
\citep{2015MNRAS.447.2803L}.

Note that at difference to the simulations of \sne\ by some other authors, 
\citet{2012MNRAS.422.2417D,2014MNRAS.438...14D,2014ApJ...785..105M,2015ApJ...807..105S,0004-637X-821-1-67} took as initial 
models for exploration of possibility of \sna\ the ones obtained by computations of WD merger, instead of 
taking an equilibrium hydrostatic model or a model, obtained by the accretion of He onto CO WD.  
It is assumed initially that WDs rotate synchronously 
\citep{2012MNRAS.421..426F,2013MNRAS.433..332B}. In the case of non-synchronous rotation, disruption of the
lower-mass WD occurs on shorter time-scale ($\sim$10\,\porb) and is more 'violent'  \citep{2011ApJ...737...89D}.
In the case of non-synchronous initial rotation, nuclear burning starts at larger densities and it is more probable 
that \sna\ occurs (e.g., \citet{pakmor_violent12,2013ApJ...770L...8P}).
As noted by \citet{2014MNRAS.438...14D}, it is likely that  in the latter case detonation happens in 
the centre of the merger product, while in the former case, at the core surface. As a result, there may be 
differences in the amount of unburned matter, ejectum velocity and ejectum  asymmetry. However, the issue of initial 
conditions still remains controversial.

A special case is the still badly explored mergers involving ONe WD. If the disrupted dwarf is He-rich, 
one may expect He detonation followed by core collapse, if the remaining mass exceeds \mch. Such an 
SN would be classified as a SN~Ib \citep[e. g., ][]{2006A&A...450..345K}. 
Whether He-detonation can in this case robustly trigger close-to-centre detonation remains 
an unsolved problem \citep{2014ApJ...785...61S}. 
\citet{2015A&A...580A.118M}
speculated that thermonuclear runaway in the core of ONe WD \textit{may} be triggered externally
and explored such a detonation. They concluded that observationally such explosions will be 
quite similar to \sne\ produced by detonations of similar mass CO~WD.   

To summarize, a clear apprehension of the nature of progenitors of \sne\ in the 
DD-scenario and processes occurring during the merger of WDs and in the post-merger stage is currently lacking.

\section{The model}
\label{sec:model}

\begin{figure}                 
\hspace{-0.8cm}
\includegraphics[width=0.54\textwidth,trim= 0. 0. 0. 1.5cm, clip]{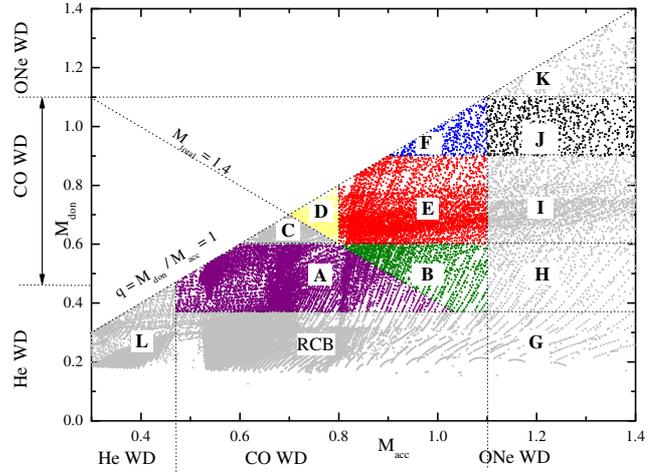}
\caption{Breakdown of the \mamd\ diagram into subregions. See the text for a detailed description. 
The density of dots in the particular regions of the diagram is proportional to the 
number of pairs of WDs that merge over 10~Gyr after an instantaneous star-formation 
burst for simulation with Z=0.02, \al=2, and tidal effects taken into account.}
\label{fig:danmap}
\end{figure}

For clarity, we reproduce in Fig.~\ref{fig:danmap} slightly modified \mamd\ diagram for  merging WDs
\citep{2014MNRAS.438...14D} which we split into 13 subregions (zones) where the pairs of WD
of different mass and chemical composition merge and different outcomes of merger are possible, 
including \sne.  We set the lower mass limit for possibly detonating 
He WD at 0.37\,\ms; above this limit detonations in the stream and  at the contact are possible 
\citep{2011ApJ...737...89D,2012MNRAS.422.2417D}, though, according to \citet{2013ApJ...771...14H},   
virialised He~WD even with $M_{He} \geq 0.24$\,\ms\ may detonate. 
Despite the fact that maximum mass of degenerate helium cores of stars slightly decreases with stellar mass 
\citep{1978ApJS...36..405S}, we use  the `canonical' value  of 0.47\,\ms\ as the lower limit for
the masses of CO WD produced by AGB-stars. We consider also `hybrid' WD. Like for  He WD, the lower
boundary of the mass range of exploding hybrid WD is set to 0.37\,\ms.

Since we aim to evaluate  the upper limit
of  \sne\ rate provided by merging of WDs, we consider as \sne\ all mergers of CO and He or hybrid  WD 
in which at least one detonation event is expected, though, as it is shown in the previous section, 
this issue is still open.    
it is unsolved, whether all such events will be identified with \sne. 

We expect the following events in the zones marked in Fig.~\ref{fig:danmap}.\\

{\bf A.} Merger of He or hybrid WD with a CO WD resulting in He detonation in the envelope 
and possible subsequent detonation of carbon in the core.\\
{\bf B.} Merger of He or hybrid WD and CO WD resulting in detonation of both WD.\\ 
{\bf C.} Merger of CO WDs with $\mt < \mch$ with the formation of a massive single WD.  \\
{\bf D.} Merger of comparable mass CO WDs with an explosion in the post-merger process.\\
{\bf E.} Merger of massive CO WDs leading to \sna\ in the post-merger stage.\\
{\bf F.} Merger of  similar mass CO WDs with an \sna\ explosion in the merger process.\\
{\bf G.} Merger of a massive He WD or a hybrid WD with ONe WD with subsequent (AIC). \\
{\bf H.} Merger of an ONe WD with a hybrid WD with subsequent AIC. \\
{\bf I.} Merger of a CO and an ONe WD with subsequent AIC of the ONe core.\\
{\bf J.} Violent merger of an ONe WD with a CO WD resulting in C detonation with subsequent detonation of the ONe WD
\citep{2015A&A...580A.118M}.  \citet{2012MNRAS.422.2417D} have shown that C-transferring systems do not detonate at contact.\\ 
{\bf K.} Merger of ONe WDs. 
Thermonuclear explosion is possible, if the envelopes contain He. Such an event may be classified  
rather as a SN~Ib than a weak \sna. \\
{\bf L.} Merger of low-mass He or hybrid WD leading to formation of helium subdwarfs. No \sna\ are produced.\\
{\bf RCB.} Merger of He and CO WDs in which the necessary conditions for He detonation are not met. Hypothetically, such mergers 
may lead to the formation of R~CrB stars \citep{web84,ity96}; evolution of the latter is defined by competition of 
core growth due to He-shell
 burning and mass loss by stellar wind. The growth of the core mass to \mch\ and explosion are not very 
likely, but if they happen, the event would probably be observed as a peculiar SN~Ib.

In assigning \sna\ status to the outcome of merger, we used, primarily, numerical results and qualitative
considerations presented in \citet{2014MNRAS.438...14D,2015MNRAS.454.4411D}, as well as data from literature 
discussed in \S~\ref{sec:merger}. Accordingly, we consider as \sne\ results of mergers occuring in the zones  
{\bf A}, {\bf B}, {\bf D}, {\bf E}, {\bf F}, and {\bf J}, although we clearly recognize that all 
of them are still hypothetical. We note the following.  
If detonation of He in the pairs merging in the zones \za\ and \zb\ does not initiate CO-core detonation, 
He explosions may be still identified with subluminous \sne\ \citep{2010ApJ...715..767S,2011ApJ...738...21W}. 
In zone {\bf C} mergers  do not result in \sne\ according to Eq.~(\ref{eq:sato16}); furthermore,  
$M_{\rm acc}$ should exceed about 0.8\,\ms.
In zone {\bf D} undisrupted CO WD accumulates \mch\ via self-regulating accretion in 
$\sim 10^6$\,yr  \citep{piersanti+03}.
To mergers in zones {\bf E} and {\bf F} the status of \sna\ is assigned according to results of
\citet{2015ApJ...807..105S} and \citet{2014ApJ...785..105M,2015ApJ...800L...7K}, respectively.
As concerns possible explosions of ONe WD, \citet{2015A&A...580A.118M} have shown that the detonations 
of these objects are possible, given that there is an external trigger. Appropriate computations are absent, but it is clear 
that the `trigger' dwarf has to be massive or such events may not happen at all. 
Thus, our limiting $M_{\rm don}=0.9$\,\ms\ may be too optimistic. Note, however, that the 
contribution of zone {\bf~J} to the rate of \sne\ in the case \al=2, which 
we consider as giving the  best agreement  with observations, does not ecxeed 10 per cent 
(see \S~\ref{sec:res}).      

\subsection{Population synthesis}
\label{sec:bps}
\noindent

For the modelling of the population of merging WDs, we applied the
code BSE \citep{hpt00,htp02,2008MNRAS.388..393K}.
The advantage of BSE is that it is based on a homogeneous system of evolutionary tracks for single stars  
with metallicity from ${\mathrm Z} = 0.0001$\ to  0.02 \citep{pol+98},
which  was used to construct analytical approximations describing stellar evolution. 
The shortcoming of BSE is that the evolution of binaries is treated on the base of assumptions originating
from the work of other authors or (educated) guesses for never calculated evolutionary transitions.

Our results are based on the modelling of $10^7$ initial systems and for each set of initial parameters represent 
one realization of the model. Hence, they are subject to Poisson noise.  

Taking into account recent suggestions on the dependence of the binarity rate on the mass of stars, we 
approximated this rate as \citep{2013A&A...552A..69V} 
\begin{equation}
\label{eq:bin_fra}
f_{\rm b} = 0.50+0.25\log(M_{1}/\ms).
\end{equation} 
 
\textit{Common envelopes.}
The problem of common envelopes (CE) is the most acute one in the theory of  evolution of close binaries; 
see \citet{2013A&ARv..21...59I} for a recent detailed discussion. In the context of the problem we address in 
the present study,
CE  form when accretor is not able to swallow all the matter supplied by the donor or the system is subject 
to the Darwin instability.  

There is no  unanimous opinion on the form of the equation describing the evolution of stars in the CE.
We apply an equation suggested by \citet{tap+97}, included in BSE as an option, because in  our opinion,
it better accounts the expenditure of energy on the expulsion of the CE than the
  `standard' \citep{web84,dek90} equation, if both components have clear-cut cores and envelopes. 
However, test
runs with `standard' equation show that the difference in the rates of \sne\ does not exceed 10 per cent.
The basic problem of both formulations of the CE equation is that its solution for the ratio of the initial and 
final separations of the components, upon which it is decided whether the components merged, depends on the product 
of two parameters -- $\alpha_{\rm ce}\lambda$. The `common envelope efficiency' $\alpha{\rm ce}$ describes 
the efficiency with which the spiralling-in cores of components transfer
orbital energy to the envelope and disperse the latter; it is expressed in
fractions of the orbital energy of the system. Parameter $\lambda$
characterizes the binding energy of the donor envelope.

A problem related to the  outcome of the CE was noted by \citet{2011MNRAS.417.1466K}: while
CE equation(s)  may formally imply that the system remains detached at the end of the CE stage, in fact
some matter of the envelope may not reach escape velocity and remains bound to the system and
form a circumbinary disc. Angular momentum loss owing to interaction with the disk may result in
further reduction of the binary separation and the merger of components. This may influence, among  other 
populations, \sne, but this problem remains unsolved. On the other hand, the influence of this phenomenon 
is partially compensated by uncertainty in the treatment of CEs in general.

Both parameters describing the CE are still highly uncertain. For \ace,\ the most important issue is whether there are other sources than
orbital energy for expulsion of CEs, that is, whether $\ace > 1$\ is possible. In fact, \ace\ is a specific
parameter of any CE. The  use of it is forced, because for a given system with components at different evolutionary
stages and particular combination of $M_1$, $M_2$, $a$, the outcome of the CE may be found only by 3D hydrodynamical calculations, which
still have insufficient resolution or do not account for all physical processes occurring in CEs on different time scales \citep{2016ApJ...816L...9O}. 
It is evident that $\lambda$ should continuously vary in the course of stellar evolution; position of the core-envelope boundary
remains uncertain -- the value of $\lambda$ may be uncertain by a factor of 10 depending on the definition 
of the core-envelope boundary \citep{td01}. Regarding  `observational' estimates, one should bear in mind
severe selection effects which may restrict the observed population to fractions of per cent of the intrinsic one,
resulting in poor statistics \citep[see e. g.][]{2014A&A...566A..86C}.   

Nevertheless, we consider it justified to use as a constant parameter 
the value(s) of \ace\ found for sufficiently representative 
samples of close binaries that evolved through the CE stage or were derived by the study of particular binaries. However, it appears,
for instance, that for low-mass systems -- post-CE binaries  (PCEB) with $M_2 \aplt 0.8$\,\ms\ from SDSS -- 
the estimates  
of \ace\ range (within error bars) between 0.02 and 10 \citep{2010A&A...520A..86Z}. Within this range, for the 
systems with AGB progenitors ,\ace\ values cluster below 0.2, while for those with FGB progenitors almost 
all \ace\  values are $\apgt 0.2$. 
Thus, the quite common claim that $\ace \approx 0.2 - 0.3$ seems to be not well justified.
The population of well-studied binary WDs, after accounting for observational selection, is rather well reproduced
with \al=2 \citep{nyp+01,2012A&A...546A..70T}. Therefore, we performed simulations for \al\ in the range
from 0.25 to 2, but consider the version with \al=2 as the main one; see \S~\ref{sec:res}. In a similar 
way,  in their studies of \sne, \citet{2010A&A...515A..89M} parametrized CEs by 
\al=1, while \citet{2012A&A...546A..70T,2014A&A...562A..14T} used \al=2.

\textit{Stability of mass-exchange.} An under-researched topic in binary star evolution is the stability of mass loss by 
stars with deep convective envelopes and He-stars.\\
For hydrogen-rich stars we kept the stability criteria accepted in BSE. For helium stars -- the remnants of components of binaries with mass $\apgt(2.0 - 2.5)$\,\ms, which experienced RLOF in the 
hydrogen-shell burning stage (case B of mass-exchange), stability criteria were modified. 
\citet{pac_he71,it85} have shown that He stars with masses below (0.8 -- 0.9)\,\ms\ do not 
expand after He exhaustion in their cores and evolve straight into hybrid WD. More massive He stars expand in 
the He-shell-burning stage; the extent of the expansion depends on the mass of the star and RLOF may restart if 
the components are close enough.  

Based on trial computations of semidetached binaries with He donors (D. Kolesnikov et al., \textit{in prep.}) 
we consider mass loss unstable if stellar radius becomes $\geq 5$\,\rs\ and $q\geq 0.78$.
If $R_{He}$ remains below 5\,\rs, the critical $q=3.5$.  

Angular momentum loss via gravitational waves radiation may bring low-mass He stars
into contact with WD prior to He exhaustion in their cores. In the systems where conditions for stable mass exchange are fulfilled,
initially, $\mdot \sim 10^{-8}$\,\myr\ \citep{skh86,it91,2008AstL...34..620Y}. At these \mdot, surface
detonation of He is possible \citep{taa80,nom82b}. We exclude such systems from further consideration,
because they are not of interest in the present study. Such binaries are suggested to be progenitors of double-
detonation \sna, but the real efficiency of this
scenario is under debate \citep{2014MNRAS.445.3239P,2015MNRAS.452.2897P,2015ApJ...807...74B}.
 
For WDs, the stability of mass exchange depends on the mass ratio of components and efficiency of spin/orbit coupling 
\citep{npv+01,mns04}. As well, it is possible that tidal interaction  allows stable mass transfer, but the rate of the latter
is super-Eddington and a CE  may  form. We considered mass transfer between WDs as stable,
irrespective of the efficiency of spin/orbit coupling,  if its rate
remains sub-Eddington and $M_{\rm don}$ (in solar units) satisfies the following approximation  
based on fig. 1 of \citet{2011ApJ...737...89D}:
\begin{equation}
M_{\rm don} \leq \begin{cases}
 0.2286\,M_{\rm acc},       & \text{if~} M_{\rm acc} \leq 0.875; \cr 
 0.133\,M_{\rm acc}+0.0833, & \text{if~} 0.875 < M_{\rm acc} \leq 1.1; \cr
-0.033\,M_{\rm acc}+0.2667, & \text{if~} 1.1 <M_{\rm acc} \leq 1.44. \cr
\end{cases}
\end{equation}

\textit{Tidal effects}. Tidal effects are significant predominantly for giant stars. The main evolutionary consequence
of tides is earlier RLOF, if the tides are taken into account. For instance, RLOF in RG-stage instead of AGB-stage
may occur and a He or hybrid WD to form instead of a CO WD.  On the other hand, for later RLOF, formation of 
a CE and merger of components inside them is more probable. 
There is no consensus on the account of tides in binary population synthesis (BPS); see 
comparison of assumptions in table~1 of \citet{2014A&A...562A..14T}. For instance, they were not taken into account 
in the latest simulations of \citet{2014A&A...563A..83C}.  We used the algorithm for the account
of circularisation of the orbits and synchronization of rotation  as it is incorporated in BSE:
according to \citet{1975A&A....41..329Z} for stars with radiative envelopes and according to 
\citet{1981A&A....99..126H} for stars with convective envelopes.

We implemented in the code modifications to the algorithm of computation of mass transfer rate
that make it more stable \citep[][Eq.~(11)]{2014A&A...563A..83C}. 
   
\textit{Star-formation rate.}  
For the estimate of \sne\ rate in the Galaxy, 
we accepted that the star-formation rate ($SFR$) in the bulge and thin Galactic disk can be described by a function
combining exponentially declining and slow constant components 
\citep{2010A&A...521A..85Y}: 

\begin{equation}
SFR(t)=11\exp(-(t-t_0)/\tau)+0.12(t-t_0)\myr~\text{for}~t>t_0.
\label{eq:sfr}
\end{equation}
Here, time $t$ is in Gyr, $\tau$=9~Gyr, $t_0$=4~Gyr, Galactic age is 14~Gyr. We neglect halo 
and thick Galactic disk stars with total mass of only several per cent of the mass of the bulge and thin disk. In 
the bulge and thin disk $SFR(t)=0$ at $t \leq t_0$.   
Current Galactic $SFR$=4.82\,\myr, well within the range of modern estimates -- from 
 $\sim1$ to $\sim10$\,\myr\ \citep{gilm00}. The total mass of the bulge and the disk is then   
$7.2 \times10^{10}$\,\ms, close to the estimate of \citet{2002ApJ...573..597K} --  $7.0\times10^{10}$\,\ms. \\

We neglect in the models stellar winds of He stars, because 
extrapolation of the rates for WR stars or Reimers-type winds assumed in BSE 
is not justified and, even then, winds are  hardly  evolutionary meaningful, because of 
extremely short lifetimes of He stars.  
 
\section{Results}
\label{sec:res} 

\subsection{Scenarii for the formation of merging WDs}
\label{sec:scen}

Population synthesis calculations generate hundreds of evolutionary scenarios.
However,  the dominant fraction of merging WDs of interest usually  form,  via
 only a few channels, differing mainly by the stage in which initially more massive component overflows 
its Roche lobe.  Despite these scenarii were analysed many times, starting from semi-analytical studies by 
\citet{ty81,it84a,web84} and,  most recently,  by for example 
\citet{2009ApJ...699.2026R,2010A&A...515A..89M,2014A&A...562A..14T,2014A&A...563A..83C}, they still deserve 
attention.
 
The fraction of stars evolving via certain channels depends, mainly, on the accepted criteria of 
the stability of mass loss, the treatment of CE stage(s), the  mass and momentum loss from the system,
treatment of accretion onto compact stars and criteria according to which the stars are considered as `merged'.
For instance, compare the simulations of \citet{2010A&A...515A..89M} and \citet{2012A&A...546A..70T} which differ in 
the assumed radii of stellar remnants and \al\ values -- 1 in  \citeauthor{2010A&A...515A..89M} and 
2 in \citeauthor{2012A&A...546A..70T}. In the first study,  80 per cent of CO WD 
and CO cores of donors merge in the CE, while in the second one only 45 per cent do.  

\begin{table*}
\centering
\caption{Main evolutionary scenarios leading to formation of merging WD and possible \sna.
First row: number of the sequence;
second row:  evolutionary stage preceeding the first common envelope episode. 
Brace by component identifier mean that this star overflows Roche lobe. 
Absence of brace means that the system is detached.
The stages with similar outcome are joined by horizontal braces. 
`CE1' and `CE2' denote the first and the second common envelope episodes, respectively.
For the rest of the notation see the text. Down-arrow symbols indicate that in a particular scenario 
certain evolutionary stages are absent.\vspace{-1mm} }
\tabcolsep 0.0 mm          
{\scriptsize
\begin{tabular}{lccclccp{1.2cm}cl}
\hline
& 1 & \multicolumn{3}{c}{2}       & \multicolumn{3}{c}{3}         & \multicolumn{2}{c||}{4}	\\ 
\hline																	
& COWD+HG & \multicolumn{3}{c}{COWD+GB}  & \multicolumn{3}{c}{COWD+EAGB} & \multicolumn{2}{c}{HeMS+GB} 	\\ 	
\hline															
1 & MS+MS & \multicolumn{3}{c}{MS+MS}     & \multicolumn{3}{c}{MS+MS} & \multicolumn{2}{c||}{MS+MS} \\ 
2 & HG+MS & \multicolumn{2}{c}{$\swarrow$} & $\searrow$ & \multicolumn{3}{c}{HG+MS} & \multicolumn{2}{c}{HG+MS} \\ 
3 & \{HG+MS	& \multicolumn{2}{c}{\{HG+MS} & HG+MS                     &\multicolumn{3}{c}{\{HG+MS}  & \multicolumn{2}{c}{\{HG+MS}       \\ 
4&        & \multicolumn{2}{c}{\{GB+MS} & \{GB+MS                   & \multicolumn{3}{c}{\{GB+MS} & \multicolumn{2}{c}{\{GB+MS} \\ 
5&	  \{GB+MS	                 & \multicolumn{3}{c|}{\hspace{3em}$\underbrace{~~~~~~~~~~~~~~~~~~~~~~~~~~~~~~~~~~~~~~}_{}$}& & 
$\swarrow$~$~~~~~~~~~~~~~~\searrow$ &   & \multicolumn{2}{c}{$\Downarrow$}  \\
6& HeMS+MS	& \multicolumn{3}{c|}{\hspace{2em}HeMS+MS}	 & HeMS+MS & & \{GB+MS     &\multicolumn{2}{c}{HeMS+MS}  	\\ 
7 & COHe+MS & \multicolumn{2}{c}{\hspace{4em}$\swarrow$} & $\hspace{-2em}\searrow$ & $~~~~~\downarrow$~~~~~~~~~$\searrow$ & & \hspace{-.5cm}
$\swarrow$  &\multicolumn{2}{c}{HeMS+HG} \\ 
8 & \{COHe+MS & \multicolumn{2}{c}{\{COHe+MS} & \hspace{-4em}COHe+MS  & \{COHe+MS	& HeMS+MS  & \hspace{1.32em}$\Downarrow$& 	$\hspace{2em}\swarrow$ & 
$\hspace{2em}\searrow$	\\
9 &      	& \multicolumn{2}{c}{\{COHe+MS}  &           &   COHe+MS      &           &\hspace{1em}  $\Downarrow$     &\{COHe+GB & HeMS+GB    	\\
10 & & \multicolumn{3}{c|}{\hspace{1em}$\underbrace{~~~~~~~~~~~~~~~~~~~~~~~~~~~~~~~~~~~~~~~}_{}$} & 
\multicolumn{3}{c}{$\underbrace{~~~~~~~~~~~~~~~~~~~~~~~~~~~~~~~~~~~~~~~~~~~~}$} & & HeMS+GB\}	\\ 
[-6mm] 
\multicolumn{8}{c}{ } & \multicolumn{2}{c}{\hspace{-2em}$\underbrace{~~~~~~~~~~~~~~~~~~~~~~~~~~~~~~~~~~~}$}  \\
11 &	COHe+MS	&\multicolumn{3}{c|}{}&\multicolumn{3}{c}{} &\multicolumn{2}{c}{}		\\
12 &	COWD+MS	&	\multicolumn{3}{c|}{COWD+MS}	&\multicolumn{3}{c}{COWD+MS} & \multicolumn{2}{c||}{\textbf{CE1}} \\	
13 &	COWD+HG &	\multicolumn{3}{c|}{COWD+HG}	&	\multicolumn{3}{c}{COWD+HG} &  &\\
14 &	COWD+HG\} &	\multicolumn{3}{c|}{COWD+GB}	&	\multicolumn{3}{c}{COWD+GB} & \multicolumn{2} {c}{HeMS+HeMS}\\	
15 &	&	\multicolumn{3}{c|}{COWD+GB\}} &	\multicolumn{3}{c}{COWD+CHeB} &  &\\
16 &		&	\multicolumn{3}{c|}{	 }      &	\multicolumn{3}{c} {		 COWD+EAGB			} &\multicolumn{2} {c}{COHe+HeMS}\\
17 &		&	\multicolumn{3}{c|}{}	        &\multicolumn{3}{c}{COWD+EAGB\}}&&\\		 
18 & \textbf{CE} &	\multicolumn{3}{c|}{\textbf{CE1}}&\multicolumn{3}{c}{\textbf{CE1}}&\multicolumn{2} {c}{COWD+HeMS}\\	 
19 &		&	\multicolumn{1}{c} {$~~~~~~~~~~~\swarrow$} &\multicolumn{2}{c|}{$~~~~\searrow$} &\multicolumn{3}{c}{}& &\\
20 & COWD+HeMS  & COWD+HeWD& \multicolumn{2}{c|}{COWD+HeMS} & \multicolumn{3}{c}{} &  & \\
21 & COWD+COHe& & \multicolumn{2}{c|}{COWD+COHe} & \multicolumn{3}{c}{COWD+COHe}&\multicolumn{2}{c|}{COWD+COHe}~\\
22 &	&	& $~~~\swarrow$ & $~~~~~\searrow$ &	\multicolumn{2}{c} {$\swarrow$	} &\hspace{-2em}$\searrow$ &\multicolumn{2}{c|}{COWD+COHe\}}\\	
23 & COWD+ COHe\}	&	$\Downarrow$ & COWD+COHe~&~COWD+COHe\}& \multicolumn{3}{c}{COWD+COHe\}~~~ COWD+COHe} &$~~~~~~~\swarrow$ & $~~~~~\searrow$      \\
24 &  & & & \multicolumn{1}{c|}{\textbf{CE2}}&\multicolumn{2}{c}{\textbf{CE2}}& &~~~ $\Downarrow$ &~~~~~~~~~\textbf{CE2}\\		
25 &$\Downarrow$ & $\Downarrow$ & \multicolumn{2}{c|}{$\underbrace{~~~~~~~~~~~~~~~~~~~~~~~~~~~~~~~~~~~~~~~}_{}$} & 
\multicolumn{3}{c}{$\underbrace{~~~~~~~~~~~~~~~~~~~~~~~~~~~~~~~~~~~~~~~}_{}$} & &\\[-1em]
26 &	&	& \multicolumn{2}{c|} {	$\Downarrow$}	& \multicolumn{3}{c} {	$\Downarrow$}& & \\
27 & COWD+COWD	& COWD+HeWD	& \multicolumn{2}{c|}{COWD+COWD} & \multicolumn{3}{c}{COWD+COWD} &~COWD+COWD~~ &~~COWD+ONeWD\\
28 &\{COWD+COWD & COWD+HeWD\}&\multicolumn{2}{c|}{COWD+COWD\}} & \multicolumn{3}{c}{~COWD+COWD\}}&\{COWD+COWD~~&~~COWD+ONeWD\}\\

29 & \multicolumn{9}{c}{$\underbrace{~~~~~~~~~~~~~~~~~~~~~~~~~~~~~~~~~~~~~~~~~~~~~~~~~~~~~~~~~~~~~~~~~~~~~~~~~~~~~~~~~~~~~~~~~~~~~~~~~~~~~~~~~~~~~~~~~~~~~~~~~~~~~~~~~~~~~~~~~~~~~~~~~~~~~~~~~~~~~~~~~~~~~~~~~~~~~~~~~~~~~~~} $}\\
30 & \multicolumn{9}{c}{\bf Merger}\\
\hline\hline

\hline\hline
	&	5	&	6	&	~~~~~~~~~~~~~~~~	&	\multicolumn{2}{c}	{	7	}		&	~~~~~~~	&	\multicolumn{3}{c}	{	8	}		\\[-1mm]	
\hline																				
																																							
	&	       EAGB+ CHeB	&	      TPAGB+ CHeB	&	~~~~~~	&	\multicolumn{2}{c}	{	       EAGB+MS	}		&	~~~~~~~	&	\multicolumn{3}{c}	{	      TPAGB+MS	}		\\
\hline																		
1	&	\multicolumn{2}{c||} {		        MS+MS	}&	~~~~~~~~~~~~~~~~	&	\multicolumn{2}{c}	{MS+MS}		&	~~~~~~~	&	\multicolumn{3}{c}	{	       MS+MS	}		\\
2	&	\multicolumn{2}{c||}{HG+MS}&	~~~~~~~~~~~~~~~~	&	\multicolumn{2}{c}	{	        HG+MS	}		&	~~~~~~~	&	\multicolumn{3}{c}	{	        HG+MS	}		\\
3	&	\multicolumn{2}{c||} {			}&	~~~~~~~~~~~~~~~~	&	\multicolumn{2}{c}	{	        GB+MS	}		&	~~~~~~~	&	\multicolumn{3}{c}	{	        GB+MS	}		\\
4	&	\multicolumn{2}{c||} {			}&	~~~~~~~~~~~~~~~~	&	\multicolumn{2}{c}	{	      CHeB+MS	}		&	~~~~~~~	&	\multicolumn{3}{c}	{	      CHeB+MS	}		\\
5	&	$\swarrow$	&	$\searrow$	&	~~~~~~~~~~~~~~~~	&	\multicolumn{2}{c}	{	      EAGB+MS	}		&	~~~~~~~	&	\multicolumn{3}{c}	{	      EAGB+MS	}		\\
6	&	         GB+MS	&	         HG+HG	&	~~~~~~~~~~~~~~~~	&				&		&	~~~~~~~	&		&		&		\\
7	&		&		&	~~~~~~~~~~~~~~~~	&	\multicolumn{2}{c}{\{EAGB+MS}&	~~~~~~~	&	\multicolumn{3}{c}	{	      TPAGB+MS	}		\\
8	&	\multicolumn{2}{c||} {		$\underbrace{ ~~~~~~~~~~~~~~~~~~~~~~~~~~~~~~~~~~~~~~~~~~~~~~~~~~~~~~~~    }_{}$	}&	~~~~~~~~~~~~~~~~	&	\multicolumn{2}{c}	{		}		&	~~~~~~~	&	\multicolumn{3}{c}{\{TPAGB+MS}		\\
9	&		&		&	~~~~~~~~~~~~~~~~	&	\multicolumn{2}{c}	{	 \textbf{CE1} 	}		&	~~~~~~~	&	\multicolumn{3}{c}	{	 \textbf{CE1} 	}		\\
10	&	\multicolumn{2}{c||}{GB+HG}&	~~~~~~~~~~~~~~~~	&	\multicolumn{2}{c}	{	      COHe+MS	}		&	~~~~~~~	&		&		&		\\
11	&	\multicolumn{2}{c||} {GB+GB}&	~~~~~~~~~~~~~~~~	&			$~~~~~~~~~~\swarrow$	&	$\hspace{-0.5cm}\searrow$	&	~~~~~~~	&	\multicolumn{3}{c}	{	$\Downarrow$	}		\\
12	&		&		&	~~~~~~~~~~~~~~~~	&			      \{COHe+MS	&		&	~~~~~~~	&		&		&		\\
13	&		&		&	~~~~~~~~~~~~~~~~	&			&	$\Downarrow$	&	~~~~~~~	&	\multicolumn{3}{c}	{	$\Downarrow$	}		\\
14	&	\multicolumn{2}{c||}{CHeB+GB	}&	~~~~~~~~~~~~~~~~	&			      COHe+MS	&		&	~~~~~~~	&		&		&		\\
15	&	\multicolumn{2}{c||}{CHeB+CHeB}&	~~~~~~~~~~~~~~~~	&			      COWD+MS	&		&	~~~~~~~	&	\multicolumn{3}{c}	{	      COWD+MS	}		\\
16	&	$\swarrow$	&	$\searrow$	&	~~~~~~~~~~~~~~~~	&			      COWD+HG	&	$\Downarrow$	&	~~~~~~~	&	\multicolumn{3}{c}	{	      COWD+HG}		\\
17	&	      EAGB+CHeB	&	      EAGB+CHeB	&	~~~~~~~~~~~~~~~~	&				&		&	~~~~~~~	&		&		&		\\
18	&	   \{EAGB+CHeB	&  TPAGB+CHeB	& & & & & $ \hspace{5em}\swarrow$ & \hspace{1em}$\downarrow$	& $\searrow$	\\
19	&		&		&	~~~~~~~~~~~~~~~~	&			      COWD+HG\}	&		&	~~~~~~~	&	      COWD+HG\}	&		&		\\
20	&		&		&	~~~~~~~~~~~~~~~~	&				&	      COWD+GB\}	&	~~~~~~~	&		&	 COWD+GB\}	&	      COWD+GB	\\
21	&		&	    \{TPAGB+CHeB	&		&				&		&	&		&		&	      COWD+CHeB	\\
22	&		&		&	~~~~~~~~~~~~~~~~	&				&		&	~~~~~~~	&\hspace{0.5cm}$\Downarrow$		&	$\Downarrow$	&	      COWD+EAGB	\\
23	&		&		&	~~~~~~~~~~~~~~~~	&				&		&	~~~~~~~	&		&		&	      COWD+EAGB\}	\\
24	& \multicolumn{2}{c}{$\underbrace{~~~~~~~~~~~~~~~~~~~~~~~~~~~~~~~~~~~~~~~~~~~~~~~~~~~~~~~~~}_{}$} & & \multicolumn{2}{c}{$\underbrace{~~~~~~~~~~~~~~~~~~~~~~~~~~~~~~~~~~~~~~~}_{}$} & & \multicolumn{3}{c}{$\underbrace{~~~~~~~~~~~~~~~~~~~~~~~~~~~~~~~~~~~~~~~~~~~~~~~~~~~}_{}$}\\[-0.5em]
25	&	\multicolumn{2}{c||} {		 \textbf{CE1} 	}&	~~~~~~~~~~~~~~~~	&	\multicolumn{2}{c}	{	 \textbf{CE2} 	}		&	~~~~~~~	&	\multicolumn{3}{c}	{	 \textbf{CE2} 	}		\\
26	&	\multicolumn{2}{c||} {			}&	~~~~~~~~~~~~~~~~	&				&		&	~~~~~~~	&	$~~~~~~~~~~~~~~~~~~~~~~~~~~~\swarrow$	&		&	$  \searrow$	\\
27	&	\multicolumn{2}{c||} {		      COWD+HeMS	}&	~~~~~~~~~~~~~~~~	&	\multicolumn{2}{c}	{	      COWD+HeMS	}		&	~~~~~~~	&	\multicolumn{2}{c}	{	      COWD+HeMS	}&		\\
28	&	\multicolumn{2}{c||}{COWD+COHe	}&	~~~~~~~~~~~~~~~~	&	\multicolumn{2}{c}	{	      COWD+COHe	}		&	~~~~~~~	&	\multicolumn{2}{c}	{		}&		\\
29	&		&		&	~~~~~~~~~~~~~~~~	&	\multicolumn{2}{c}	{	      COWD+COHe\}	}		&	~~~~~~~	&	\multicolumn{2}{c}	{	      COWD+COHe\}	}&	      COWD+COHe	\\
30	&	\multicolumn{9}{c}{\hspace{-1.5em}$\underbrace{~~~~~~~~~~~~~~~~~~~~~~~~~~~~~~~~~~~~~~~~~~~~~~~~~~~~~~~~~~~~~~~~~~~~~~~~~~~~~~~~~~~~~~~~~~~~~~~~~~~~~~~~~~~~~~~~~~~~~~~~~~~~~~~~~~~~~~~~~~~~~~~~~~~~~~~~~~~~~~~~~~~~~~~~~~~~~~~~~~~~~~~~}$} \\[-0.5mm]

31	&		&		&	~~~~~~~~~~~~~~~~	&	\multicolumn{2}{c}	{	      COWD+COWD	}		&	~~~~~~~	&	\multicolumn{3}{c}{}		\\
32	&		&		&	~~~~~~~~~~~~~~~~	&	\multicolumn{2}{c}	{	      COWD+COWD\}	}		&	~~~~~~~	&	\multicolumn{3}{c}	{		}		\\[-1mm]
33 &  \multicolumn{9}{c}{\hspace{-1.5em}$\underbrace{~~~~~~~~~~~~~~~~~~~~~~~~}$}\\[-0.5mm]
34 & \multicolumn{9}{c}{\hspace{-1.em}\bf Merger}\\
\hline
\end{tabular}
}
\label{tab:scen}
\end{table*}

Eight main scenarios, which in our simulations result in the formation of not less than about 90 per cent of all 
merging pairs of WDs,  considered as possible precursors of \sna\ are shown in Table~\ref{tab:scen}. 
The table corresponds to Z=0.02, \al=2 and tides are taken into account.
We apply with slight modification the notation accepted in BSE and widely used in the literature:
MS, main-sequence star, HG, Hertzsprung gap star, GB, first giant-branch star,
CHeB, a star with central He burning, EAGB and TPAGB, early and thermally-pulsating AGB stars, respectively,
HeMS, helium-burning remnant of a star, HeWD, COWD, ONeWD, helium, carbon-oxygen and
oxygen-neon white dwarfs, respectively. 
We consider He-shell-burning `helium Hertzsprung-gap' (HeHG in BSE) stars and
`helium-giants' (HeG in BSE) as similar objects with CO-cores and helium-burning shells, because
they differ only in the extent of expansion of He envelopes. They are identified in 
Table~\ref{tab:scen} and in the text as `COHe'. 
For the pairs of WD produced by a certain scenario, we indicate at pre-merger  stage the most 
common combination of components: for instance, scenario 1 is marked as forming a 
CO WD+CO WD pair, while in fact, about 25 per cent of pairs contain ONe components, owing to a `change 
of the roles' during RLOF when initially less-massive secondaries accumulate large mass; we do 
not consider pairs with ONe WD for computation of \sne\ rates etc.  In Appendix A, in 
Figs.~\ref{fig:scen1} -- \ref{fig:scen8}, we show for these scenarios positions of initial 
systems in $M_{1,i}-M_{2,i}$, $M_{1,i}-a_i$ and pre-merger \mamd\ diagrams. The plots are for 
the \al=2 case in which the mergers of WD occur most efficiently (see below).  

\begin{table*}    
\caption{Relative number of mergers, possible precursors of SNe Ia, occurring in particular zones of \mamd\ 
diagram over 10 Gyr, for various values of \al.
Column~(1), \al.
Column~(2), the number of scenario according to Table~\ref{tab:scen}. 
Column~(3), evolutionary stage of the system preceding the first CE. 
Columns~(4) to (9) with headers A to J indicate the relative numbers (in per cent) of WD pairs formed  via 
scenarii listed in column (2) and merging in particular zones of the \mamd\ diagram (Fig.~\ref{fig:danmap}) over 
10\,Gyr.
Column~(10), relative input of a particular scenario into the total rate of WD+WD mergers for a given \al. 
Column~(11), the ratio of the number of mergers of possible precursors of \sna\  and total number of merging WD 
pairs for every channel and \al\ value. The absence of data on some scenarios for certain \al\ 
means that for this \al\ the code does not generate such a scenario at all.
}
\footnotesize
\tabcolsep 3.2 mm
\begin{tabular}{ccccccccccc}
\hline
        & Sc.&	Stage	&  	\multicolumn{6}{c}{Zone of \mamd\ diagram}	&	SNIa  &$N_{\textrm{SNIa}}/$\\
 	\al	& &	 before CE1	&	A	&	B	&	D	&	E	&	F	&	J	&	 & 	$N_{\textrm{WD2}}$ \\
\hline																	(1)&(2)&(3)&(4)&(5)&(6)&(7)&(8)&(9)&(10)&(11)\\
\hline																					
\hspace{2mm}																					
0.25	&	2	&	COWD+GB	&	2.1	&	 -     	&	 -     	&	 -     	&	 -     	&	0.2	&	2.3	&	5.60E-03	\\
0.25	&	3	&	COWD+EAGB	&	31.8	&	12.3	&	1.3	&	39.4	&	1.3	&	8.1	&	94.1	&	2.30E-01	\\
0.25	&	4	&	HeMS+GB	&	 -     	&	 -     	&	 -     	&	 -     	&	 -     	&	0.6	&	0.6	&	1.40E-03	\\
0.25	&	5	&	EAGB+CHeB	&	 -     	&	 -     	&	 -     	&	 -     	&	 -     	&	0.1	&	0.1	&	1.60E-04	\\
\hline																					
0.25	&		&	sum	&	33.8	&	12.3	&	1.3	&	39.4	&	1.3	&	8.9	&	97	&	2.40E-01	\\
\hline																					
\hline																					
0.5	&	2	&	COWD+GB	&	1.4	&	0.2	&	 -     	&	1.4	&	 -     	&	2.8	&	5.8	&	9.20E-03	\\
0.5	&	3	&	COWD+EAGB	&	50.3	&	2.9	&	5.4	&	24.3	&	0.7	&	2.2	&	85.7	&	1.30E-01	\\
0.5	&	4	&	HeMS+GB	&	 -     	&	 -     	&	 -     	&	0.5	&	0.2	&	0.9	&	1.6	&	2.60E-03	\\
0.5	&	5	&	EAGB+CHeB	&	0.8	&	 -     	&	 -     	&	 -     	&	 -     	&	0.2	&	1	&	1.60E-03	\\
0.5	&	6	&	TPAGB+CHeB	&	0.2	&	 -     	&	 -     	&	 -     	&	 -     	&	0.02	&	0.3	&	4.20E-04	\\
\hline																					
0.5	&		&	sum	&	52.7	&	3.1	&	5.4	&	26.2	&	0.9	&	6.3	&	94.5	&	1.50E-01	\\
\hline																					
\hline																					
1	&	1	&	COWD+HG	&	 -     	&	 -     	&	 -     	&	 -     	&	 -     	&	0.2	&	0.2	&	2.10E-04	\\
1	&	2	&	COWD+GB	&	5.7	&	1.3	&	2.4	&	10.2	&	0.1	&	3.2	&	23	&	2.50E-02	\\
1	&	3	&	COWD+EAGB	&	23.4	&	0.4	&	2.9	&	6.8	&	0.6	&	2	&	36.1	&	4.00E-02	\\
1	&	4	&	HeMS+GB	&	5.1	&	1.1	&	3	&	14.2	&	0.1	&	 -     	&	23.4	&	2.60E-02	\\
1	&	5	&	EAGB+CHeB	&	4.5	&	 -     	&	0.01	&	0.04	&	0.04	&	0.3	&	4.8	&	5.40E-03	\\
1	&	6	&	TPAGB+CHeB	&	2.4	&	0.1	&	0.1	&	1.4	&	 -     	&	0.03	&	4	&	4.40E-03	\\
\hline																					
1	&		&	sum	&	41	&	2.9	&	8.5	&	32.6	&	0.8	&	5.7	&	91.5	&	1.00E-01	\\
\hline																					
\hline																					
2	&	1	&	COWD+HG	&	 -     	&	 -     	&	0.2	&	0.5	&	0.4	&	1.1	&	2.1	&	5.00E-03	\\
2	&	2	&	COWD+GB	&	9.2	&	0.5	&	1.7	&	5.4	&	1.2	&	1.7	&	19.7	&	4.60E-02	\\
2	&	3	&	COWD+EAGB	&	2.9	&	 -     	&	0.6	&	1.3	&	0.2	&	0.5	&	5.5	&	1.30E-02	\\
2	&	4	&	HeMS+GB	&	16.6	&	0.3	&	4.8	&	7.1	&	0.004	&	 -     	&	28.8	&	6.80E-02	\\
2	&	5	&	EAGB+CHeB	&	5.8	&	 -     	&	0.5	&	0.5	&	0.02	&	0.1	&	7	&	1.60E-02	\\
2	&	6	&	TPAGB+CHeB	&	2.8	&	0.04	&	0.2	&	0.8	&	 -     	&	0.02	&	3.9	&	9.20E-03	\\
2	&	7	&	EAGB+MS	&	2.7	&	2.2	&	 -     	&	5	&	 -     	&	 -     	&	9.9	&	2.30E-02	\\
2	&	8	&	TPAGB+MS	&	5.6	&	3.3	&	0.04	&	2.1	&	 -     	&	 -     	&	10.9	&	2.60E-02	\\
\hline																					
2	&		&	sum	&	45.6	&	6.3	&	7.9	&	22.9	&	1.8	&	3.3	&	87.8	&	2.10E-01	\\
\end{tabular}
\label{tab:scen_frac}
\end{table*}

In scenarii 1 to 4, the first CO WD forms via stable RLOF starting in the Hertzsprung gap or in the first RG-branch 
(case B of mass exchange). This results in the formation of a 
He star. As already mentioned,  the latter may evolve straight into a CO WD, if their mass is $\aplt(0.8 - 0.9)$\,\ms\
or refill Roche lobe and lose some mass  (in small-separation systems)  for a short time when the
 star expands on the thermal time scale in the helium-shell burning stage. Because the first RLOF is stable, 
the secondaries typically accumulate mass larger than $\approx 2\,\ms$ and after mass loss become He-stars.   
He WD appear only in the side branch of scenario 2 after the CE, if  $M_2 \aplt 2.25$\,\ms.
Most merging CO WD+He WD pairs are actually formed by several scenarii that are not among the most
prolific ones\footnote{In BSE,  as He WD are labelled also descendants of He-burning stars which fill Roche lobes and, 
because of stable mass loss, have mass reduced below conventional threshold for He-burning of 0.32\,\ms. In 
fact, in such stars nuclear burning is frozen almost immediately 
after  RLOF and their interiors may vary from almost pure He to a C/O mixture, 
depending on the degree of exhaustion of He in the centre at the instant of RLOF. Their entropy is 
lower than that of He WD -- former degenerate cores of red giants \citep{2008AstL...34..620Y}. These stars 
may be donors in systems experiencing single He-detonation or double-detonation.}.

Scenario 1 involves stars more massive than $\simeq4$\,\ms.
Because the masses of components are relatively comparable (Fig.~\ref{fig:scen1}) both resulting WDs are 
quite massive and feed, predominantly zones  {\bf E}, {\bf F} and {\bf J}.
   
 Systems with $3 \aplt M_1/\ms \aplt8$ evolve via scenario 2, but on average these systems have
less massive secondaries than in scenario 1. 
This scenario is to a significant extent similar
to scenario 1, but because of the larger span of the initial masses of components 
it contributes merging pairs of CO WD to virtually all zones of \mamd\ plane 
(Fig.~\ref{fig:scen2}).  Because the secondaries in some of the initial systems evolving via scenario 2 have 
initial masses as small as almost 1\,\ms, certain fraction of former secondaries
after mass loss in the CE becomes He WD. However, scenario 2 is not the main channel of He WD formation, most 
of them form, as mentioned above, via non-common routes. In total, in scenario 2, the fraction of merging He 
and CO WD pairs feeding zone {\bf A} in Fig.~\ref{fig:danmap}  among all merging pairs is 7 per cent for 
\al=0.25, 8 per cent for \al=0.5, 0.3 per cent for \al=1.0 and
4.5 per cent for \al=2.0. Most CO and He WD merge in zone {\bf L} where they produce hot subdwarfs
or in zone {\bf RCB}, see Fig.~\ref{fig:danmap}.
 
Binaries with $2.5 \aplt M_1/\ms \aplt 8$\,\ms\ and less massive 
secondaries than in the previous scenarios --  $M_2 \aplt 4$\,\ms\ -- evolve via scenario 3.   After the first
stable mass exchange these systems become so wide that former initial secondaries overflow Roche lobe only in the 
EAGB-stage. Their C/O-cores are small, however, and after mass loss in the CE the stars still have 
relatively massive He envelopes and their evolution is similar to the evolution of He-stars. 
This scenario contributes to the regions of the \mamd\ diagram, where single detonation of He ({\textbf A}) or
merger of CO WDs with $\mt \geq \mch$ ({\textbf E, F}) is possible (Fig.~\ref{fig:scen3}). 

Scenario 4  is followed by systems  with $2\,\aplt M_1/\ms\,\aplt 7$, with initial separations 
of components on average smaller than in scenario 3 (Fig.~\ref{fig:scen4}). For this reason, prior to 
the CE , secondaries are still in the RGB-stage of evolution and after mass loss produce 
He stars. The least massive of He stars evolve straight into hybrid CO WD, while more massive   
ones expand and the system passes through a second CE (Table~\ref{tab:scen}).     
If $\al \leq 0.5$, scenario 4 produces also some ONeWD+COWD pairs and 
makes a small contribution to zone {\bf J} (Table \ref{tab:scen_frac}). 

In scenarios 5 -- 8, at difference to scenarios 1 --4, the first RLOF in the system is accompanied by 
a CE. These systems are wide, the primaries overflow 
their Roche lobes in EAGB and TPAGB stages of evolution (case C of mass exchange),
mass-loss is dynamically unstable.
Scenarios 5 and 6 (Figs.~\ref{fig:scen5}, \ref{fig:scen6}) are quite similar. Through these scenarios
evolve systems that have  similar masses of components. 
At the time of the first RLOF, companions to mass-losing stars are in the core He-burning stage 
(Table~\ref{tab:scen}) and the compact cores of both stars appear to be immersed in a  `double common 
envelope'. The result of the CE-stage is formation of a COWD+HeMS binary. Merging pairs contain 
either a CO WD and a hybrid WD or two CO WDs. The  outcome of the merger may be single detonation 
(zone {\bf A}), formation of a massive CO WD (zone {\bf C})
 or the merger of (super)-\mch\ CO WDs (zones {\bf D} and {\bf E}).

Through scenarios 7 and 8 (Figs.~\ref{fig:scen7}, \ref{fig:scen8}) evolve the most wide close binaries with secondaries within
quite a large range from 2.5\,\ms\ to almost 5\,\ms. The primaries in these stars
overflow Roche lobes in EAGB or TPAGB stage, while their companions still remain MS stars.         
A CE stage follows. In scenario 7 the donor with small He core first becomes 
a COHe-star and later -- a CO WD.  In scenario 7, the  separation of components is such that 
former secondaries overflow their critical lobes in HG or GB stages and turn into He stars. 
After exhaustion of He in the cores they evolve into CO WD. Scenario 7 feeds `massive' part of 
the sub-\mch\ zone {\bf B} where both WDs may explode at contact and zone {\bf E}, where (super)-\mch\ CO WDs 
merge. In scenario 8, with the wider initial separation of components, a CO WD is formed straight after the
first CE, while the former secondary may, depending on the initial mass and separation,
overflow the critical lobe in HG, GB or EAGB stages. In all cases, a CE  forms and  He stars
are formed, which later evolve into CO WDs. Like for scenario 7, scenario 8 feeds massive WD part of 
zone {\bf A} and zone {\bf E}.     
    
It is worthwhile to note that, while the precursors of most observed binary WD apparently form via two stages of CE 
\citep{2012A&A...546A..70T}, most progenitor binaries of merging pairs of WD in our simulations
have stable first mass-exchange episode and an unstable (with a CE) second one, as 
also found by \citet{2010A&A...515A..89M,2013MNRAS.429.1425R}. In the simulations of 
\citet{2012A&A...546A..70T}, the fraction of systems which have only one CE stage is close to 50 per cent.

Above, we presented scenarios dominating in the \al=2 case. Some scenarios do not realize for all \al. 
Furthermore, their relative input varies with \al.
In Table~\ref{tab:scen_frac} we compare the fractions of systems evolving via scenarios  listed 
in Table~\ref{tab:scen}, depending on \al\ and their input to the particular regions of the  
\mamd\ diagram. Because of the uncertainty of the value of \al\ we studied the demography of the
\mamd\ diagram for \al = 0.25, 0.5, 1.0 and 2.0. 

For \al=0.25  essentially only scenario 3 can result in the formation of a merging pair of WDs and a \sna.
Because \al\ is low, this may be understood as a result of the merger of components in CE1 or CE2 in other scenarios.
Two branches of evolution can be distinguished, leading, predominantly,  
to the filling of regions {\bf A} and {\bf E} of the 
\mamd\ plane. In the first case, a CO WD merges with a hybrid or a He WD,  while in the second one, 
two CO WDs with $\mt \geq \mch$ merge. The contribution to the two zones is comparable. 
For \al=0.25 no systems evolve via scenario  1, that is, mergers of CO and ONe WD do not occur and 
the `triggering' of explosions of ONe WD simulating \sna\ \citep{2015A&A...580A.118M} is infeasible. 
For other values of \al, significance of this scenario is vanishingly small.  
     
Scenario 3 dominates for \al=0.5 and 1 and feeds predominantly zones {\bf A} and {\bf E} of the 
\mamd\ diagram. With an increase of \al\ contribution of this scenario to zone \za\
first increases, because a larger fraction of low-mass CO WDs avoids merger. However, with a 
further increase of \al\ over 1, the separation of components  after the second CE episode remains so large, 
that binaries never merge.  

For \al=1, scenarii 2, 4, 5 become significant. In general, as expected, with an increase of \al, the 
contribution of systems in which RLOF occurs at later phases, i. e., wider at the zero-age main sequence,  increases. 

For \al=2 scenarii 2 and 4 become dominant, and scenarios 5 and 8 provide a significant contribution to 
zones {\bf A} and {\bf E}, It is important that in all scenarios at least one 
of the merging WDs is a hybrid one and probably retains some He in the envelope up to the beginning of 
merger and, in principle, may detonate. Formation of hybrid WD occurs because AGB evolution is typically 
aborted in a quite early stage, when the CO core is still not well 
developed and the post-RLOF star continues its evolution as 
a star with thick He envelope (COHe in our notation). Similarity between the outcome of cases B and 
early C of mass exchange was noticed by \citet{iben86}, but with a caveat that in the case B the stellar 
remnants are less massive. If RLOF occurred in TPAGB stage, the mass of the nascent WD is also 
higher than for RLOF in EAGB case, because of the dredge-up event, which occurred in the course of 
evolution between two stages.   

The trends observed in Table~\ref{tab:scen_frac} are a relatively even (within a factor 1.5)
fractional population of regions {\bf A} and {\bf E}  of \mamd\ diagram irrespective of \al\ and systematic 
decrease of the population of regions {\bf F} and {\bf J}, associated with the most massive WDs, with an 
increase of \al. Most mergers of WDs occur in zones {\bf A} and {\bf E} (Table~\ref{tab:scen_frac}); that is, 
as mergers of relatively massive He or hybrid WDs and CO WDs or pairs of CO WDs.  

Table~\ref{tab:scen_frac} clearly shows the influence of the combined parameter \al\ on the merger rate 
of WDs. In scenarios 1--4, most systems after the first, stable RLOF become so wide that, if the expulsion of 
the matter from the system in CE is efficient (\ace\ is high) merger of 
components is avoided and they form pairs of WDs which are close enough 
to merge in Hubble time. With an increase of \ace\ the fraction of merging pairs of WD  decreases.
Figure~\ref{fig:scen3}, as an example,  illustrates variation of the initial parameters of systems
contributing to scenario~3 for two extreme cases of \al\ and the influence of this parameter on 
masses and types of merging WD. 

We do not present results for the runs with  $\al>2$ because such high values of \al\  
currently seem unrealistic.  

\begin{figure*}     
\includegraphics[width=0.49\textwidth]{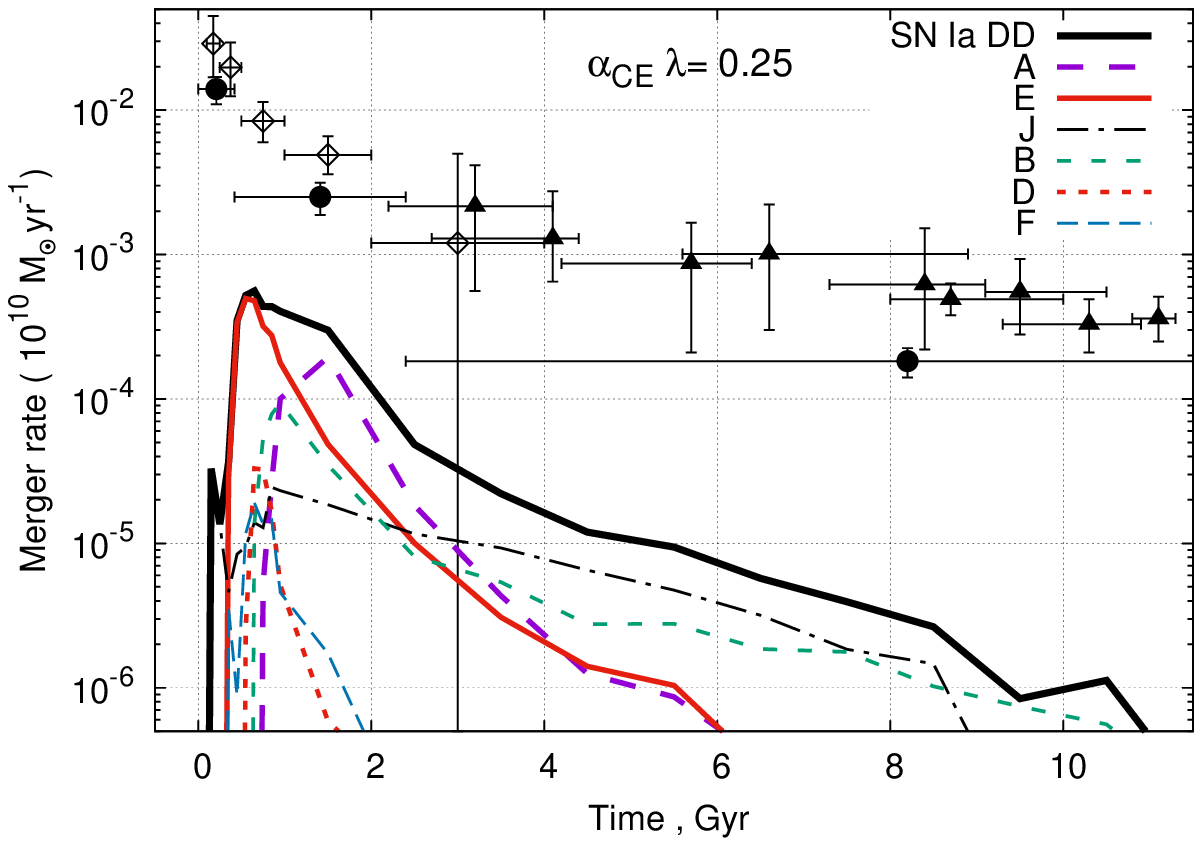}
\includegraphics[width=0.49\textwidth]{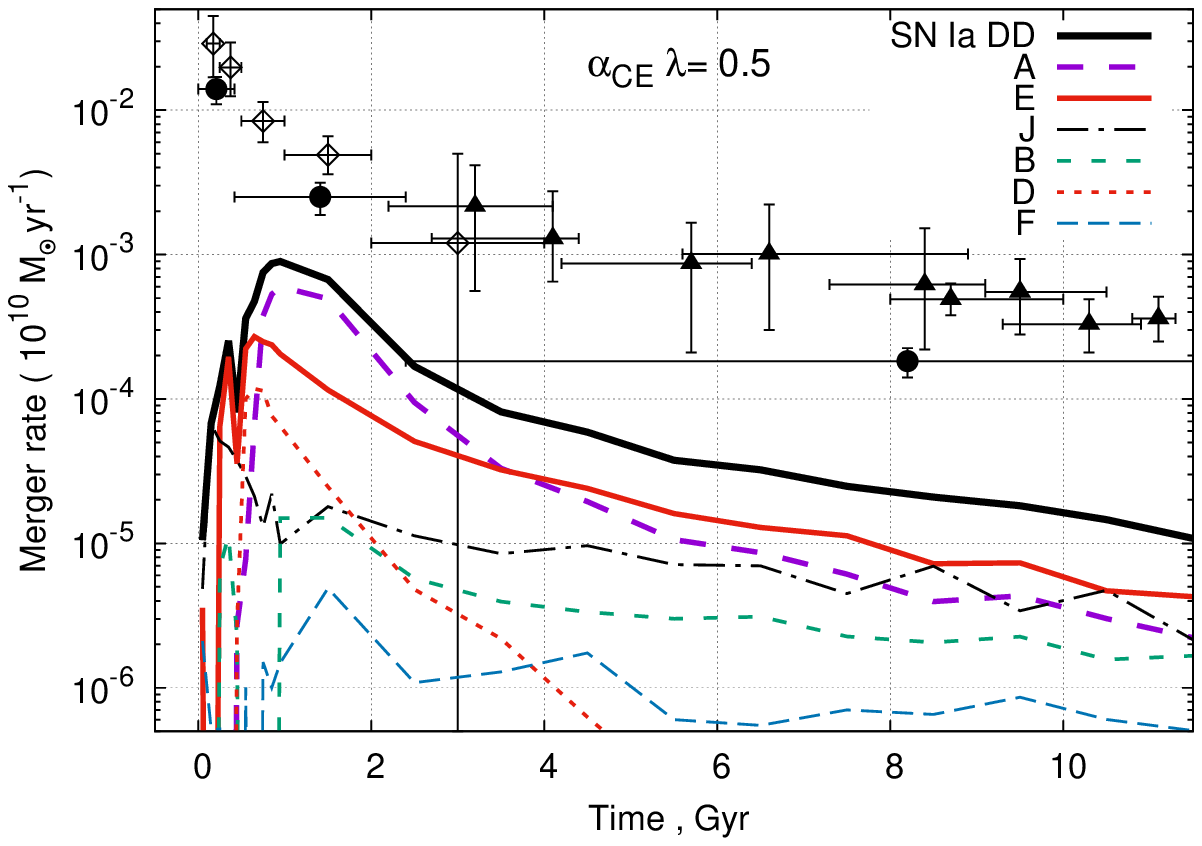}
\includegraphics[width=0.49\textwidth]{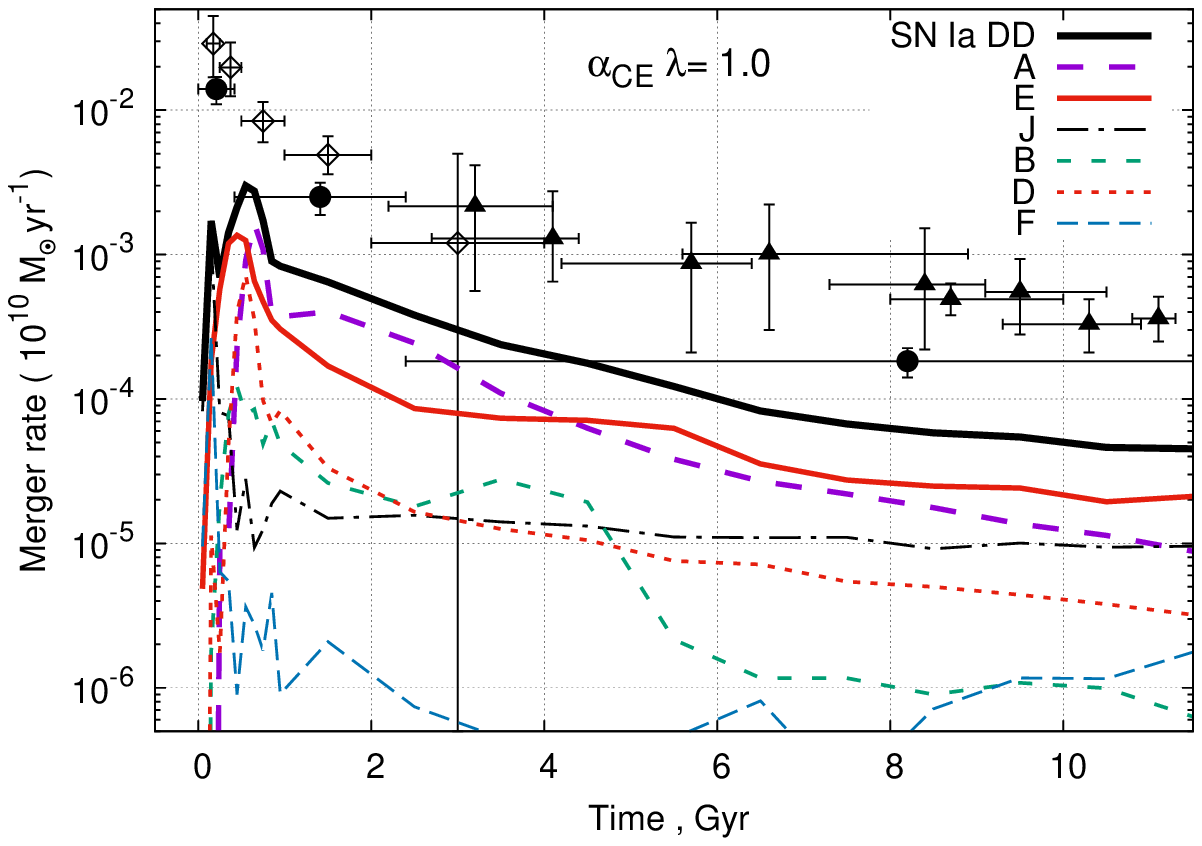}
\includegraphics[width=0.49\textwidth]{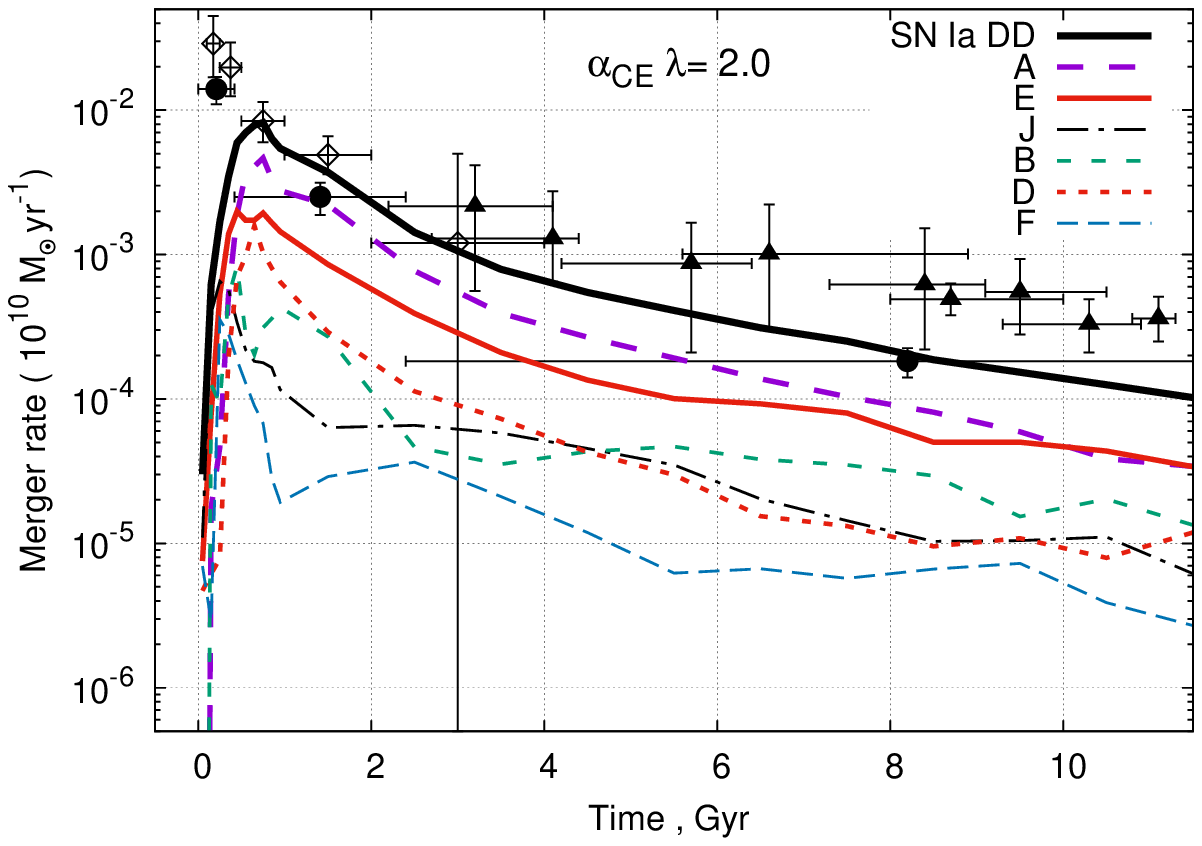}
\caption{DTD for the model WD+WD mergers potentially producing  \sne. Different line styles 
represent  mergers occurring in different zones of the \mamd\ diagram, annotated as in Fig.~\ref{fig:danmap}. 
The thick solid line (\sna) represents the sum of the models. 
The models correspond to metal abundance Z=0.02. Symbols with error bars: `observational' DTD from Subaru/XMM 
Survey  \citep{2008PASJ...60.1327T},  diamonds; 
galaxy clusters \citep{2010ApJ...722.1879M}, triangles; 
a sample of field galaxies from  the SLOAN~II Survey \citep{2012MNRAS.426.3282M}, heavy dots.  }
\label{fig:dtd_ton}
\end{figure*}

\begin{figure*}    
\begin{minipage}{0.49\textwidth}
\includegraphics[width=\textwidth]{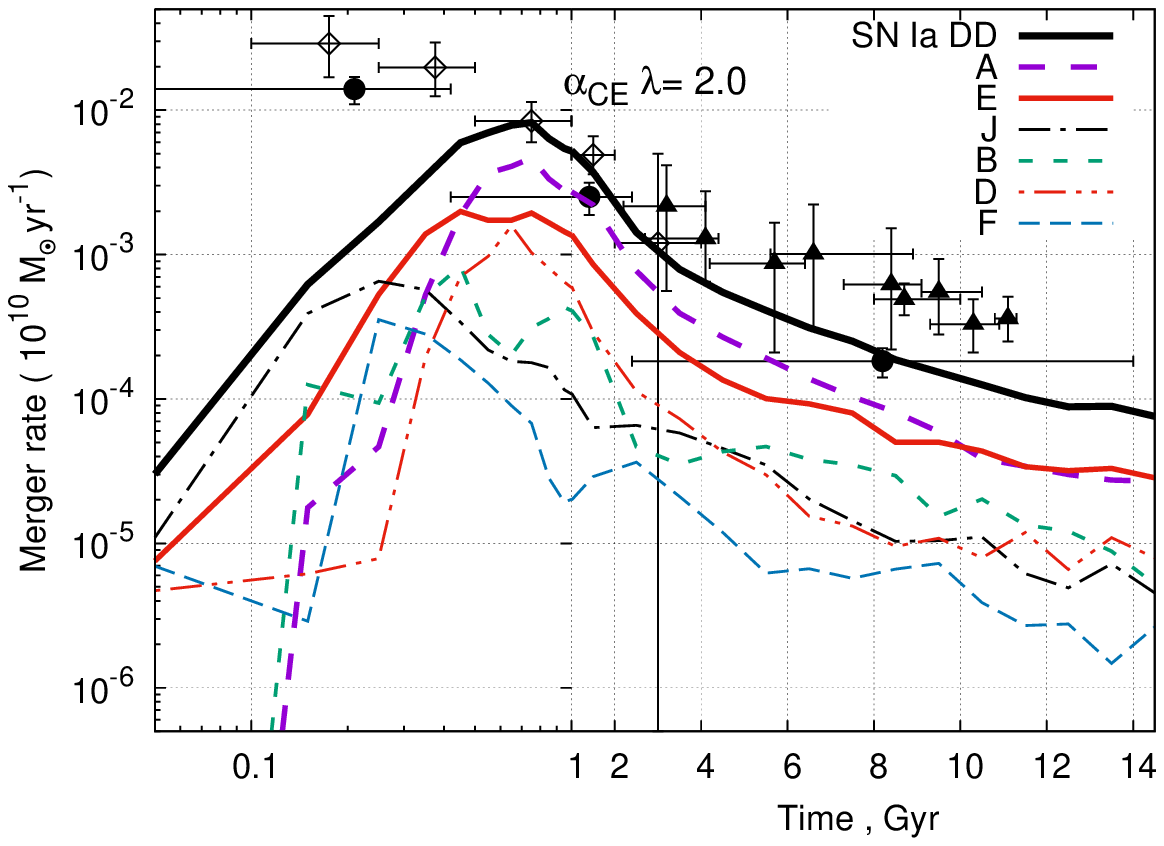}
\caption{DTD for the model WD+WD mergers potentially producing  \sne\ in the case \al=2 (similar to the lower right
panel of Fig.~\ref{fig:dtd_ton}), but showing the DTD for the first 0.05--2~Gyr on a log-scale. Annotation is similar 
to Fig.~\ref{fig:dtd_ton}.  }
\label{fig:dtd_log}
\end{minipage}
\hfill
\begin{minipage}{0.49\textwidth}
\vspace{ -4mm}
\includegraphics[width=\textwidth]{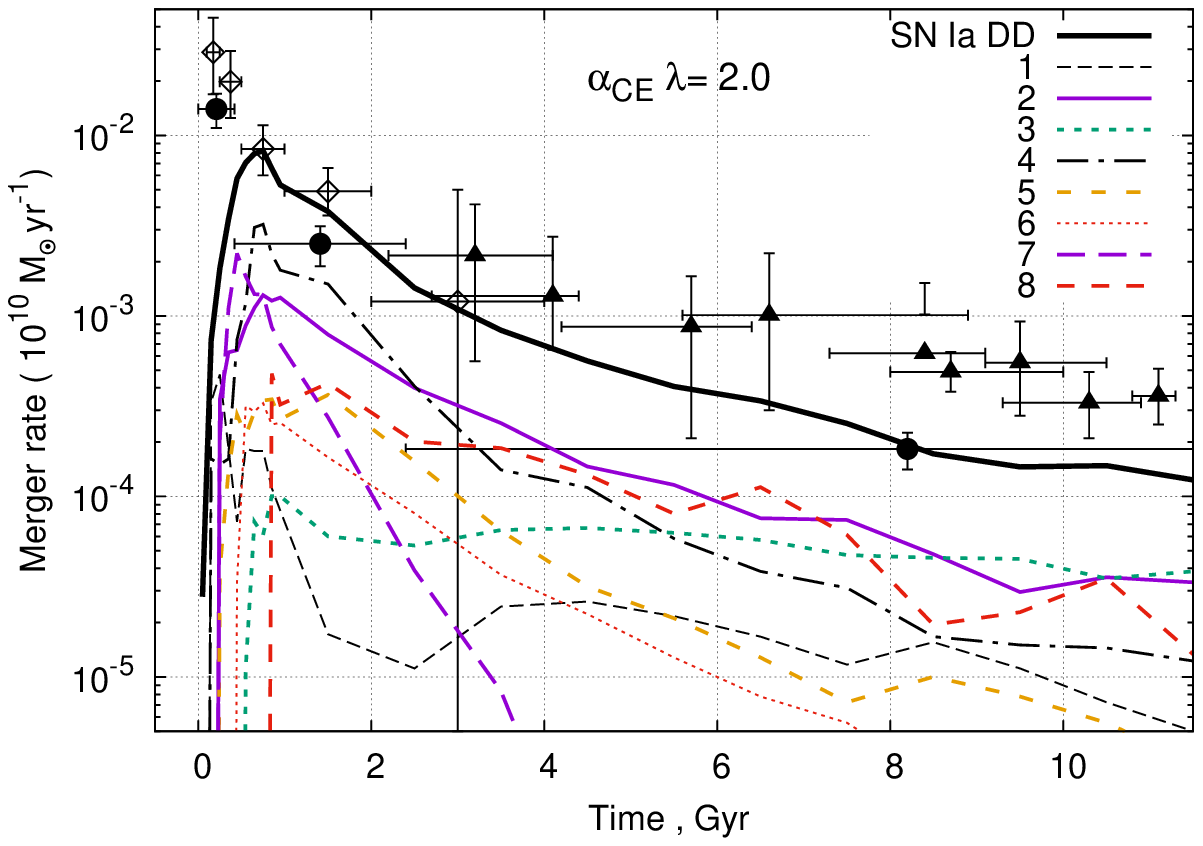}
\caption{DTD for the WD mergers following scenarios listed in Table~\ref{tab:scen} for the case \al=2.  Symbols with error bars: `observational' DTD like in Fig.~\ref{fig:dtd_ton}. }
\label{fig:dtd_scen}
\end{minipage}
\end{figure*}

\begin{table*}     
\caption{The rate of binary WD mergers  occurring at 10\,Gyr after an instantaneous 
star-formation burst in different regions of \mamd\ diagram as a function of \al\
(per $\mathrm{10^{10}\,\ms\,yr^{-1}}$).   
The second column indicates whether \sne\ are potentially possible. Notation  $x(y)$ means 
$x \times 10^{y}$.}
   \tabcolsep 1.2 mm
\begin{tabular}{cccccc}
\hline
	Zone   & DD &	\multicolumn{4}{c}{\al}\\
	 $M_{accr}-M_{don}$  & SN Ia &	0.25 &	0.5 &	1.0 &	2.0 \\
\hline
A	&	Y	&	4(-8)	$\pm$	4(-8)	&	3.7(-6)	$\pm$	5(-7)	&	1.3(-5)	$\pm$	1(-6)	&	4.9(-5)	$\pm$	2(-6)	\\
B	&	Y	&	7(-7)	$\pm$	3(-7)	&	2.0(-6)	$\pm$	3(-7)	&	1.0(-6)	$\pm$	3(-7)	&	1.8(-5)	$\pm$	1(-6)	\\
C	&	N	&	-			&	6.8(-8)	$\pm$	6(-8)	&	5.6(-6)	$\pm$	7(-7)	&	7.8(-6)	$\pm$	9(-7)	\\
D	&	Y	&	-			&	2.3(-8)	$\pm$	3(-8)	&	4.1(-6)	$\pm$	6(-7)	&	9.5(-6)	$\pm$	9(-7)	\\
E	&	Y	&	-			&	6.0(-6)	$\pm$	6(-7)	&	2.2(-5)	$\pm$	1(-6)	&	4.7(-5)	$\pm$	2(-6)	\\
F	&	Y	&	-			&	7.3(-7)	$\pm$	2(-7)	&	1.2(-6)	$\pm$	3(-7)	&	5.6(-6)	$\pm$	7(-7)	\\
G	&	N	&	-			&	6.3(-7)	$\pm$	2(-7)	&	9.7(-6)	$\pm$	1(-6)	&	2.2(-5)	$\pm$	1(-6)	\\
H	&	N	&	2(-7)	$\pm$	1(-7)	&	-			&	-			&	1.4(-5)	$\pm$	1(-6)	\\
I	&	N	&	2.2(-6)	$\pm$	5(-7)	&	8.5(-6)	$\pm$	7(-7)	&	1.5(-5)	$\pm$	1(-6)	&	4.0(-5)	$\pm$	2(-6)	\\
J	&	Y	&	3(-7)	$\pm$	2(-7)	&	4.1(-6)	$\pm$	5(-7)	&	9.7(-6)	$\pm$	1(-6)	&	1.1(-5)	$\pm$	1(-6)	\\
K	&	N	&	-			&	4.8(-6)	$\pm$	5(-7)	&	1.1(-5)	$\pm$	9(-7)	&	1.5(-5)	$\pm$	1(-6)	\\
L	&	N	&	2.9(-5)	$\pm$	2(-6)	&	1.6(-4)	$\pm$	3(-6)	&	5.1(-4)	$\pm$	6(-6)	&	7.7(-4)	$\pm$	8(-6)	\\
R CrB	&	N	&	3.4(-5)	$\pm$	2(-6)	&	2.5(-4)	$\pm$	3(-6)	&	4.5(-4)	$\pm$	6(-6)	&	5.2(-4)	$\pm$	7(-6)	\\
\hline																			
\multicolumn{2}{c}{SN Ia}	&	9(-7)	$\pm$	3(-7)	&	1.7(-5)	$\pm$	1(-6)	&	5.1(-5)	$\pm$	2(-6)	&	1.4(-4)	$\pm$	4(-6)	\\
\multicolumn{2}{c}{WD2 Merger}	&	6.6(-5)	$\pm$	2(-6)	&	4.3(-4)	$\pm$	4(-6)	&	1.1(-3)	$\pm$	9(-6)	&	1.5(-3)	$\pm$	1(-5)	\\
\label{tab:dtd_inst}
\end{tabular}
\end{table*}

\begin{table*}    
\caption{The rate of mergers  of binary WD formed via  different 
evolutionary scenarii at 10\,Gyr after an 
instantaneous star-formation burst  as a function of \al\ 
(per $\mathrm{10^{10}\,M_\odot\,yr^{-1}}$).}
\tabcolsep 1.2mm
\begin{tabular}{lcccc}
\hline
Scen.	    &	\multicolumn{4}{c}{\al}\\
	    &	0.25 &	0.5 &	1.0 &	2.0 \\
\hline
1	&			-	&			-	&			-	&	9.8(-6)	$\pm$	2(-6)	\\
2	&			-	&	3.9(-6)	$\pm$	7(-7)	&	1.3(-5)	$\pm$	2(-6)	&	3.6(-5)	$\pm$	4(-6)	\\
3	&	9(-7)$\pm$	4(-7)&	1.3(-5)	$\pm$	1(-6)	&	2.5(-5)	$\pm$	3(-6)	&	4.3(-5)	$\pm$	5(-6)	\\
4	&			-	&			-	&	1.1(-5)	$\pm$	2(-6)	&	1.6(-5)	$\pm$	3(-6)	\\
5	&			-	&	1.1(-7)	$\pm$	5(-8	&	1.4(-6)	$\pm$	7(-7)	&	7.2(-6)	$\pm$	2(-6)\\
6	&			-	&			-	&	4.7(-7)	$\pm$	4(-7)	&	2.1(-6)	$\pm$	1(-6)\\
7	&			-	&			-	&			-	&				\\
8	&			-	&			-	&			-	&	3.1(-5)	$\pm$	4(-6)\\
\hline																	
SN Ia &	9(-7)$\pm$	4(-7)&	1.7(-5)	$\pm$	1(-6)	&	5.1(-5)	$\pm$	4(-6)&	1.4(-4)$\pm$	9(-6)\\
\label{tab:dtd_type}
\end{tabular}
\end{table*}

\subsection{Delay-time distribution}
\label{sec:dtd}
 
The fundamental characteristic of \sne\  is the `delay-time distribution' (DTD); that is,
distribution over time-intervals between the formation of a close binary and \sna\ explosion, because
every scenario of \sna\ has a typical time scale \citep{ty94,1995ApJ...447L..69R,1997ApJ...486..110J,yl00}.  
The empirical DTD is, as a rule, derived from the rate of \sne\ in the samples of galaxies at large  
redshifts ($z\sim1$), but may also be derived for individual galaxies, see \citet{2014ARA&A..52..107M}.
Theoretically, it is a model of the dependence of the rate of \sne\ with different precursors on the time 
elapsed from an instantaneous burst of star-formation that formed a unit of stellar mass. It is evident, that 
hypothetical \sne\ associated with short-lived objects, for example, He stars, should have short delays 
($\aplt$1\,Gyr). On th other hand, if particular scenario is associated with potentially 
long-living objects, like double-degenerates, delays for them may, in principle, overlap with entire lifetime 
of a galaxy, because most massive WD start to merge in 
several tens of Myr after a star-formation burst and the upper limit is the Hubble time.
Both experimental and theoretical estimates of DTD are overburdened by numerous 
uncertainties. For empirical estimates, uncertainty may reach an order of magnitude, depending on the 
sample of \sne\ under study and possible systematic errors; see \citet{2014ARA&A..52..107M} for a 
detailed discussion and Fig.~\ref{fig:dtd_ton}. The scatter in the theoretical estimates results mainly 
from the difference in the treatment of  evolutionary transformations of binaries 
in different BPS codes \citep{2014A&A...562A..14T}.

Figure~\ref{fig:dtd_ton} shows the model DTD for the mergers of WDs potentially leading to
\sne\ and empirical data for elliptical galaxies from Subaru/XMM-Newton Deep Survey 
\citep{2008PASJ...60.1327T}, galaxy clusters \citep{2010ApJ...722.1879M}, and a sample of 
galaxies from  SLOAN~II Survey \citep{2012MNRAS.426.3282M}. Clearly, none of the models fit observations at 
very early epochs ($\aplt500$\,Myr). Only models for \al=2 fit  points at $\approx$1 and 8~Gyr of DTD 
derived for SLOAN~II Survey which has very large time-bins and error bars. 
If we consider the DTD for galaxy clusters, at  $\approx$(7 -- 10)\,Gyr, the difference approaches a factor 
close to 3-4. The main fraction of mergers occurs in the zones {\bf A} and {\bf E} of the \mamd\ 
diagram. For illustration, at the request of the referee, in Fig.~\ref{fig:dtd_log} we replot DTD for the 
\al=2 case, showing data for the first (0.05--2)\,Gyr on a log-scale. Note that, while the lines for 
particular scenarios are quite irregular, the summary line shows a gradual growth of the rate of SNe Ia,
mainly as a result of the smooth increase of 
mergers occurring in zone \ze\ (merger of CO WDs). Recall also, that at $t\aplt$2\,Gyr a significant 
contribution to \sne\ rate may provide a SD-chanell, associated either with hydrogen or helium transfer 
\citep[see e. g.][]{bours_retention_13,2009ApJ...701.1540W}.  
 
Figure~\ref{fig:dtd_scen} shows DTD for eight scenarios listed in Table~\ref{tab:scen} for the most 
prolific combination \al=2. While at very early times, $t\aplt$1\,Gyr, scenarios 2, 4 and 7 dominate,  
later, at $t\approx$(3 - 10)\,Gyr, comparable contribution is, crudely, provided by scenarii 2, 3 and 8 
(see also Table~\ref{tab:dtd_inst}). Scenario 7 is associated with 
massive WDs, and there is only a very narrow `gap' of initial separations for binaries with 
$M_1 \aplt 4.5$\,\ms\ which just enables mergers in less than about 4\,Gyr. It is important that in 
all scenarios one of the merging components is either a He WD or a CO WD which descended from 
a He star. Thus, the envelopes of WD always have certain amount of He, which \textit{may} experience  
detonation and, under favourable conditions, trigger a detonation in an accreting WD. 

As a complement to Fig.~\ref{fig:dtd_ton}, in Table~\ref{tab:dtd_inst} we present the rate of WD 
mergers occurring at 10\,Gyr after star formation 
burst in different regions of the \mamd\ diagram, while the rates of WD mergers at 10\,Gyr after the  
burst as a function of \al\  are presented in Table~\ref{tab:dtd_type}. 
 
Figures~\ref{fig:danmap} -- \ref{fig:dtd_scen} and Tables~\ref{tab:dtd_inst} and 
\ref{tab:dtd_type} suggest that a satisfactory reproduction of the
extant data on the DTD at $t \apgt$ several 100 Myr requires comparable contributions of mergers of pairs of 
CO+CO WDs with $\mt \apgt \mch$ and mergers of $\mt \aplt \mch$\ pairs with CO WD accretors and very massive 
He or hybrid WD donors. Basically, this agrees 
with  the proportions of $\sim$\mch\ and sub-\mch\ \sne\ inferred from consideration of the
solar abundance of manganese \citep{2013A&A...559L...5S} and from the analysis of the mass of \sne\ 
ejecta  \citep{2014MNRAS.440.1498S,2014MNRAS.445.2535S,2015MNRAS.454.3816C}, see Discussion section. 
Note  also that in the lower right panel of Fig.\ref{fig:dtd_ton} the model results for zone \ze,
namely mergers of CO+CO WD pairs with $M_1 \geq 0.8$\,\ms\ and $M_2\,\geq 0.6$\,\ms\ taken alone, also fit, within errors, 
observational data of \citep{2012MNRAS.426.3282M} for the 500~Myr--2.5~Gyr and 2.5~Gyr--12~Gyr time bins.  

\begin{figure}   
\includegraphics[height=0.3\textheight,width=0.49\textwidth]{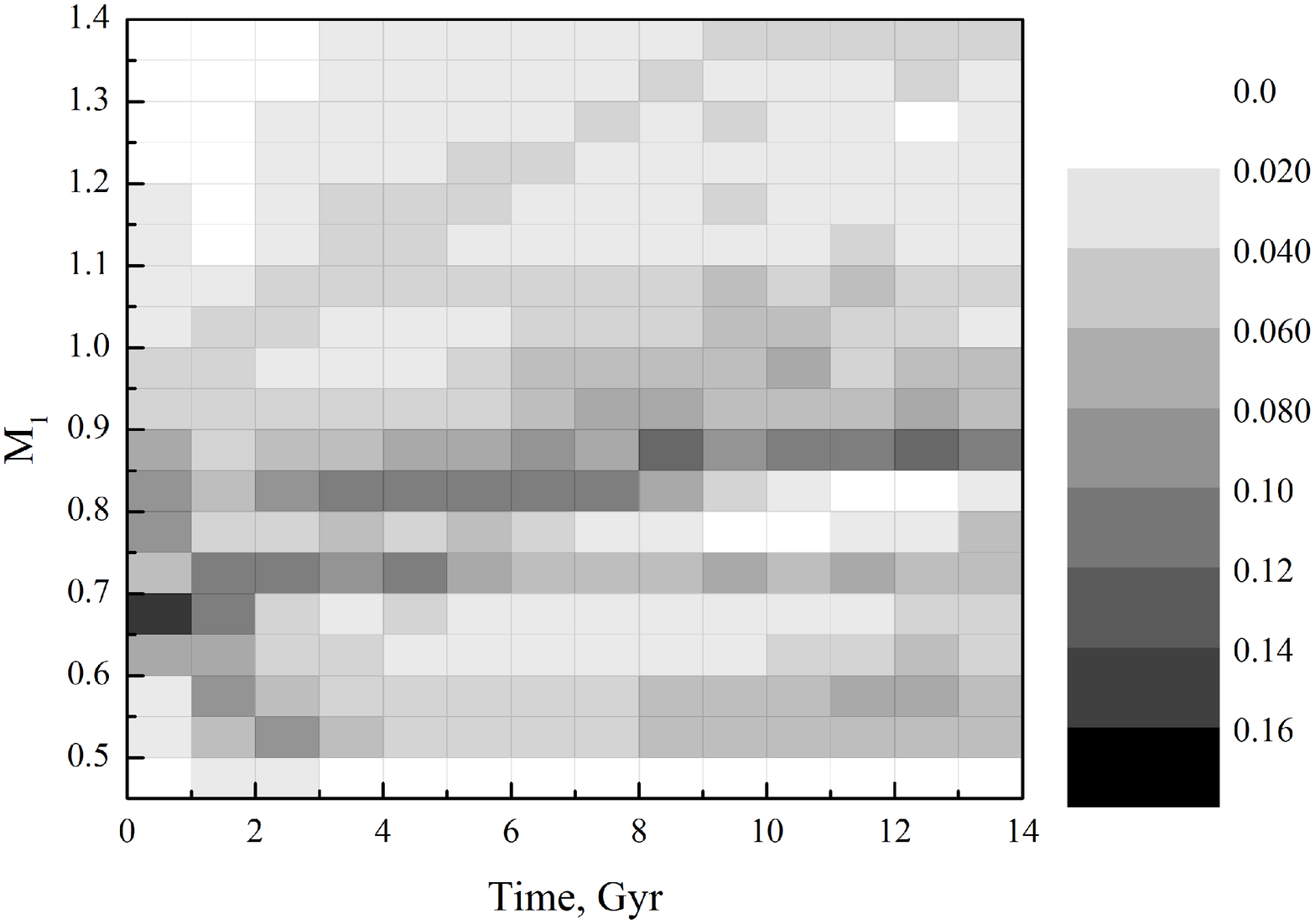}
\includegraphics[height=0.3\textheight,width=0.49\textwidth]{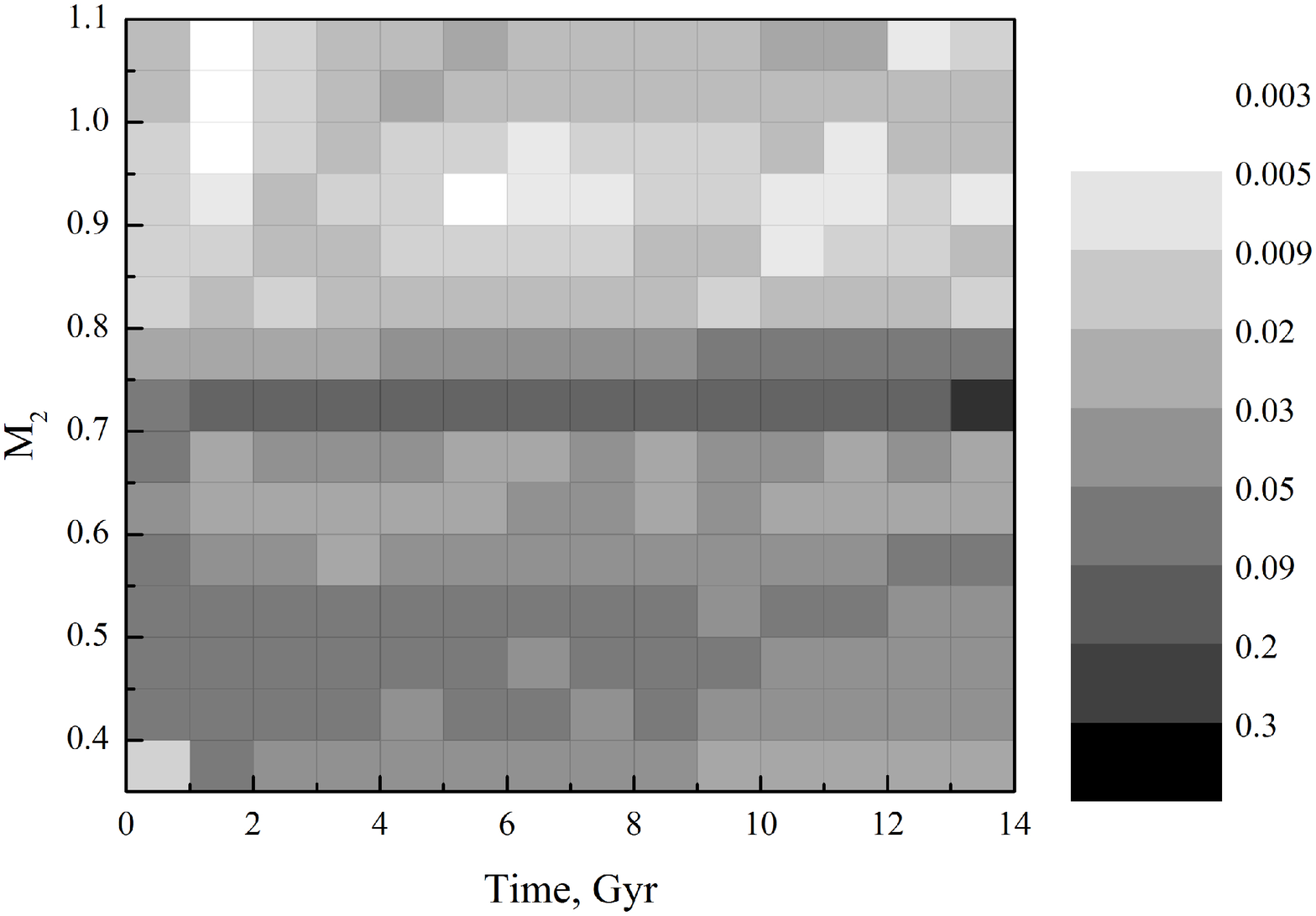}
\includegraphics[height=0.3\textheight,width=0.49\textwidth]{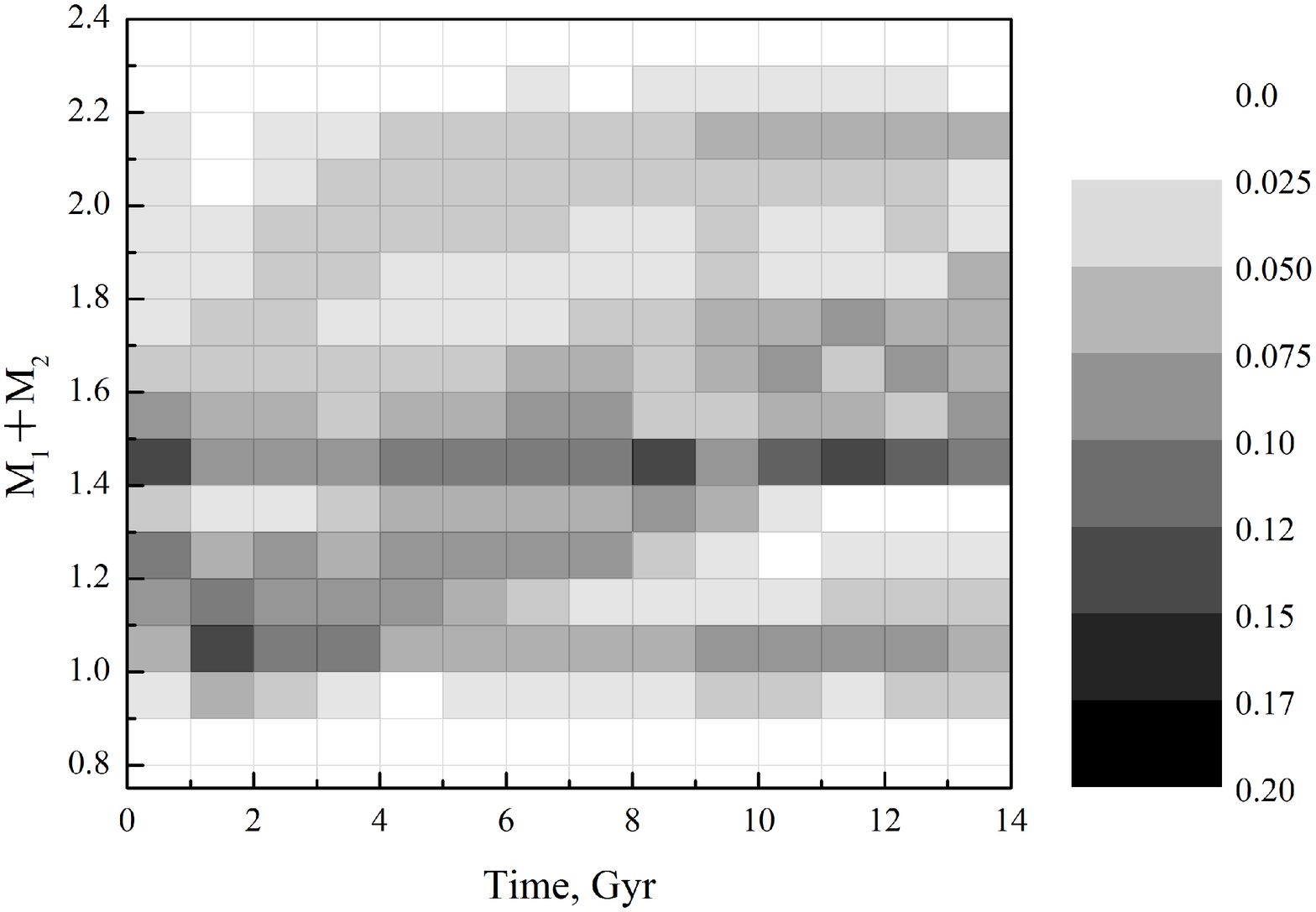}
\caption{Distribution of masses of accretors (upper panel), donors (middle panel) and total mass
(lower panel) of merging WD versus time after an instantaneous burst of star formation. Every time-bin is normalized to 1.} 
\label{fig:m1_m2}
\end{figure}

\begin{figure}   
\includegraphics[height=0.3\textheight,width=0.49\textwidth]{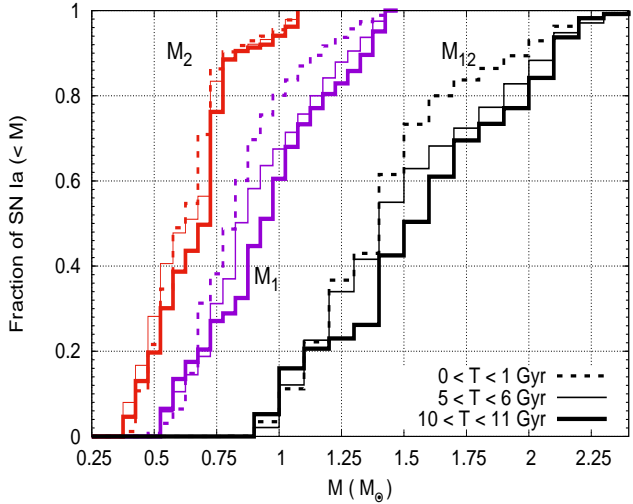}
\caption{Distribution of $M_1$, $M_2$, and \mt\ (annotated as M12) at three different epochs of the 
evolution of the `star-burst' galaxy, shown in Fig.~\ref{fig:m1_m2}. 
Dotted lines: $0\leq t \leq 1$ Gyr,
thin solid lines: $5\leq t \leq 6$ Gyr,
thick solid lines: $10\leq t \leq 11$ Gyr. }
\label{fig:m1_m2_hist}
\end{figure}
   
However, mergers of sub-\mch\ pairs  occur predominantly in zone {\bf A} of the \mamd\ diagram. 
Surface detonations are likely to occur only if $M_1 \apgt 0.8$\,\ms\ 
\citep{2010ApJ...709L..64G,2012MNRAS.422.2417D}.

If the first detonation occurs in the post-merger phase, further evolution
resembles that of double-detonation systems, where detonation in the envelope plays the role of the 
trigger, as also noted by \citet{2015MNRAS.454.4411D}.
For the latter scenario, it was found that, for explosion to resemble a \sna,  $M_1$ should be
$\apgt 0.9$\,\ms\ \citep{2010ApJ...714L..52S}.
In our simulations for \al=2, in the case of an instantaneous star formation burst, the fraction of
accretors with $M_1\apgt 0.8$\,\ms\ in the systems merging in zone {\bf A} is, at 
$t\approx10$\,Gyr, close to 20 per cent. It  is close to 40 per cent at $t\approx8$\,Gyr and
much lower at other epochs. In the case of SFR described by Eq.~(\ref{eq:sfr}) it is permanently close to 
15 per cent, see below. If the stars that do not explode in the merger process also do not explode later, 
above-mentioned $M_1$ limits, if confirmed in future, may strongly reduce  possible 
contribution of zone \za\ to the rate of \sna.

Figures~\ref{fig:m1_m2} and \ref{fig:m1_m2_hist} illustrate evolution with time of the 
distribution of masses of accretors, donors and of total mass of the systems at merger for the case 
of instantaneous star-formation burst.  Figure~\ref{fig:m1_m2_hist} shows these parameters for three 
time-bins. The average mass of accretors slightly increases with time: $M_1 \apgt 0.8$\,\ms\ have 40 per cent
of them at (0--1)\,Gyr, 50 per cent at (5--6)\,Gyr, and 70 per cent at (10--11)\,Gyr. 
About 50 per cent of the donors have mass $\aplt 0.6$\,\ms\ at any epoch; the other 50 per cent are more 
massive. Recall that WD with mass $\apgt 0.6$\,\ms\ have only traces of He at the surface.
Sub-\mch\ total mass have 60 per cent of pairs at $t\aplt$6~Gyr and 40 per cent at later time.
On the other hand, because the maximum mass of He WD is close to 0.47\,\ms, Fig.~\ref{fig:m1_m2} strongly 
indicates that the majority of donors in merging systems are hybrid WD.   

\begin{table}    
  \caption{The rate of mergers  of binary WDs resulting in \sna\ formed via  different 
evolutionary channels at $t=$14\,Gyr as a function of \al\ (per $\mathrm{10^{10} M_\odot yr^{-1}}$).
The star formation rate follows Eq.~(\ref{eq:sfr}). The errors are negligibly small compared with rates.}
\tabcolsep 1.2 mm
\centering
\begin{tabular}{ccccc}
\hline
Scenario	    &	\multicolumn{4}{c}{\al}\\
  	    &	0.25 &	0.5 &	1.0 &	2.0 \\
\hline

1	&	-	&	5.3(-7)	&	9.0(-6)	&	2.8(-5)	\\
2	&	2.9(-6)	&	2.1(-5)	&	7.2(-5)	&	2.1(-4)	\\
3	&	4.3(-5)	&	9.9(-5)	&	9.2(-5)	&	5.5(-5)	\\
4	&	2.4(-7)	&	1.8(-6)	&	5.5(-5)	&	2.6(-4)	\\
5	&	4.7(-8)	&	1.3(-6)	&	1.2(-5)	&	6.3(-5)	\\
6	&	-	&	3.1(-7)	&	9.1(-6)	&	3.5(-5)	\\
7	&	-	&	-	&	-	&	8.5(-5)	\\
8	&	-	&	-	&	9.7(-7)	&	1.1(-4)	\\
\hline									
SN Ia	&	4.7(-5)	&	1.2(-4)	&	2.5(-4)	&	9.0(-4)	\\
\label{tab:fdt_type}
\end{tabular}
\end{table}
	
\begin{table}      
  \caption{The rate of mergers  of binary WDs resulting in \sne\ formed via  
different 
evolutionary channels and occurring in different regions of the \mamd\ diagram at 
$t=$14\,Gyr as a function of \al\ (per $\mathrm{10^{10} M_\odot yr^{-1}}$).
Star formation rate follows Eq.~(\ref{eq:sfr}). Row `\sna'
provides the rate of potential \sna, while row `WD2' shows the total rate of WD mergers. }
\centering
\tabcolsep 1.2 mm
\begin{tabular}{lccccc}
\hline
	Zone   & DD &	\multicolumn{4}{c}{\al}\\
	 $M_{accr}-M_{don}$  & SN Ia &	0.25 &	0.5 &	1.0 &	2.0 \\
\hline
A	&	Y	&	1.5(-5)	&	5.7(-5)	&	1.0(-4)	&	4.5(-4)	\\
B	&	Y	&	5.8(-6)	&	3.9(-6)	&	1.1(-5)	&	6.8(-5)	\\
C	&	N	&	1.4(-6)	&	1.3(-5)	&	3.2(-5)	&	8.4(-5)	\\
D	&	Y	&	6.4(-7)	&	5.2(-6)	&	2.0(-5)	&	8.4(-5)	\\
E	&	Y	&	1.8(-5)	&	3.1(-5)	&	9.1(-5)	&	2.4(-4)	\\
F	&	Y	&	5.6(-7)	&	1.2(-6)	&	3.0(-6)	&	1.9(-5)	\\
G	&	N	&	8.2(-9)	&	5.9(-6)	&	5.5(-5)	&	5.7(-5)	\\
H	&	N	&	1.2(-7)	&	1.3(-7)	&	7.7(-7)	&	3.9(-5)	\\
I	&	N	&	1.1(-5)	&	3.3(-5)	&	6.4(-5)	&	1.3(-4)	\\
J	&	Y	&	5.2(-6)	&	9.1(-6)	&	2.2(-5)	&	4.6(-5)	\\
K	&	N	&	4.8(-6)	&	2.6(-5)	&	4.4(-5)	&	4.4(-5)	\\
L	&	N	&	9.7(-5)	&	2.2(-4)	&	6.4(-4)	&	1.0(-3)	\\
R CrB	&	N	&	4.7(-5)	&	4.6(-4)	&	1.4(-3)	&	1.7(-3)	\\
\hline											
\multicolumn{2}{l}{SN Ia}		&	4.7(-5)	&	1.2(-4)	&	2.5(-4)	&	9.0(-4)	\\
\multicolumn{2}{l}{WD2}			&	2.1(-4)	&	8.6(-4)	&	2.4(-3)	&	4.0(-3)	\\
\label{tab:fdt}
\end{tabular}
\end{table}

\begin{figure}   
\includegraphics[height=0.3\textwidth,width=0.49\textwidth]{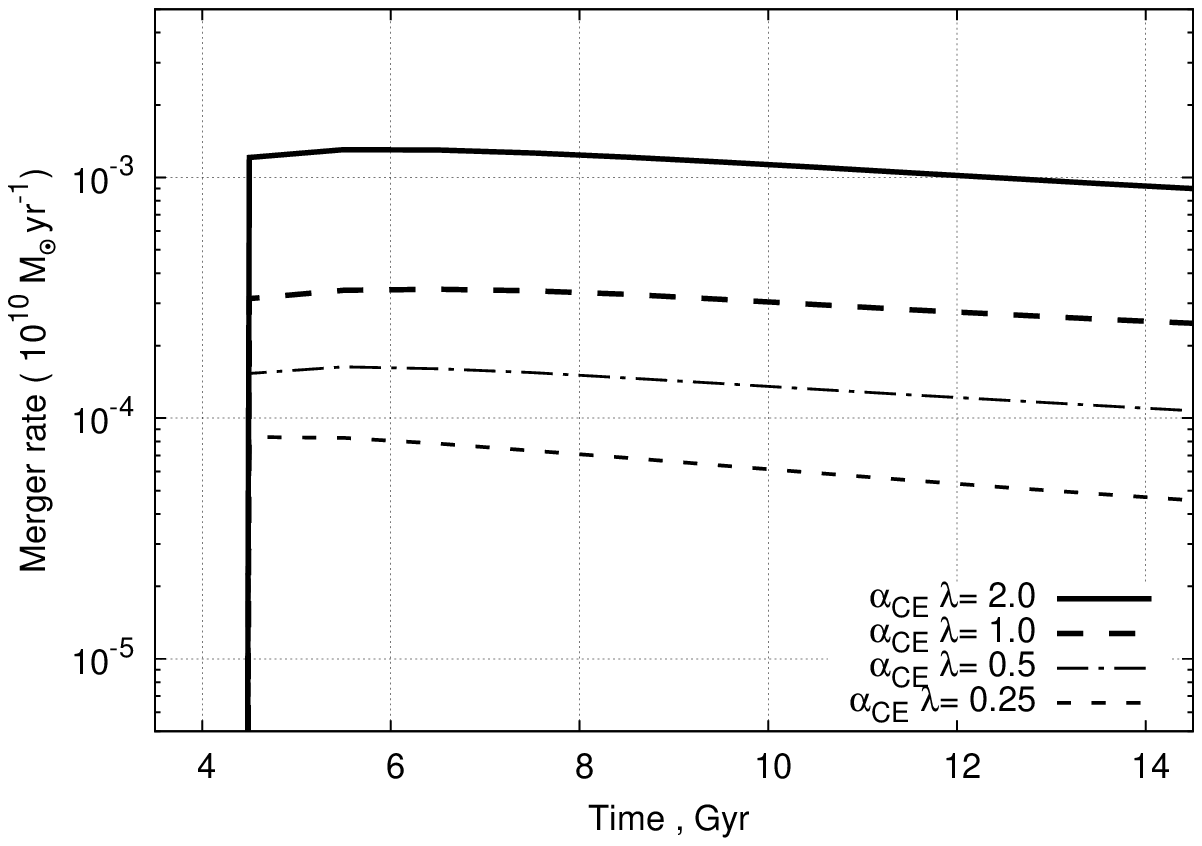}
\includegraphics[height=0.3\textwidth,width=0.49\textwidth]{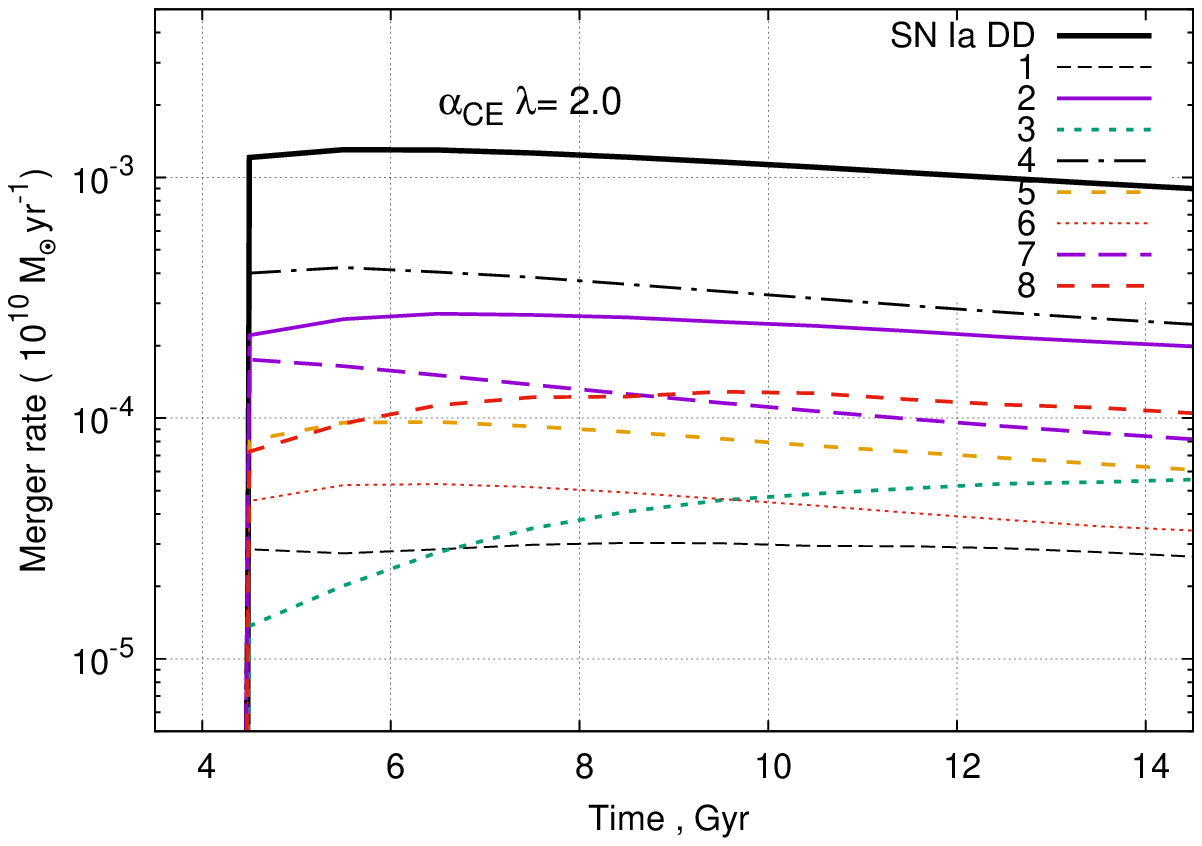}
\includegraphics[height=0.3\textwidth,width=0.49\textwidth]{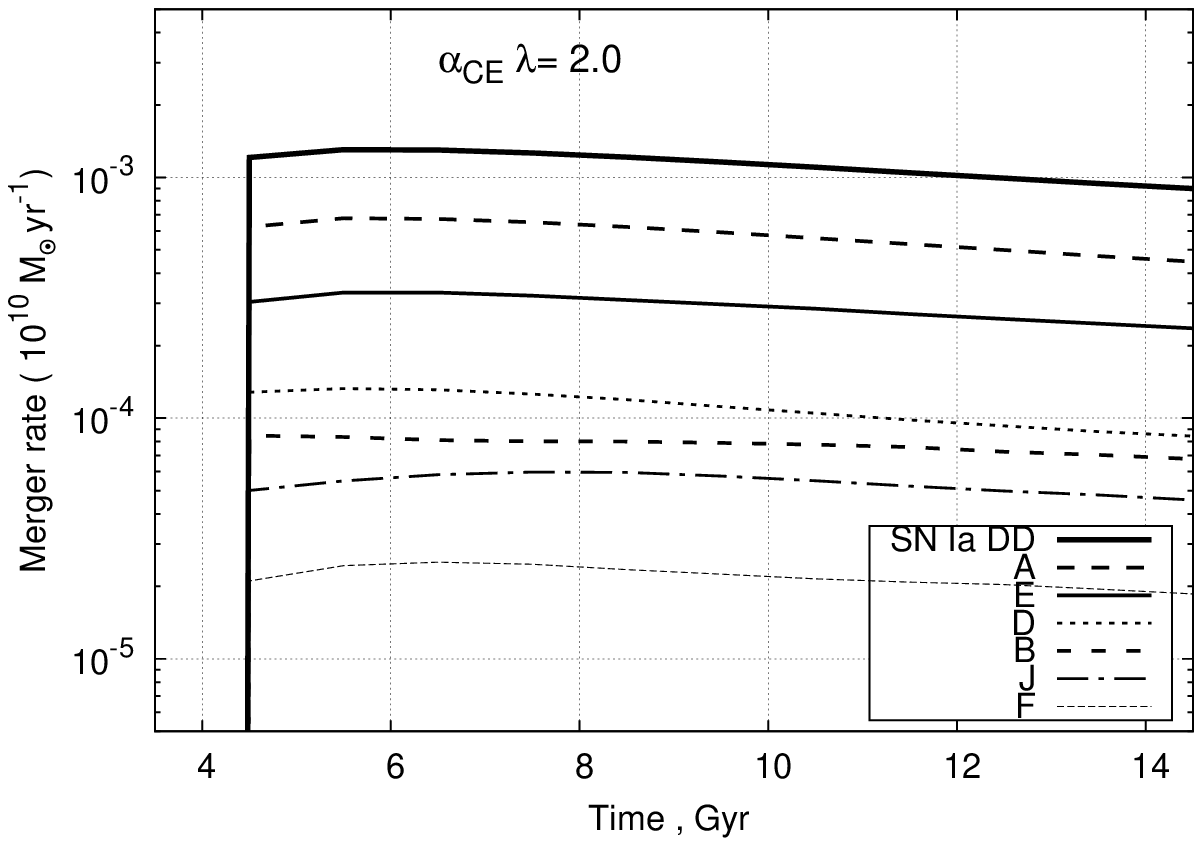}
\caption{Evolution of the rate of WD mergers potentially leading to \sne\ in the 
Galaxy, if star formation rate is set by Eq.~(\ref{eq:sfr}).
Upper panel: all potential \sne\ as a function of \al.
Middle panel: systems produced by particular scenarios (Table~\ref{tab:scen}).
Lower panel: systems feeding different regions of the\mamd\ diagram (Fig.~\ref{fig:danmap}).}  
\label{fig:ftd_ton}
 \end{figure}

\begin{figure}   
\vspace{-0.35cm} 
\includegraphics[height=0.3\textwidth, width=0.49\textwidth]{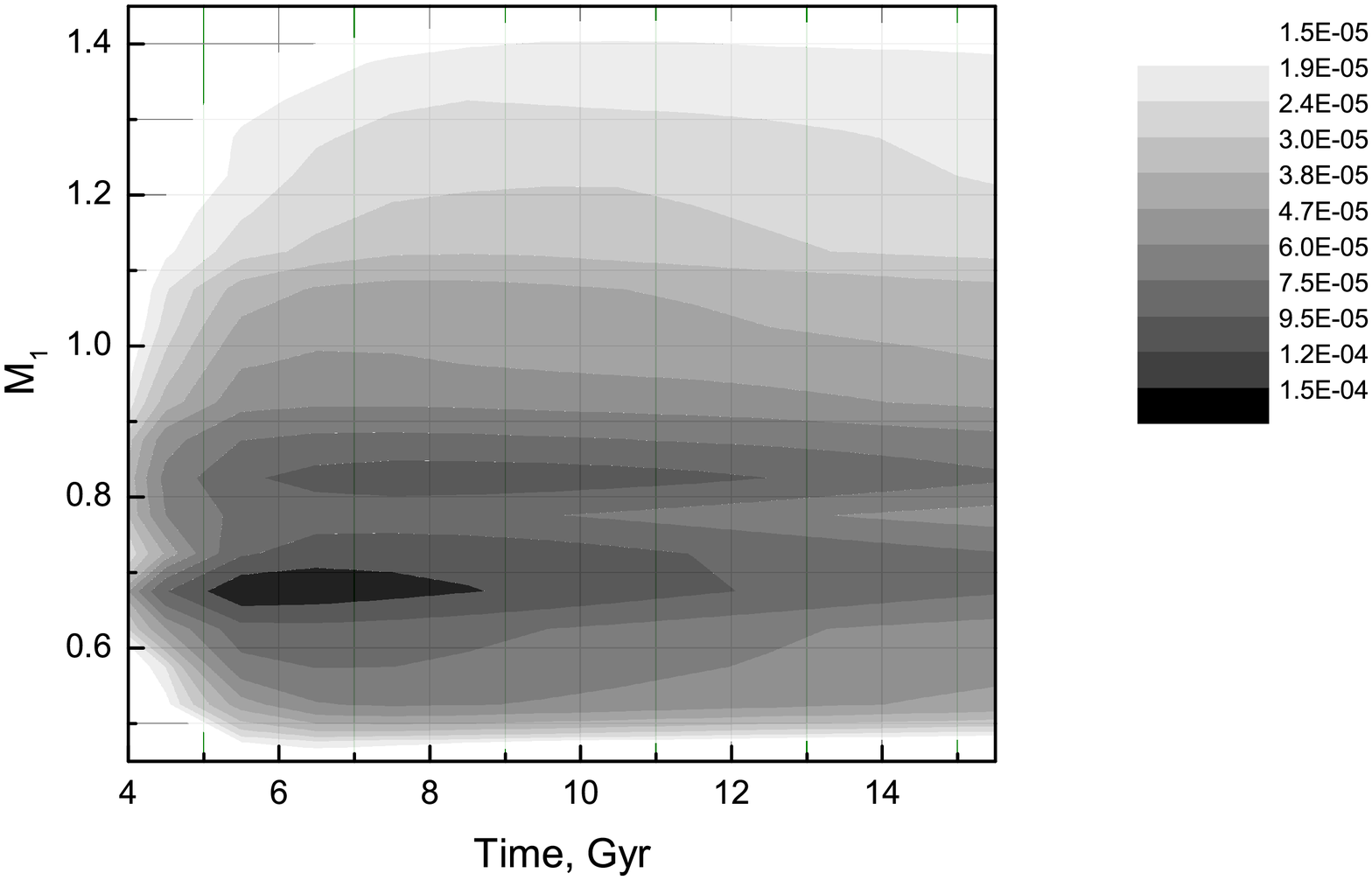}
\includegraphics[height=0.3\textwidth, width=0.49\textwidth]{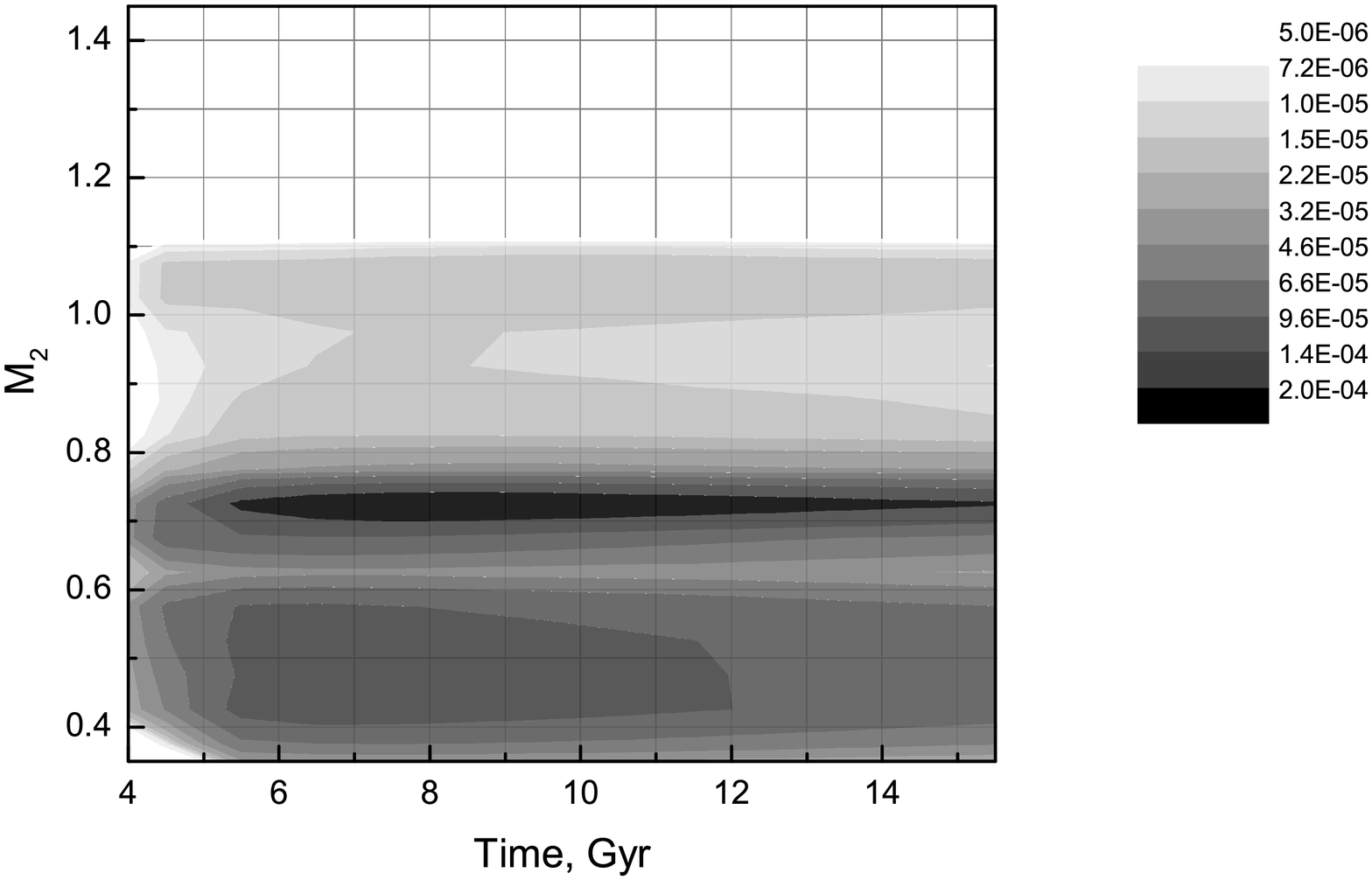}
\includegraphics[height=0.3\textwidth, width=0.49\textwidth]{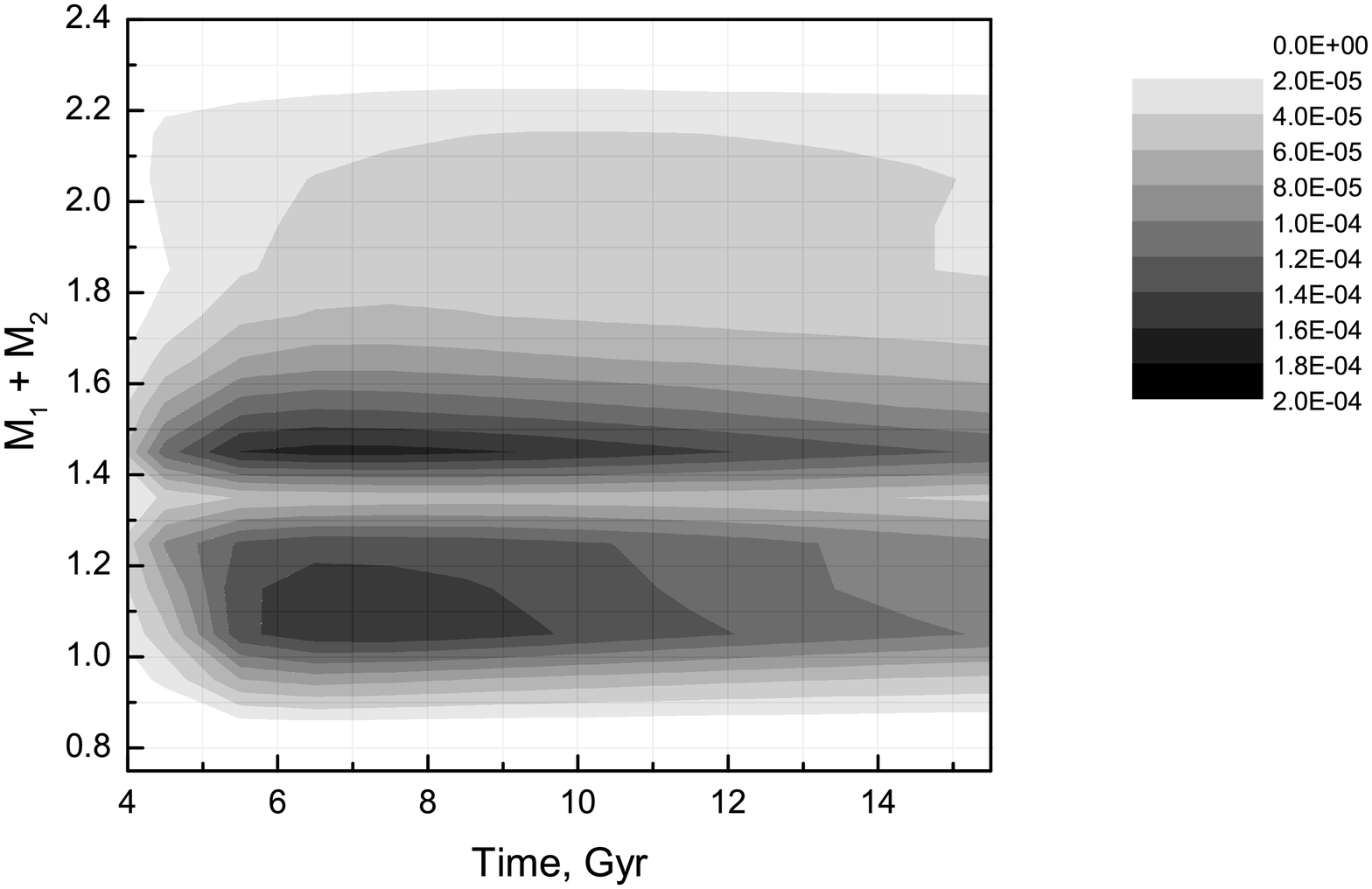}
\caption{Distribution of masses of accretors (upper panel), donors (middle panel) and total mass (lower
panel) of merging pairs of WDs potentially leading to \sne\ versus time  for a 
$10^{10}$\,\ms\ galaxy with the star formation rate set by Eq.~(\ref{eq:sfr}).}
\label{fig:fm1_m2}
\end{figure}

In Fig.~\ref{fig:ftd_ton} we show the evolution of the rate of WD mergers potentially leading to \sne\
for a galaxy mimicking the Milky Way, with $SFR$ given by Eq.~(\ref{eq:sfr}). 
Dependence of the rate of SNe Ia on $\alpha_{ce}\lambda$\ for different scenarios and zones of Fig. 1 is  
presented in tables 5 and 6, respectively. It is clear that, as
in the case of star-formation burst, the rate of mergers increases with \al, because
less systems merge in CEs. The rate of putative \sne\ slightly declines with time. This reflects declining 
$SFR$ and the fact that, in all most prolific scenarios, mergers of DD  
peak at $\aplt 1$\,Gyr and then  decline rather fast.  Our estimate of the possible 
Galactic \sne\ rate owing to DD mechanism, namely $\mathrm {6.5\times 10^{-3}\,yr^{-1}}$  (for the mass of 
the bulge and thin disk equal to 
$7.2\times10^{10}$\,\ms) is close to the latest estimate presented in the literature, namely,  
$\mathrm{(5.4\pm0.12)\times 10^{-3}\,yr^{-1}}$ (with systematic factor $\sim2$; 
\citet{2011MNRAS.412.1473L}). Recall, however, that we employ an extreme assumption that    
\textit{all} the following contribute to \sne: mergers of super-Chandrasekhar pairs of CO WDs, mergers of CO 
WDs more massive than 0.47 $M_\odot$ with hybrid or helium WDs more massive than
0.37$M_\odot$ and mergers of ONe and massive ($\geq 0.9~M_\odot$) CO WDs (Fig. 1).
An increase of these limits, will  reduce obtained rate. 

The masses of accretors have a peak close to 0.7\,\ms\ at $t=(1 - 7)$\,Gyr after beginning of star formation 
in the bulge and thin disk and range, predominantly between 0.45\,\ms and 1\,\ms\ (Fig.~\ref{fig:fm1_m2}).
Later, average $M_1$ values very smoothly become lower and at the current assumed age of the Galaxy
(14~Gyr) most of the accretor masses are between 0.6 and 1.0~\ms. Donor masses have two peaks  -- 
close to 0.7\,\ms\ (CO WD) and at 0.4\,\ms\  to 0.6\,\ms\ (most massive He WD and hybrid WD).  
The existence of two peaks in donor masses results in a double-peaked distribution of total mass of merging 
WDs, with peaks close to 1.0\,\ms\ and 1.4\,\ms. 

\section{Discussion}
\label{sec:anal}

\subsection{Tidal effects}
\label{sec:tideff}

\begin{figure}    
\includegraphics[height=0.3\textwidth,width=0.49\textwidth]{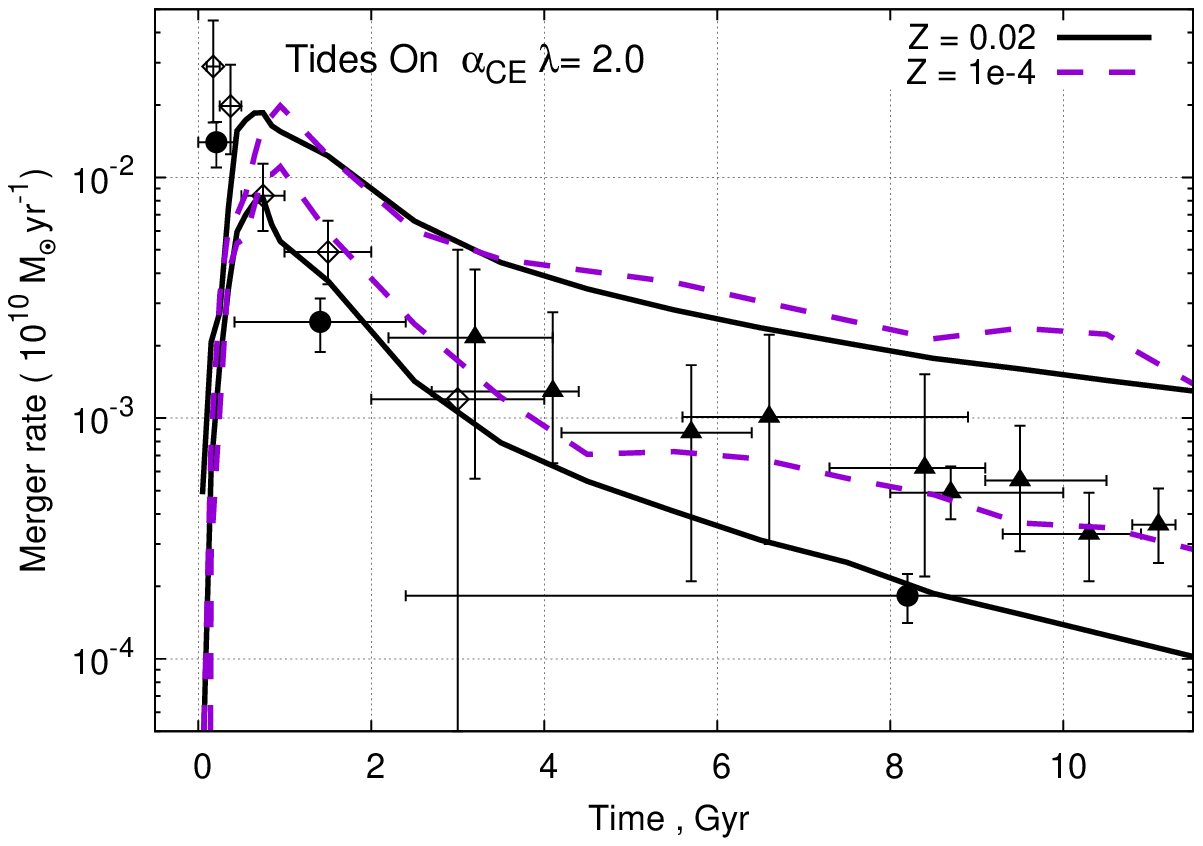}
\includegraphics[height=0.3\textwidth,width=0.49\textwidth]{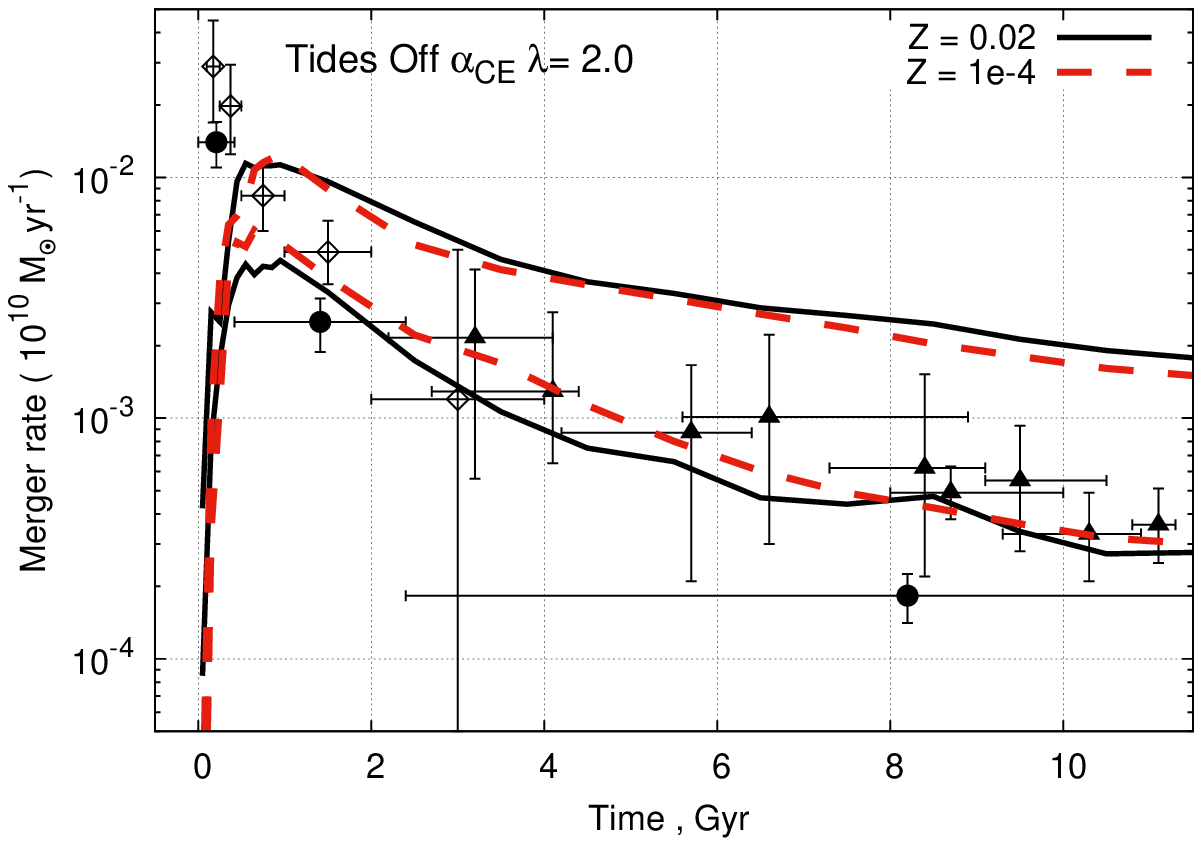} 

\caption{The effect of tides and abundance of metals on the DTD. 
Upper panel: lower solid line, DTD in the `standard' model for \al=2, Z=0.02 with tides taken into account; 
lower dashed line, DTD computed for the same model for Z=0.0001.  
The upper pair of lines shows the DTD for merger of \textit{all} WD  for two values of Z. 
Lower panel: the same distributions for the case when tidal effects are  not taken into account. 
Observational data points are the same as in Fig.~\ref{fig:dtd_ton}.}
\label{fig:z_tides}
\end{figure}

A substantial uncertainty in the results of BPS for putative precursors of \sne\
is caused by the treatment of tidal effects. As noted in \S~\ref{sec:bps}, they are not always taken into 
account, in contrast to our study.
In order to illustrate the effect of tides, we present in Fig.~\ref{fig:z_tides} a model DTD obtained using 
the same BPS code, but excluding tidal effects and compare it with the model DTD obtained `with tides' and 
with observations, as in Fig.~\ref{fig:dtd_ton}. We present only summary curves. 
It is immediately clear that in the  extreme case of absence of tides, the DTD becomes more compatible with 
observations, at least for the DTD derived from \sne\ in the Subaru/XMM Survey by \citet{2008PASJ...60.1327T} 
(diamonds) and in galaxy clusters by  \citet{2010ApJ...722.1879M} (triangles). For comparison, we also show 
that, if as an extreme assumption we suppose that \textit{all} merging WD produce 
\sne, the rate of the latter due to DD-scenario becomes even higher than observed.  
The reason for better agreement with observations may be understood as the effect of typically 
later RLOF in the same systems and production of more massive WD. Some systems experience
case C of mass exchange instead of case B. 

\subsection{Metallicity}
\label{sec:meteff}

Another source of uncertainty is stars with sub-solar metallicity: ${\mathrm Z} \sim  0.0001$ may
be typical for the first generation of stars enriched by heavy elements produced in explosions of
Pop.~III stars \citep{2015MNRAS.452.2822S}. Even in the Galaxy about 10 ultra-low-metallicity
([Fe/H] $<$-7) stars are known \citep{2014Natur.506..463K}.
Reduction of Z to an extreme value of 0.0001 results in improvement of agreement with observational DTD 
(Fig.~\ref{fig:dtd_ton}). The reason is a general 
increase of pre-contact masses of stars with decreasing Z, owing to strongly reduced stellar 
winds. However, this result should be taken with a pinch of salt: evolution of close binaries 
with non-solar Z in all BPS codes is an extrapolation of computations for Z=0.02. Because of
different masses of stellar remnants at the end of similar evolutionary stages, 
the time-scales of evolution, further evolutionary scenarios and/or their 
relative significance may be different. In addition, the chemical composition of WDs depends on their initial
metallicity and may influence the development of explosions involving  them. 

\subsection{Stability of mass-loss by the donors.}
\label{sec:stabdon}

A long-standing problem is stability of mass-loss by donors with deep convective envelopes
and in systems with high mass-ratio. In simulations, we used  
critical values of the mass-ratios of components for low-mass stars and  giants 
implemented in BSE \citep[][Eqs.~(56), (57)]{htp02}, which 
allow stable mass loss by stars with deep convective envelopes only for 
$q_{cr}=M_{don}/M_{acc} \aplt 1$, even if the donor has a condensed  core. 
However, recent studies \citep{2011ApJ...739L..48W,2012ApJ...760...90P} have shown that 
dynamical mass loss by red giants may be  avoided owing
to existence of a super-adiabatic outer layer of the giant's envelope which has a local thermal 
time-scale comparable to the dynamical time-scale and has enough time to 
readjust thermally. \citet{2015MNRAS.449.4415P} found that $q_{cr}$ varies from 1.5 to 2.2 for conservative 
mass transfer\footnote{Earlier, the possibility of $q_{cr} \simeq2$ was found by 
\citet{2008MNRAS.387.1416C} in evolutionary computations, but no 
 physical justification was provided.}.   

In scenarios 1 -- 4 (Table~\ref{tab:scen}), the first CO WD in the system forms via stable RLOF.
For revised upward  $q_{cr}$, more systems would avoid the first CE and, possibly, evolve to form a pair of 
WDs. Because typical separations of components after the second CE will be larger, a lower efficiency of 
matter ejection in CEs will be required for retaining or increase the rate of WD mergers.  

\subsection{Pre-CE core radii of the donors}
\label{sec:coreradii}

\citet{2014MNRAS.444.3209H} called attention to an uncertainty inherent to \textit{all} 
BPS codes: it is unclear what values of radii is necessary to compare 
in order to find whether stars merged in CEs, namely,  pre-CE core radii of the donors or
the radii of post-CE stripped remnants. Currently, in BSE the first option is implemented. As well,
the radii of stellar cores themselves are poorly approximated.  
This uncertainty may influence, mostly, the outcome of CE produced by RLOF in the pairs
of (i) HG and RGB, (ii) RGB and RGB stars, and (iii) in binaries harbouring HG or RGB stars with He WD 
companions. To test suggestion of \citet{2014MNRAS.444.3209H}, we implemented in the code suggested by 
them corrections for RGB and TPAGB stars. However, in our simulations we  
did not encounter events of the kinds (i) and (ii), while CEs with He WD are rather rare.
Thus, our results remained virtually unaffected.
The above-mentioned imperfections are partially outset by uncertainties in the core-envelope definition (i. 
e., envelope binding energy)  and CE ejection efficiency. Hall \& Tout noted that it is difficult to 
constrain the parameters found by them to be uncertain,  by observations, since spatial densities 
of the concerned binaries are poorly known.
All the above-mentioned problems in the treatment of binary star evolution still
require a more systematic investigation before it will be possible to quantify the respective effects to the 
degree which will allow to make the necessary corrections in BPS codes.   It is  impossible to evaluate the 
influence of all of them on the rate of formation of putative progenitors of \sne, but it may be easily
suspected that, for example, the rates of merger of WDs will change within a factor of 2 -- 3.
  
\begin{figure}     
\includegraphics[height=0.3\textwidth,width=0.49\textwidth]{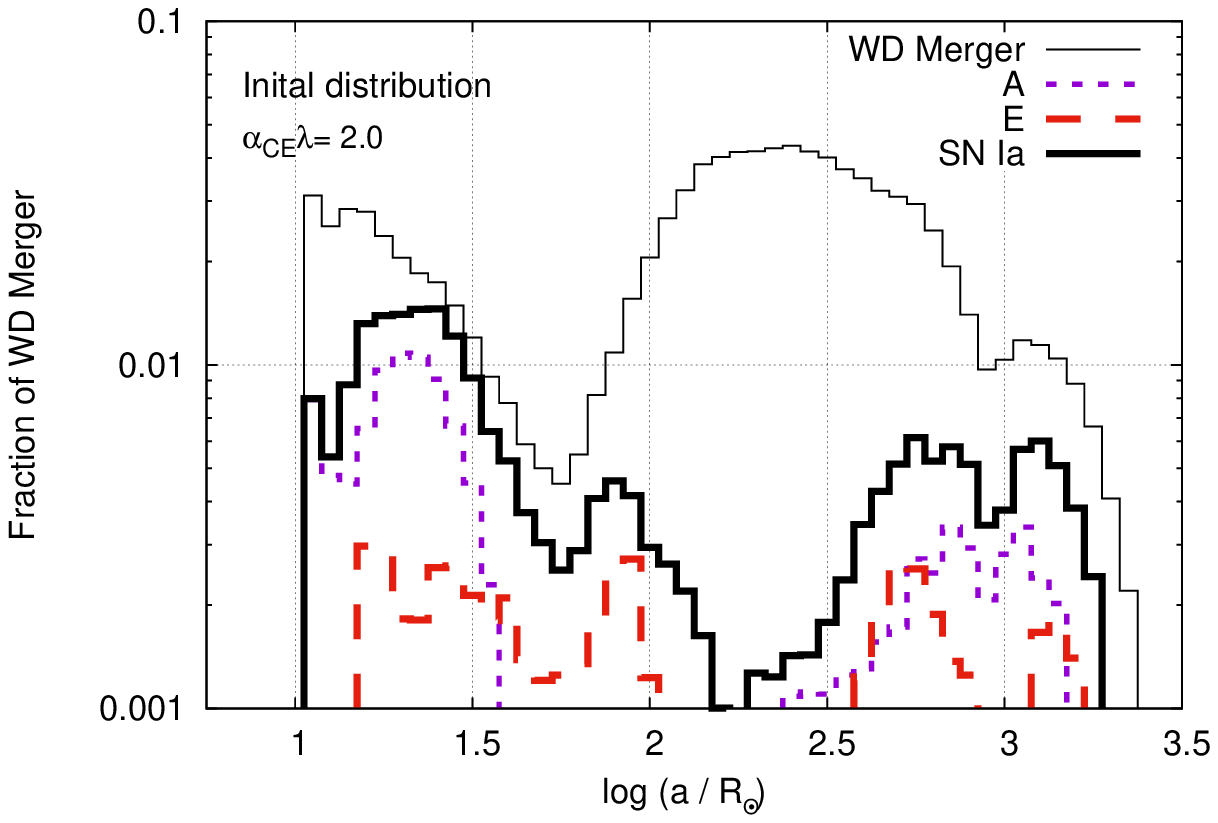}
\includegraphics[height=0.3\textwidth,width=0.49\textwidth]{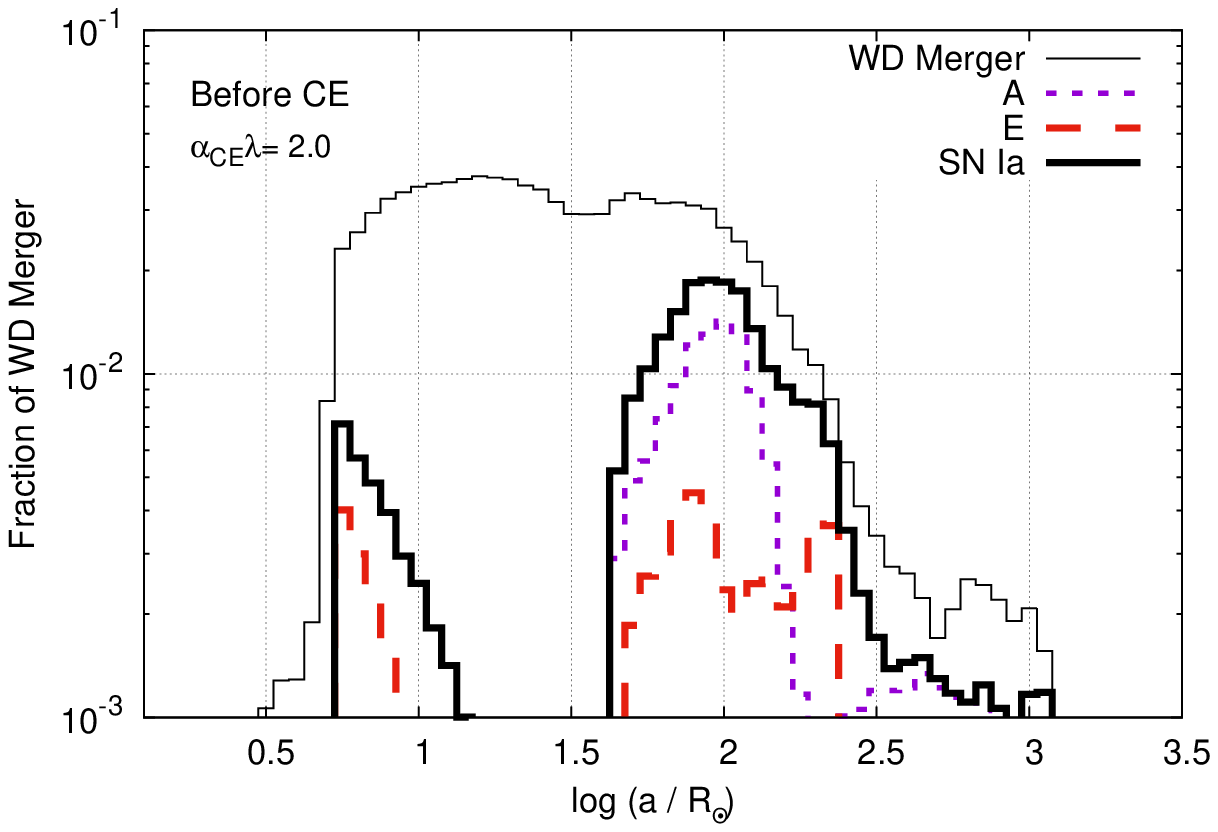}
\includegraphics[height=0.3\textwidth,width=0.49\textwidth]{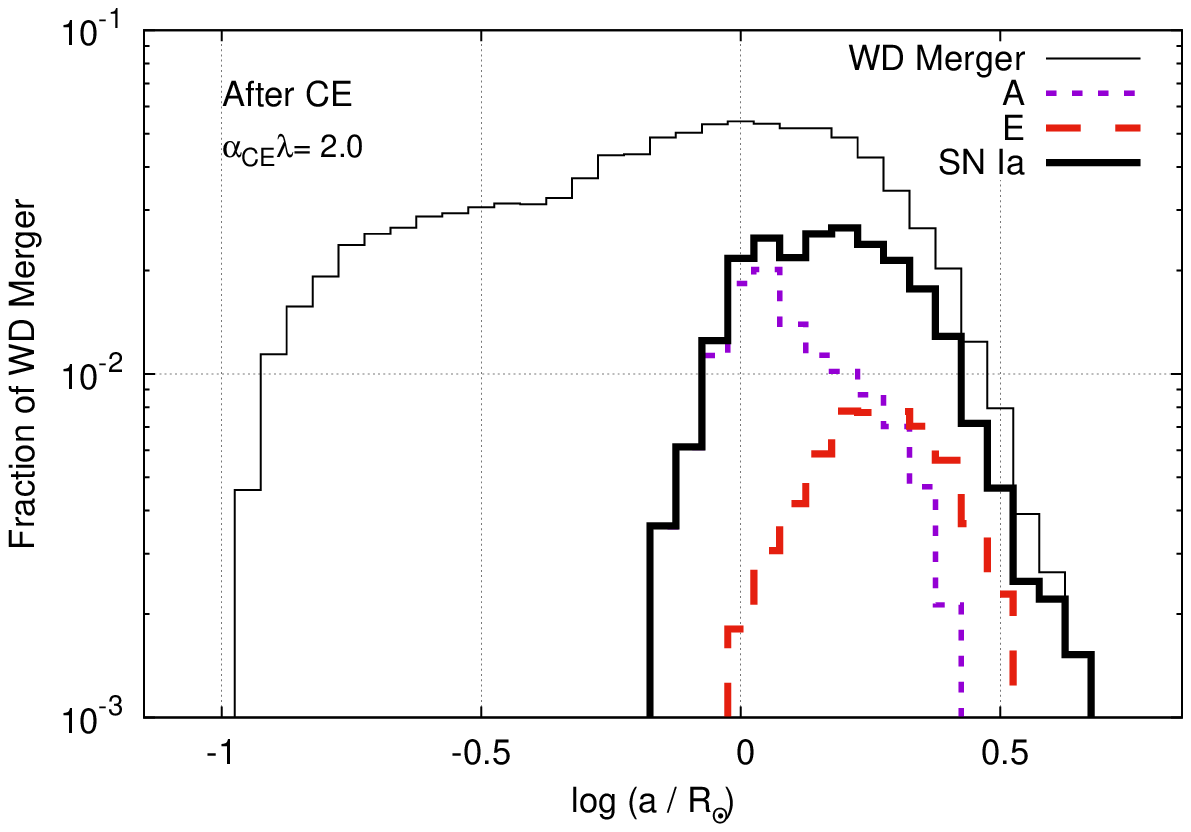}  
\caption{Distribution of separations of components of precursors of merging WD 
pairs $dN/d\log(a)$. The distribution is normalized to the total number of WD that 
merged over 10\,Gyr (WD Merger in the legend).  
Upper panel, initial distribution.
Middle panel, distribution before the last common envelope (CE) stage. 
Lower panel, distribution after the last CE.}
\label{fig:dist_ce}
\end{figure}

\begin{figure}   
\includegraphics[height=0.3\textwidth,width=0.49\textwidth]{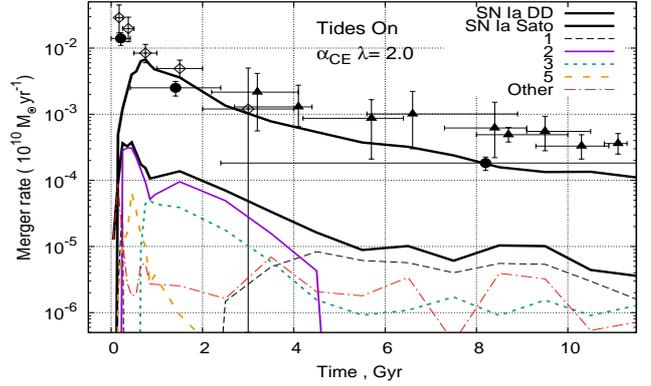}
\caption{Upper solid line, DTD in the `standard' model. 
Lower sold line, DTD for mergers of WD satisfying Eq.~(\ref{eq:sato16}). The rest of the lines shows
the contributions of particular scenarios, as in Fig.~\ref{fig:dtd_scen}.   }
\label{fig:dtd_sato16}
\end{figure}

\subsection{The slope of the DTD curve}
\label{sec:dtdcurve}

The slope of DTD curve vs. the time close to $t^{-1}$ is often considered as evidence in favour of DD-
scenario  being the main 
mechanism producing \sne. If the distance between the components $a$ after the last CE episode obeys the
power-law  $dN/da\propto a^{\epsilon}$, while the merger time depends on $a$ as 
$t\propto a^{\gamma}$, the dependence of the DTD on time should be a power law with index    
$\phi=-1+(\epsilon+1)/\gamma$. It is usually \textit{assumed} that merging pairs of WD are distributed 
over $a$ like  main sequence stars. Then $\phi=-1$, by virtue of almost `standard' assumption $\epsilon=-1$
\citep{pty82}. In Fig.~\ref{fig:dist_ce}, we show distribution of progenitors of merging WD+WD
binaries on the main sequence and before and after the last CE episode. It is clear that in the course of 
evolution distribution over $a$ experiences complicated non-linear transformation  and, separately, 
neither sub-\mch\ nor (super)-\mch\ mergers obey a $a^{-1}$ law, but the latter is, crudely,  followed by 
their combination. If other mechanisms also contribute to \sne, they may also influence the slope of the DTD 
curve. The disagreement of the shape of model time-dependence of merging double-degenerates and a simple 
power-law was also noted by \citet{2016ApJ...826...53A}.

\subsection{The rate of \sne.  }
\label{sec:ratesne}

The aim of our study was to estimate the upper bound to the input of mergers of WDs into the rate 
of \sne. For a satisfactory agreement with observations, at least some of the included model events
should be sub-\mch\ mergers, involving CO accretors and massive He WD or  hybrid WD as donors.  
Observations provide some evidence in favour of existence of sub-\mch\ \sne, through a  
significant scatter in the estimated masses of produced $\mathrm{^{56}Ni}$ and ejected mass, which 
for some samples cluster around certain values, substantially below \mch\
\citep[e. g., ][]{2006A&A...450..241S,2007Sci...315..825M,2014MNRAS.440.1498S,2014MNRAS.445.2535S,
2015MNRAS.454.3816C}. In particular, \citet{2014MNRAS.440.1498S,2014MNRAS.445.2535S} claim 
that the fraction of sub-\mch\ \sne\  may be up to 50 per cent. However, as yet, no conclusions are drawn, 
whether these features of \sne\ are related to their mass or to variations in the
explosion conditions of \mch\ WDs. On the other hand, our model sample of merging pairs of WDs
contains a substantial fraction of strongly super-\mch\ pairs. As shown by \citet{2014ApJ...785..105M},
however, the spectra of \sne\ produced in these events resemble the spectra of `normal' \sne.   

We have mentioned in comments to Figs.~\ref{fig:dtd_ton} and \ref{fig:dtd_scen} that, within observational
errors we get satisfactory agreement
with observations even if we assume that only CO+CO WD pairs with $M_1 \geq 0.8$\,\ms\
 and $M_2 \geq 0.6$\,\ms\ can produce \sne. Here, we implicitly assumed that \sne\ may 
explode either in the merger stage or in the merger product evolution stage. This  issue is, however, still 
open. If we assume that explode  pairs in which explosion conditions are met in the merger stage only and 
Eq.~(\ref{eq:sato16}) should be satisfied, the expected rate of \sne\ sharply drops. In 
Fig.~\ref{fig:dtd_sato16} we compare observational DTD with model distribution for systems 
obeying Eq.~(\ref{eq:sato16}). A reduction of the rate by a factor $\sim$20 at $t=$10\,Gyr is 
immediately seen. This may signify  serious problems either with observational estimates of DTD
or in our understanding of processes that occur during and after mergers. It is interesting that 
scenario 3, which is one of the main contributors to the \sne\ rate in the `standard' model, in `Sato et al.' 
case terminates production of massive mergers after $\approx4$\,Gyr. 

\section{Conclusion}
\label{sec:concl}

In the present study we have attempted to estimate the maximum possible contribution of merging binary WD to 
the total rate of \sne.  The main motivation of the study was the fact that, currently, merger of WDs is the 
only known (but still hypothetical)  mechanism that has a natural time-scale overlapping with 
Hubble time. In our study we did not consider relations between different types of \sne\ and 
possible combinations of components of the merging pairs, as we clearly recognize that there are `grey zones' 
in which current simulations of mergers do not demonstrate events that are similar to \sne\ of any known kind.
This may be related to inaccessibility of the physical conditions necessary for \sne\ explosions or be a 
result of an inadequate understanding of the physics, or of numerical problems. The only limits imposed on 
the components of merging pairs arose from the knowledge that merger results at least in a single detonation 
which may be treated as a transient event.

We found the most common scenarios of close binary star evolution that result in formation of merging pairs 
of WDs and studied dependence of their relative role on the still cryptic parameters of binding energy of the 
stellar envelopes ($\lambda$) and efficiency of expulsion of matter in the CE stages (\ace).
We parametrized scenarios by the product \al, which we varied from 0.25 to 2.0.
We found, in agreement with some earlier studies 
\citep[e. g., ][]{2010A&A...515A..89M,2012A&A...546A..70T,2013MNRAS.429.1425R,2014A&A...563A..83C}, that most 
merging pairs have stable RLOF 
as the first stage of mass transfer. At low \al\ the dominant scenarios are those in which the first RLOF 
occurs when the primary star overflows Roche lobe in the Hertzsprung gap or in the red giant stage. For 
larger \al, scenarios in which the first RLOF happens in EAGB or TPAGB stages start to play role. It is 
important, that in about 50 per cent of merging pairs of WDs 
at least one of components is a He or a hybrid CO WD. The latter, in fact, dominate. The presence of at least 
traces of He at the surface of WDs may facilitate explosion at merger.  With increase of \al, the role of 
different scenarios becomes more even. 

With an increase of \al\ from 0.25 to 2, the total rate of mergers increases. If $\alpha_{\rm ce}\lambda$ is 
low, a large fraction of systems merges in CE. If this product of model parameters is high, a significant 
fraction of WD+WD pairs is too wide to merge in Hubble time.

We compared the model DTD with results derived from observations of several samples of \sne. In the model, we 
consider the mergers singled out in \mamd\ diagram as precursors of \sne.  The best agreement with 
observations was obtained for high \al=2.  Within observational errors, at $1\la t \la 8$\,Gyr, the
model DTD for all mergers agrees with DTD derived by 
\citet{2008PASJ...60.1327T,2010ApJ...722.1879M,2012MNRAS.426.3282M}.
For earlier epochs, model DTD has about 3 times less events than observed DTD. For $t \approx 10$\,Gyr, the 
discrepancy with 
the data of \citet{2012MNRAS.426.3282M} is $\simeq 4$. If we take into account only 'canonical' mergers of 
CO+CO WDs with $\mt \geq 1.4$, the  model DTD  is still roughly comparable with the lowest observational DTD 
estimates for field galaxies \citep{2012MNRAS.426.3282M}, as in the studies of 
\citet{2013MNRAS.429.1425R,2014A&A...563A..83C}\footnote{Actually, this is not  surprising, since the 
codes used by \citet{2012A&A...546A..70T,2013MNRAS.429.1425R,2014A&A...563A..83C} are based on the same 
system of evolutionary tracks 
as BSE code; the difference between codes is, as mentioned in the text,  in the treatment of evolution of  binaries 
\citep{2014A&A...562A..14T}.}, but \sne\ rates are higher than in the models of \citet{2016ApJ...826...53A} 
in which a more stringent requirement, $\mt \geq 1.6$\,\ms, $q > 0.8$, is applied. 
The current rate of \sne\ in the Milky Way, if all mergers expected by us 
to result in \sne\ of some kind really produce them, is $6.5\times10^{-3} \mathrm{yr^{-1}}$, remarkably
close to the observationally inferred estimate $(5.4\pm0.12)\times10^{-3} \mathrm{yr^{-1}}$. The model 
estimate may be lower if our Eq.~(\ref{eq:sfr}) overestimates the actual Galactic $SFR$  
\citep[see e. g. ][]{2011AJ....142..197C}. 

\begin{figure}   
\includegraphics[height=0.3\textwidth,width=0.49\textwidth]{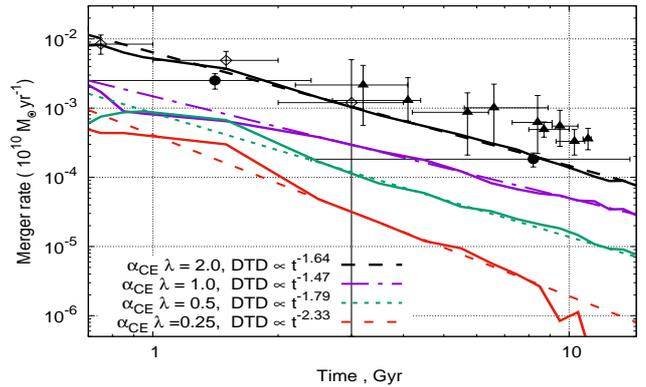}
\caption{The slope of DTD for $t\geq1$\,Gyr, as in Fig.~\ref{fig:dtd_ton}. 
Tides are accounted in the model, Z=0.02. 
Broken lines represent power-law approximations to DTD for different values of \al.}
\label{fig:dtd_al2}
\end{figure}

The transformations of the separation of components during evolution with RLOF and CEs are strongly 
non-linear. Different scenarios produce non-flat distributions over $\log(a)$ and their sum is also not flat. 
As a result, we  find that the model DTD does not depend on time like power law
with an exponent close to -1, as expected from simplified estimates, assuming that after the last CE episode 
distribution of separations of WD is flat in $\log(a)$. The slope of DTD in our models is a power law, 
depending on the assumed \al\ (Fig.~\ref{fig:dtd_al2}). The exponents of power law range from -2.33 to -1.64 
for \al\ from 0.25 to 2, respectively. The curve with the smallest absolute value of the exponent fits 
observations for $t\aplt8$\,Gyr.   

Our model does not fit \citet{2012MNRAS.426.3282M}  bin of DTD at the shortest delay time, namely, 
$t\la$420\,Myr. However, most \sne\ in this time-bin may actually be produced by double-detonations. It was 
shown by \citet{2014MNRAS.440L.101R} that the double-detonation scenario involving non-degenerate donors and 
massive CO WD accretors reaches the  peak at (200--300)\,Myr and extends to about 500\,Myr. At the peak, the 
rate of \sne\ is \ensuremath$(3 - 4)\times10^{-3}\, \textrm{per}\,\, 10^{10}\ms\,\mathrm{yr^{-1}}$. DTD of 
\citet{2014MNRAS.440L.101R} has a second peak, comparable in maximum rate and extending to several Gyr.
It is produced by the systems with long-living degenerate He-donors (AM~CVn stars). However, it was shown by 
\citet{2015MNRAS.452.2897P} that in such systems outbursts of He burning never become dynamical and, 
respectively, AM~CVn stars with He WD donors produce neither `regular' \sne\ nor lower scale SNe~.Ia. Thus, 
this scenario can not play any role in early-type galaxies with majority of stars formed in the initial spike 
of star formation. For the earliest epochs, inclusion of both SNe Ia of double-degenerate origin and double-
detonation SNe Ia may reduce the discrepancy between observations and model. 

Some evidence in favour of the existence of sub-\mch\ \sne\ is not related directly to interpretation of 
their observations. As noted above, \citet{2013A&A...559L...5S} noticed  that manganese is produced 
efficiently in explosions only of WDs close to \mch. The observed  [Mn/Fe] ratio in solar neighbourhood may 
be reproduced, if about half of \sne\ involve near-\mch\ WD, while the rest may be sub-\mch. However,
this conclusion does not restrict the mechanism of sub-\mch\ \sne. \citet{2012ApJ...749L..11B} estimated the  
merger rate of Galactic binary WD and found that it is rather similar to the inferred rate of \sne\ in the 
Milky Way-like Sbc galaxies. They concluded that there are not nearly enough super-\mch\ pairs of WDs to 
reproduce this rate and, therefore, sub-\mch\ pairs may be partially responsible for the \sne\ rate.    
  
One prediction of single-degenerate models with non-degenerate donors is sweeping of H- or He-rich material 
from the envelope of the donor by the SN ejecta. Up to now, 17 `normal' \sne\ have been surveyed for swept-up 
matter. The upper limits on the amount of the latter are inconsistent with MS/RG donor 
\citep[see ][and references therein]{2016MNRAS.457.3254M}. On the other hand, as noted by 
\citet{2013ApJ...770L..35S}, nascent He WDs have thin H-envelopes which, in the case of merging WDs,  will
be transferred stably onto CO accretor prior to tidal disruption of the He-core of the donor. This hydrogen
is likely to be ejected from the binary in Novae eruptions and sweep-up the surrounding 
ISM hundreds to thousands of years prior to a possible \sna. As found by \citet{2013ApJ...770L..35S}, it may 
create ISM profiles closely matching those, inferred from  observations of some \sne. As well, interaction of 
tidal tails with ISM may create NaI~D-line profiles similar to those observed \citep{2013ApJ...772....1R}. 
Thus, observations of narrow absorption lines from circumstellar  medium do not necessarily manifest the 
presence of a non-degenerate component in a pre-\sna\ binary.   

On the other hand, the existence of a single-degenerate channel to \sna\ may be  signified by 
enhanced brightness and blue and ultraviolet emission arising when the ejected material interacts with 
companion star \citet{2010ApJ...708.1025K}. UV-emission bursts were recently observed in the early spectra of 
several \sne\ \citep{2015Natur.521..328C,2015ApJS..221...22I,2016ApJ...820...92M}. 
However, interpretation of their observations as manifestation of SD-scenario is still a matter of debate 
\citep{2016MNRAS.459.1781L,2016MNRAS.459.4428K}. For a variety of SD-scenario -- \sne\ in symbiotic systems 
 -- \citet{2016ApJ...821..119C}, based on radio-observations, limit the fraction of \sne\ in systems with red giant components to $\aplt$10 per cent.
Strong evidence for Chandrasekhar-mass explosions is provided by observations of SN remnant  3C 397 
\citep{2015ApJ...801L..31Y} for which Ni/Fe and Mn/Fe mass ratios derived from X-ray observations are 
consistent only with nucleosynthesis processes occurring in near-\mch\ SN.    

It would be incorrect to blame population synthesis alone for the mismatch of models and observations.
As we noted in \S~\ref{sec:merger}, SPH simulations of merger process suffer from insufficient resolution. 
Increase of the latter will allow us to resolve smaller hot regions, thus allowing better understanding of 
conditions for detonation at contact, especially in the systems with He-rich donors. DTD, in turn, are 
uncertain by almost an order of magnitude, as seen in the Figures above;
almost certainly this is not an effect of different methods applied for their recovery, but  also an effect 
of dependence of samples of \sne\ under study on the environment. 

We have shown that by accounting for all WD merger events that hypothetically \textit{may}  produce \sne\ 
either during merger processes or in the course of further evolution, it is possible within reasonable limits 
to explain the DTD for \sne\ and the rate of \sna\ in the Milky Way. However, the diversity of combinations 
of chemical composition of components of merging pairs and their masses leaves open the question, why 
majority of \sne\ are so `standard'.

\vspace{2mm}
The authors acknowledge the referee for his/her valuable comments.
We appreciate helpful discussions with M. Dan, N. Chugai, G. Nelemans, K. Postnov, S. Toonen, H.-L. Chen. 
D. Kolesnikov is acknowledged for trial computations of evolutionary sequences for helium stars. 
We acknowledge J. Hurley and his co-authors for making the code BSE public. 
The study was partially supported by Basic Research Program P-7 of the Praesidium of the Russian Academy of 
Sciences and  Russian Foundation for Basic Research (contracts No. 14-12-00146 and 14-02-00657).
AGK acknowledges support from M.V. Lomonosov Moscow State University Program of Development. 
This research has made use of NASA’s ADS Bibliographic Services.

\input{preprint.bbl}
\appendix
\renewcommand\thefigure{\thesection\arabic{figure}} 
\setcounter{figure}{0}   
\newpage
\section{Scenarii}
Below, we show for scenarios 1 -- 8 the initial distributions of the masses of components of the progenitor 
systems, the initial relations between masses of primaries and the separation of the components in these 
systems and the location  of components of merging systems in the \mamd\ diagram.
Labels in the Figures correspond to Table~\ref{tab:scen}. 

In Tables A1-A8  we present for each scenario typical tracks leading to the merger, indicating 
evolutionary lifetime (T),  evolutionary status of the components (Star1 and Star2),
the filling factors of the respective Roche lobes (R1/RL1 and R2/RL2), the masses of the components (M1 and M2) and their separation A (in solar units).
If filling  factor is $<0.01$ we assign to it, for brevity, the value of 0.0. 

\begin{figure}       
\includegraphics[width=0.49\textwidth]{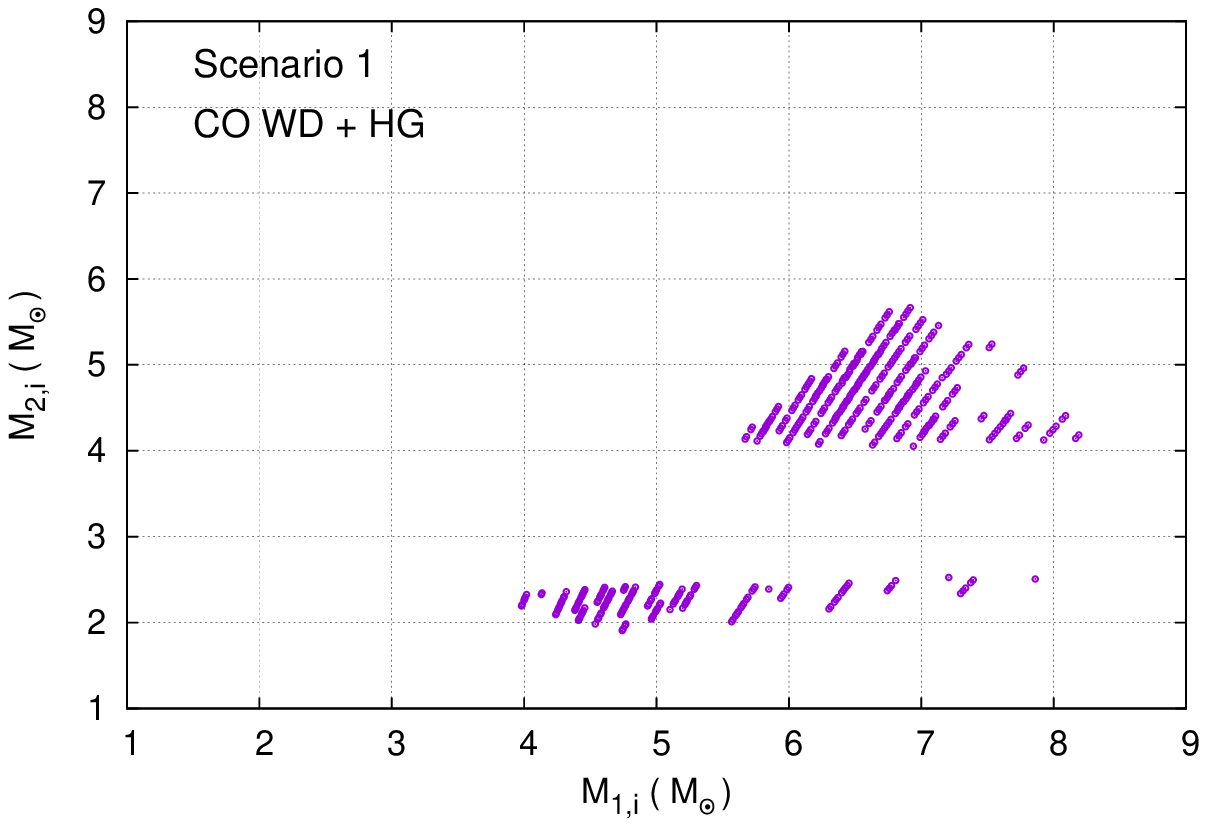}
\includegraphics[width=0.49\textwidth]{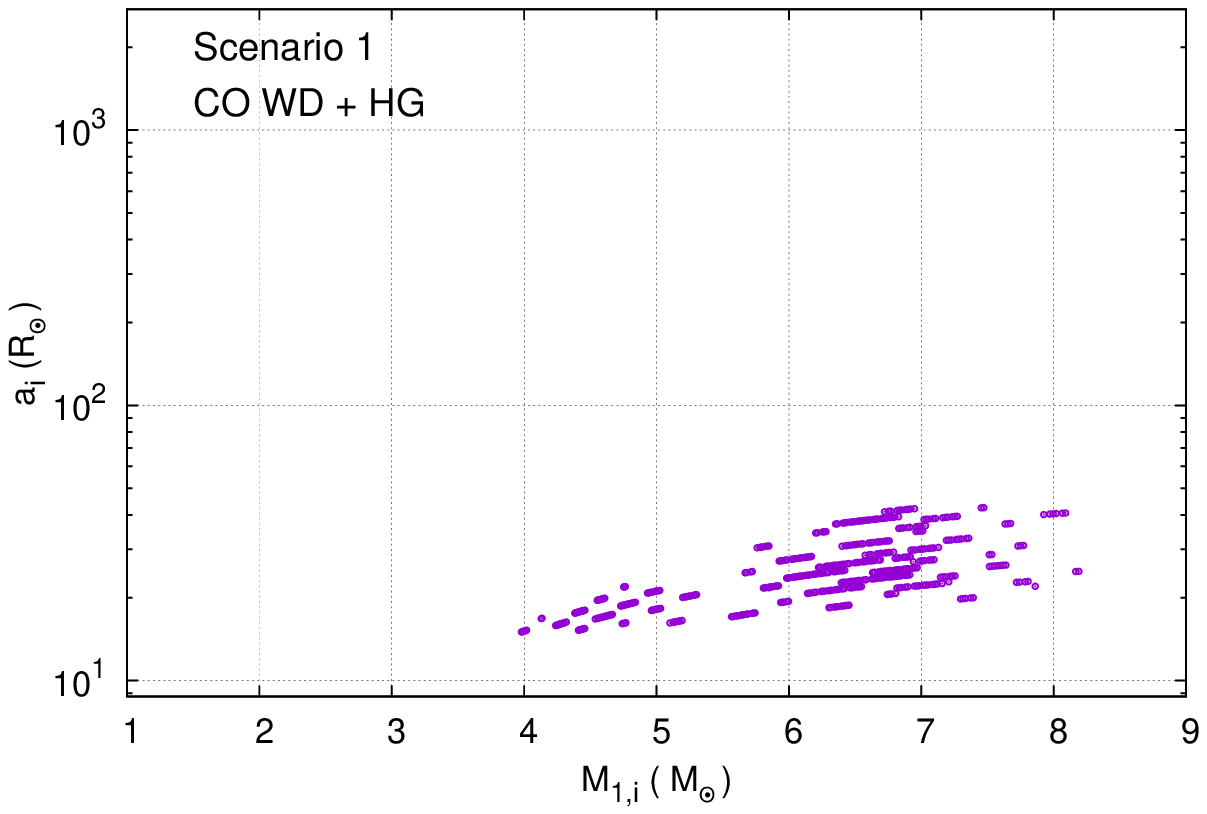}
\includegraphics[width=0.49\textwidth]{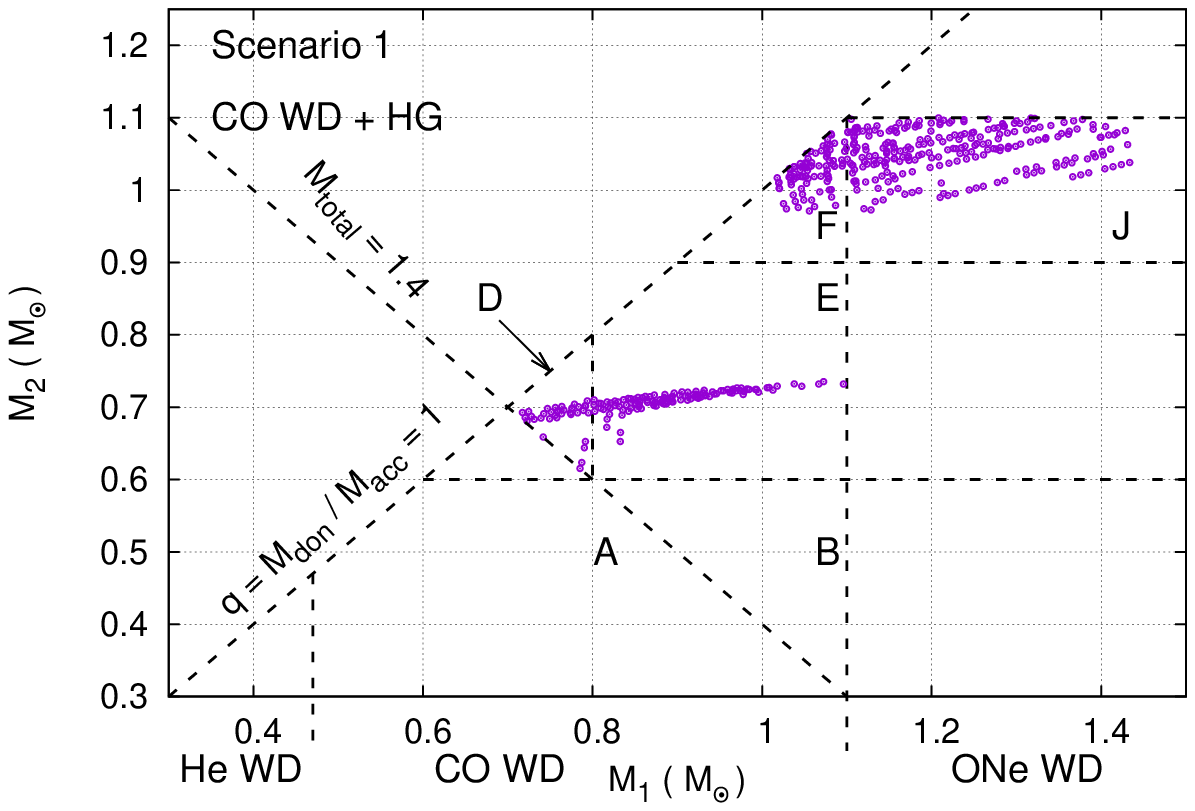}
\caption{Close binaries evolving via scenario 1.
Upper panel: initial distribution of the masses of the components.
Middle panel: initial relationship between the masses of primaries and the separation of components.
Lower panel: position of merging systems produced via scenario 1 in \mamd\ diagram.
The label in the Figure indicates evolutionary stage of the system prior to the first common envelope 
episode in the system as in Table~\ref{tab:scen}. }
\label{fig:scen1}
\end{figure}

\begin{figure}      
\includegraphics[width=0.49\textwidth]{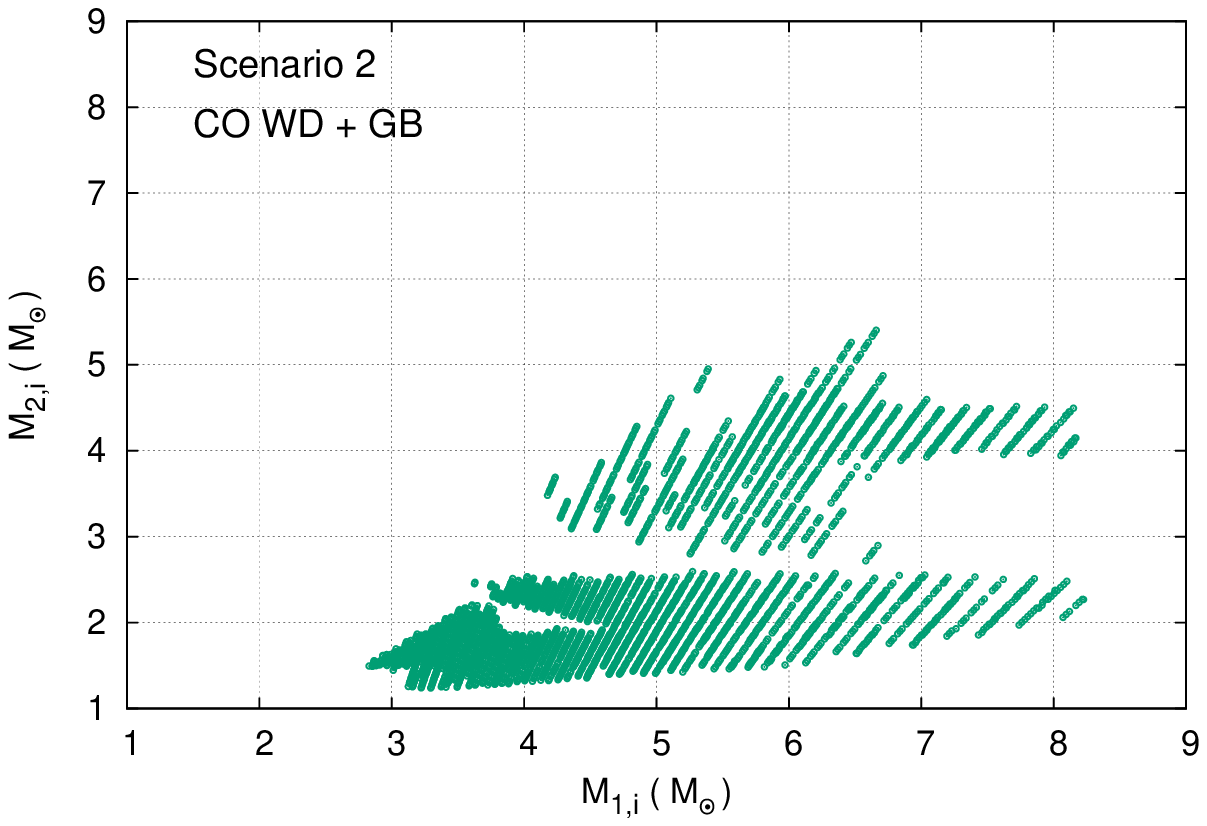}
\includegraphics[width=0.49\textwidth]{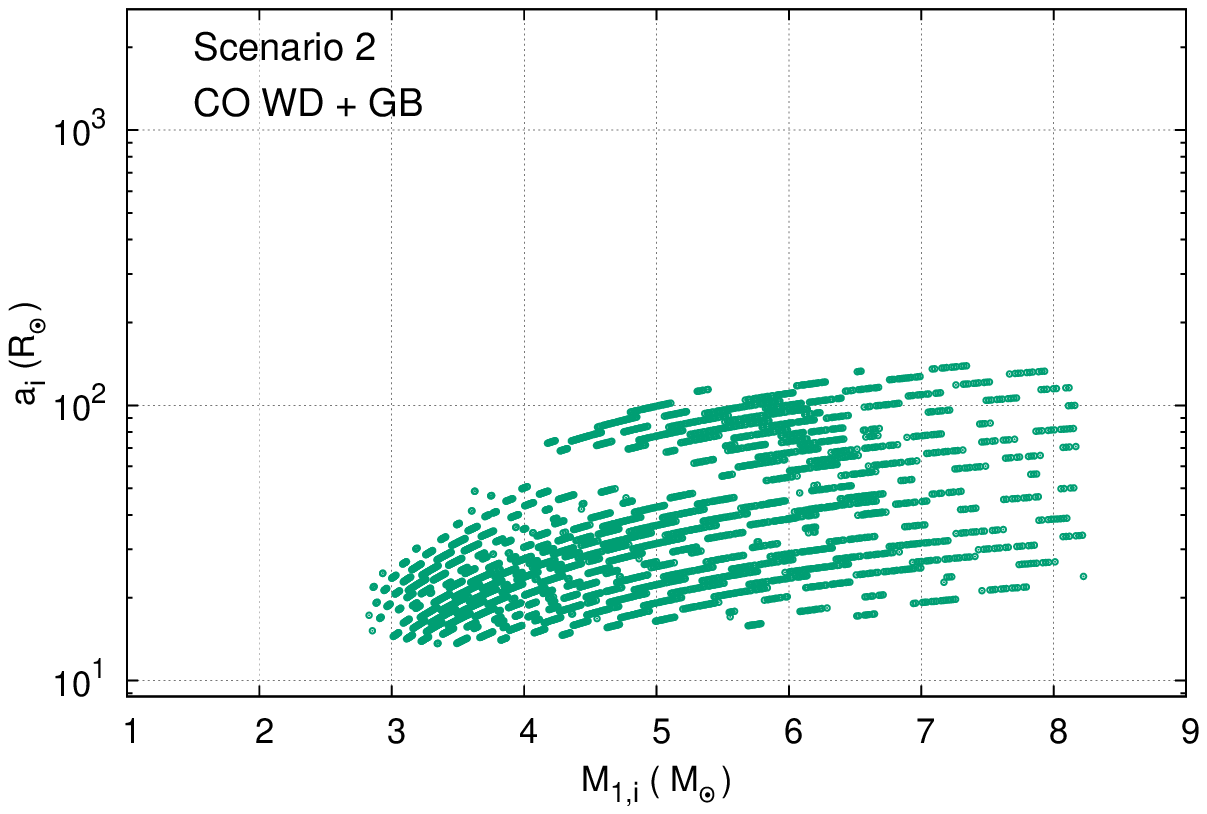}
\includegraphics[width=0.49\textwidth]{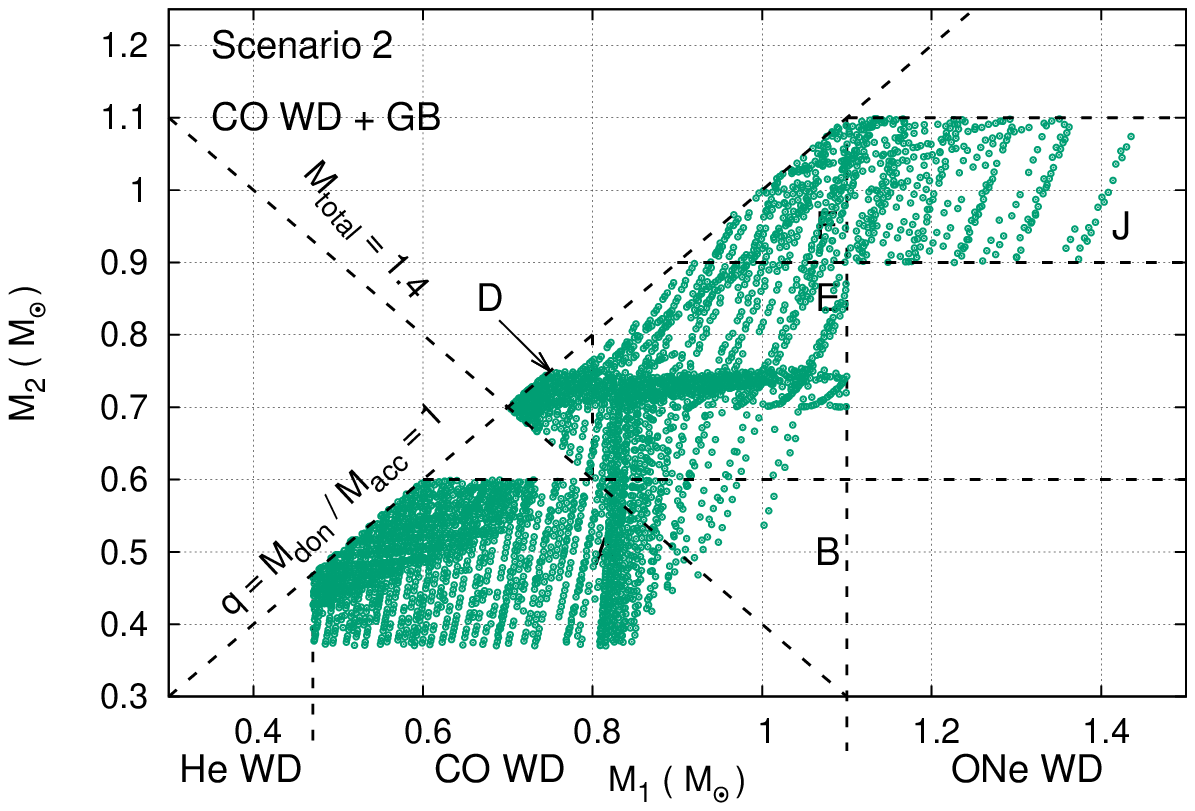}
\caption{As Fig.~\ref{fig:scen1}, but for scenario 2.}
\label{fig:scen2}
\end{figure}

\newpage

\begin{figure*}      
\includegraphics[width=0.49\textwidth]{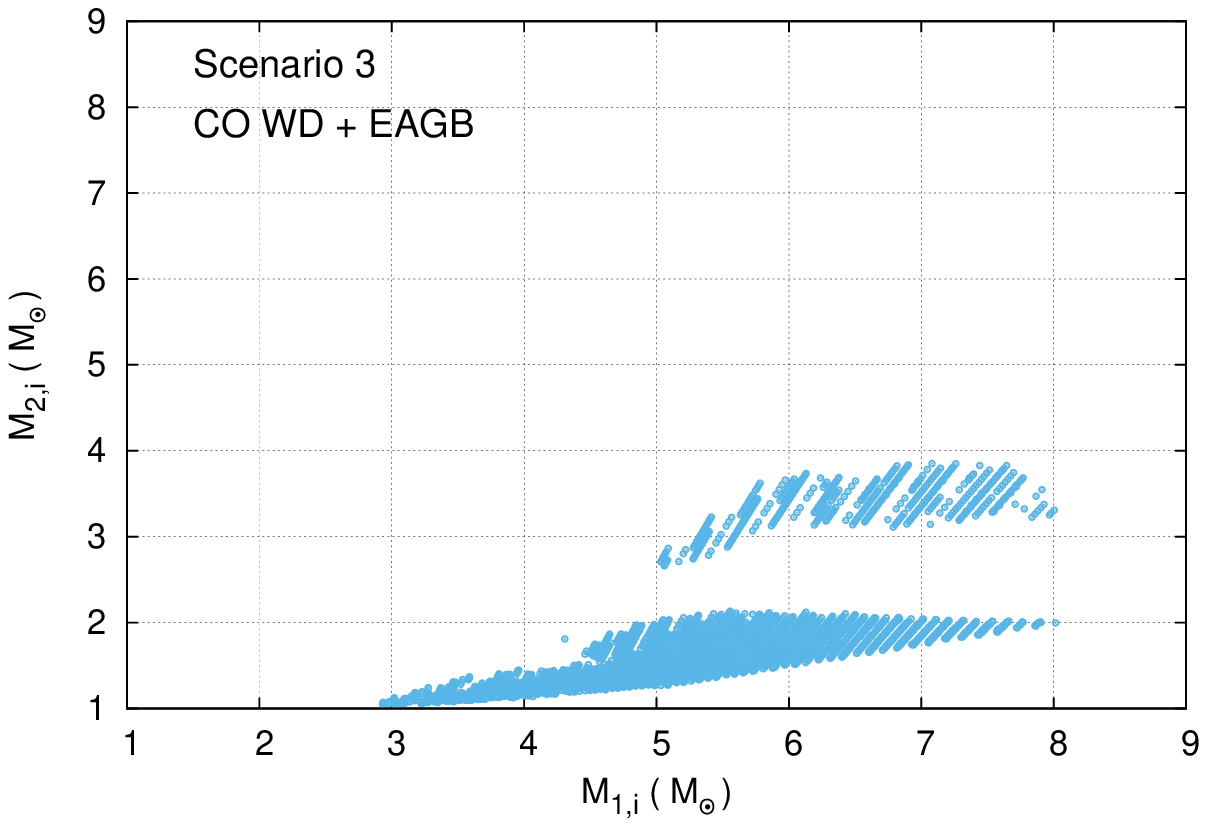}
\includegraphics[width=0.49\textwidth]{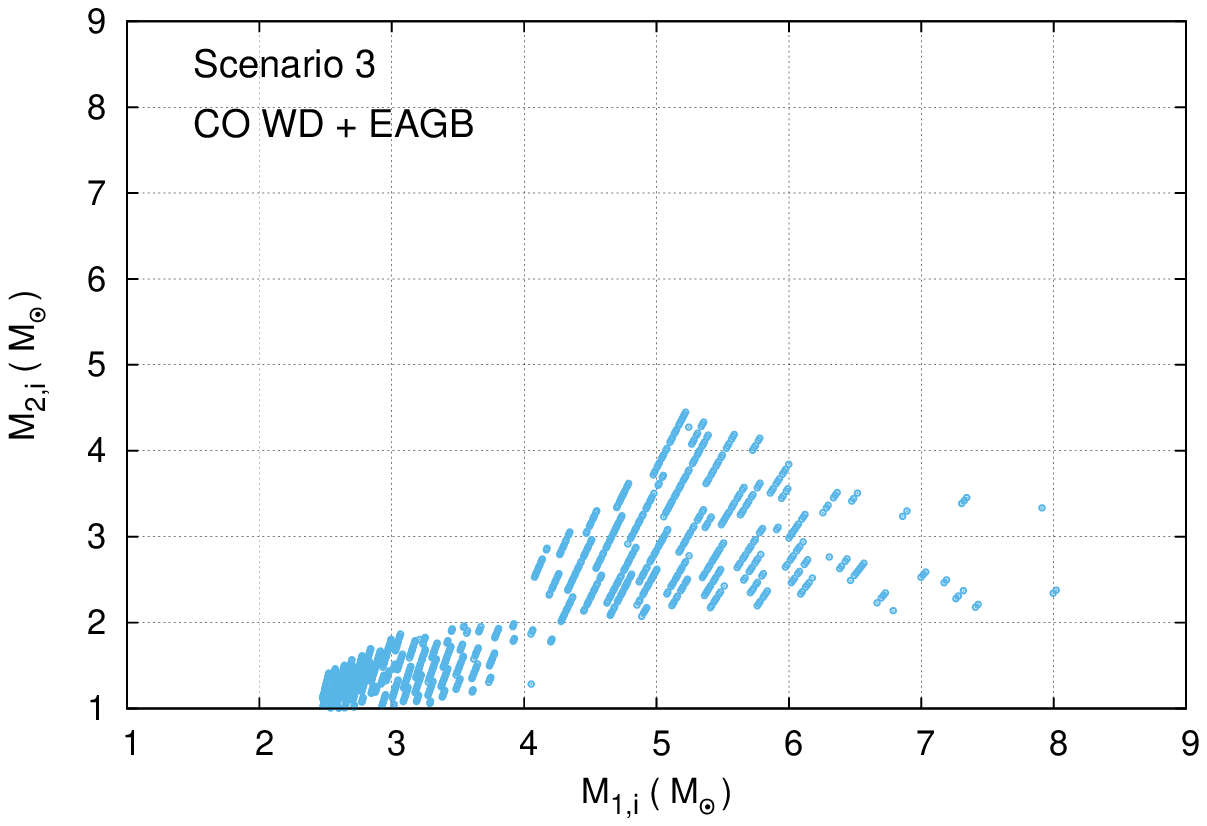}
\includegraphics[width=0.49\textwidth]{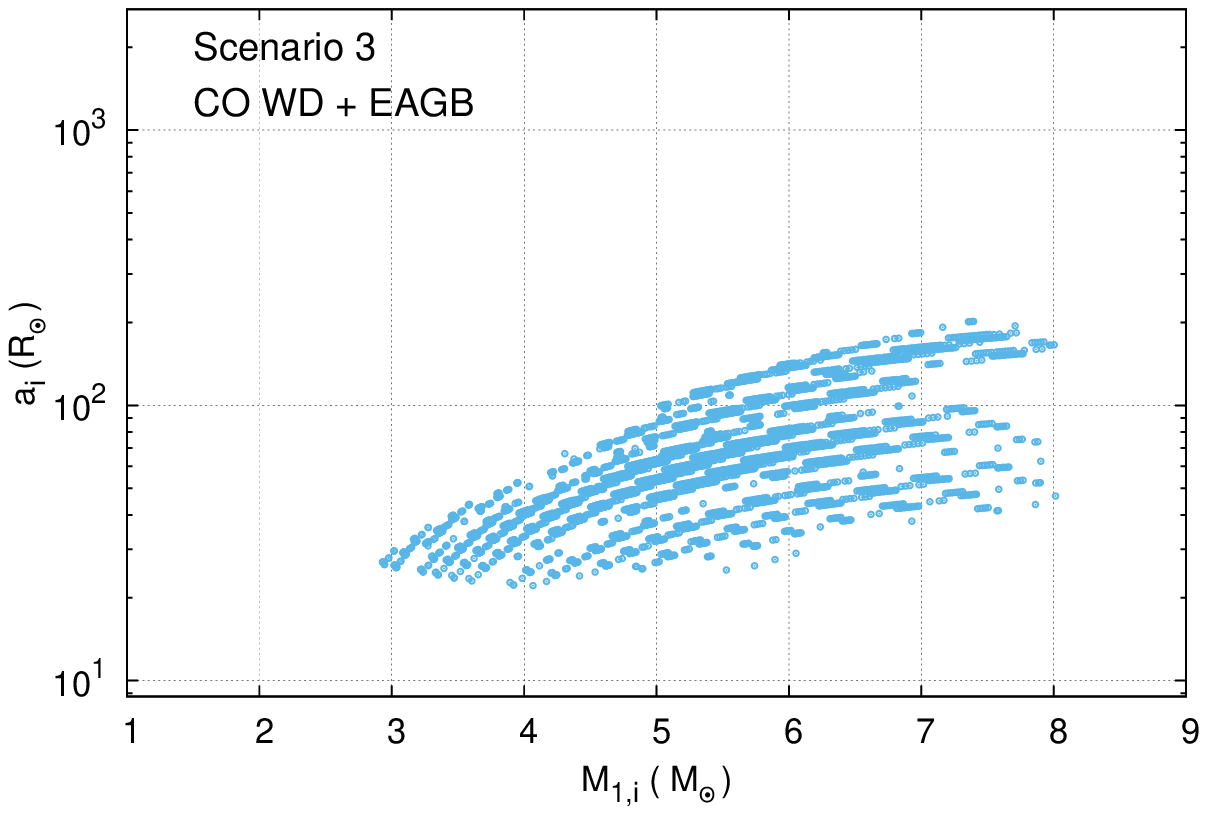}
\includegraphics[width=0.49\textwidth]{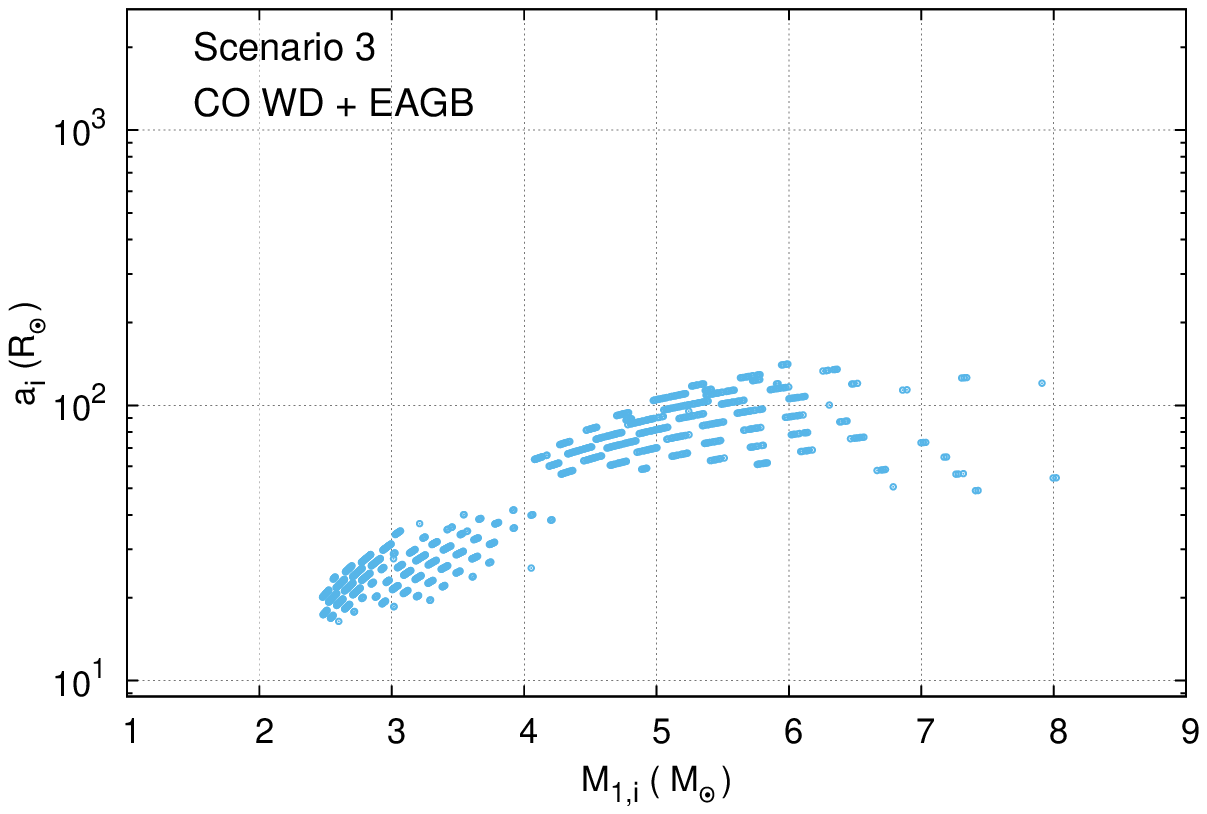}
\includegraphics[width=0.49\textwidth]{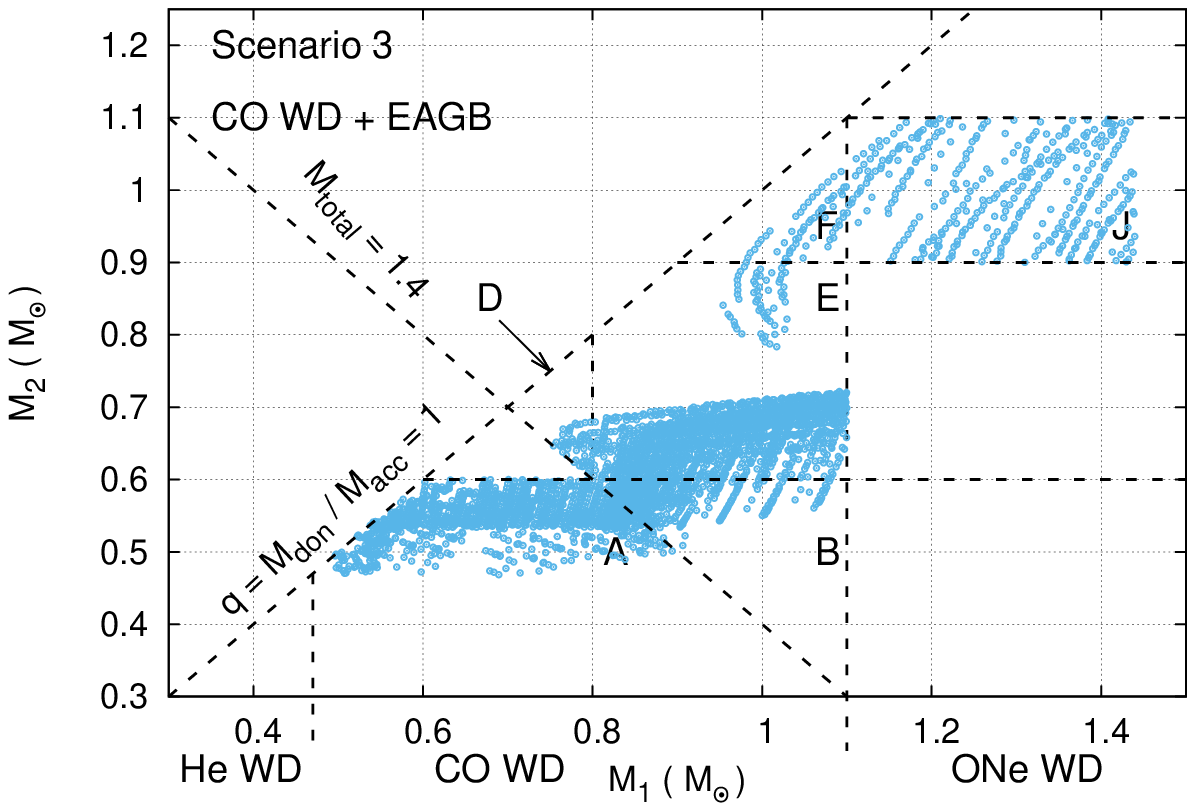}
\includegraphics[width=0.49\textwidth]{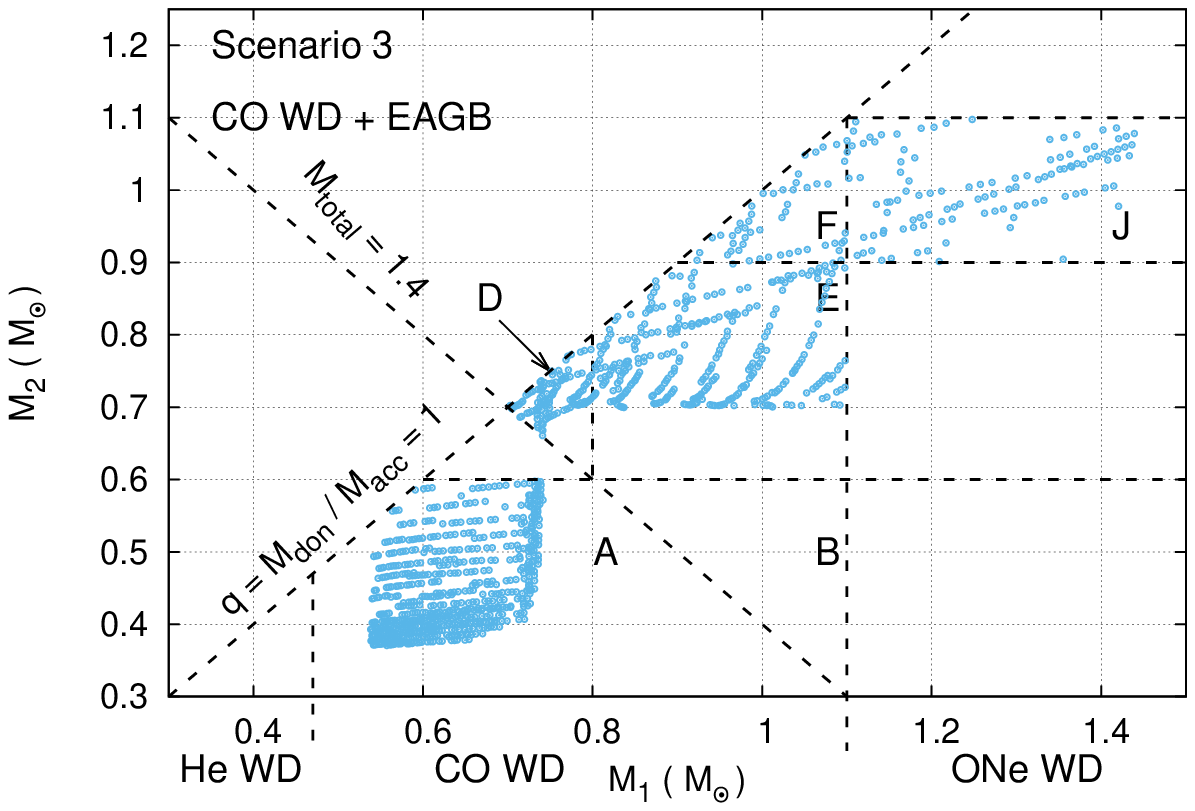}
\caption{As in Fig.~\ref{fig:scen1}, but for scenario 3 with \al=0.25\ (left column) and \al=2 (right column).}
\label{fig:scen3}
\end{figure*}

\begin{figure} 
\includegraphics[width=0.49\textwidth]{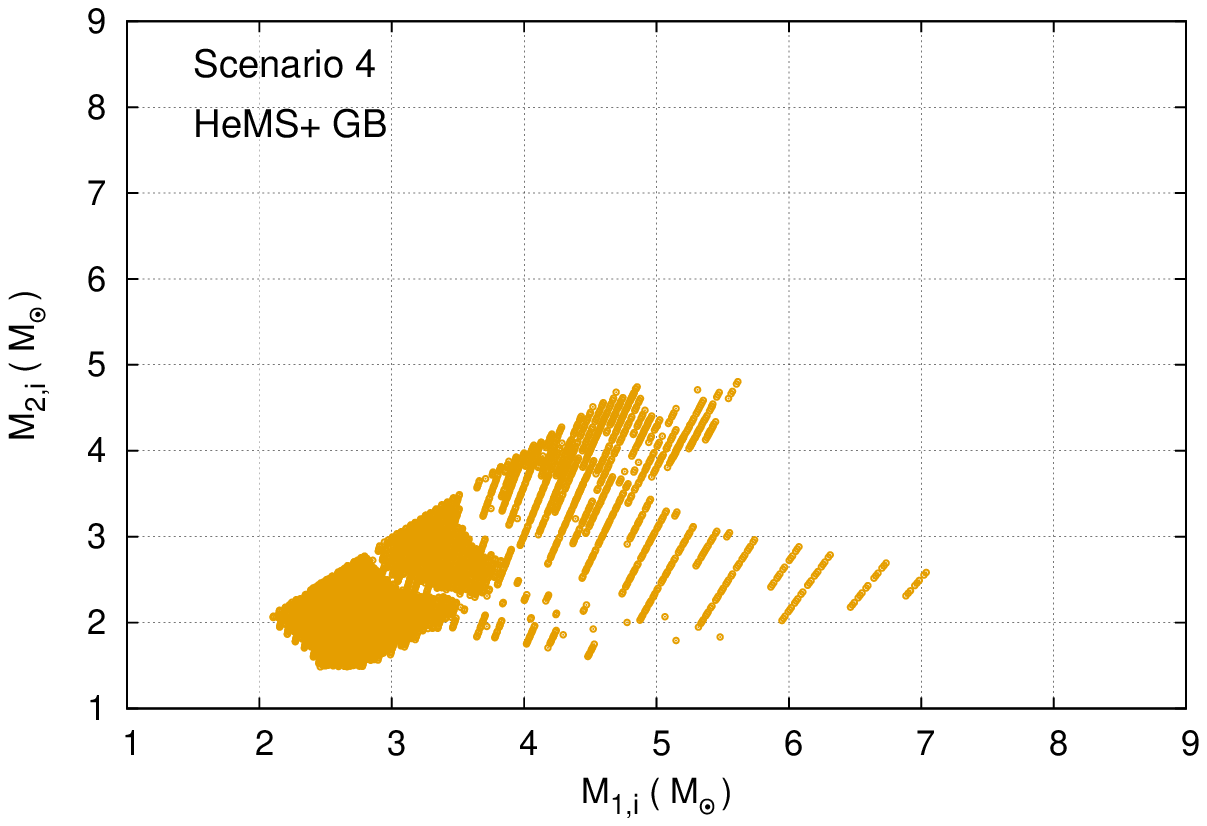}
\includegraphics[width=0.49\textwidth]{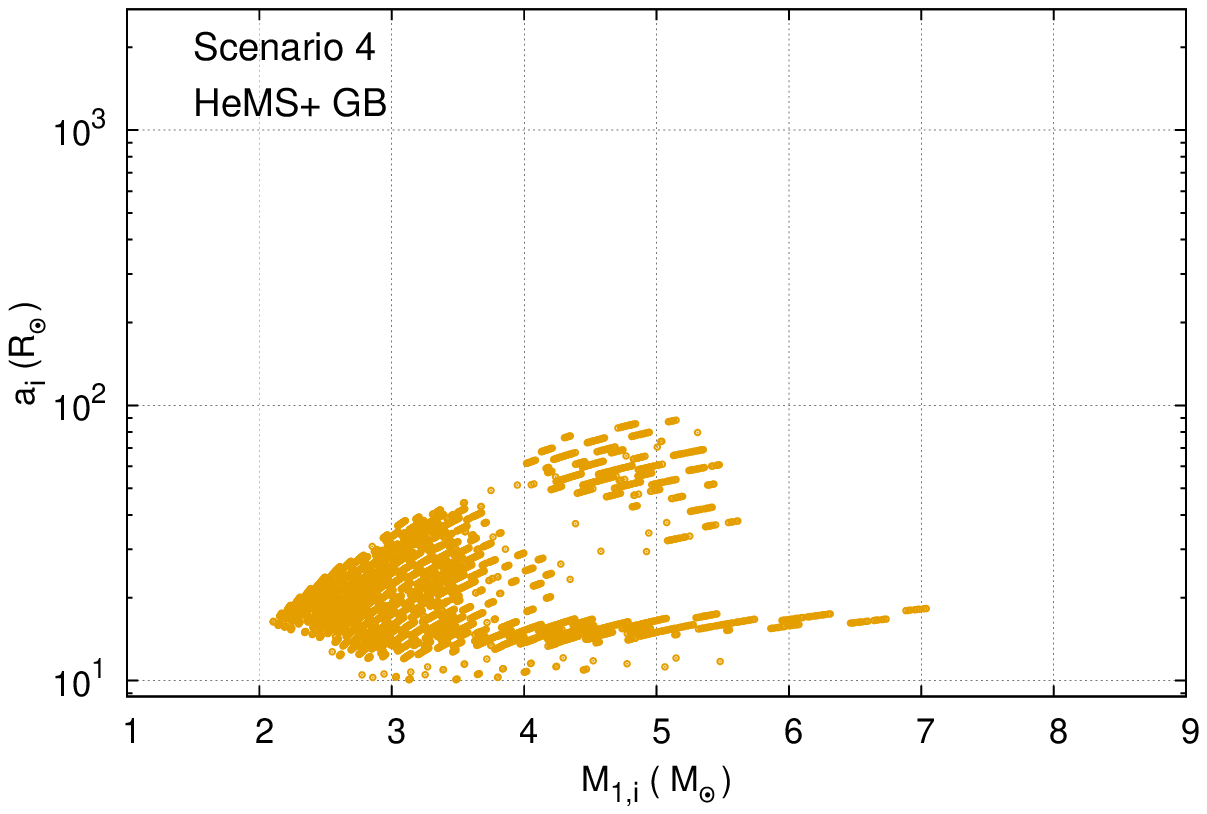}
\includegraphics[width=0.49\textwidth]{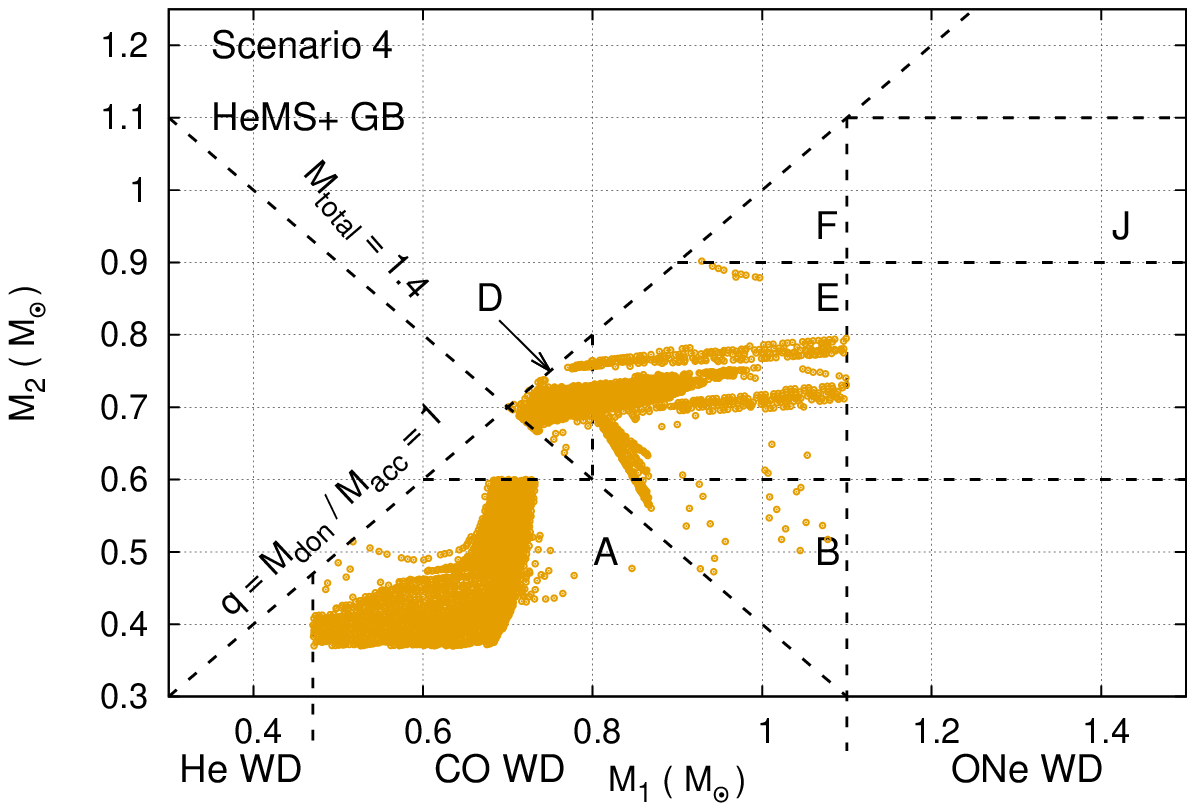}
\caption{As in Fig.~\ref{fig:scen1}, but for scenario 4.}
\label{fig:scen4}
\end{figure}

\begin{figure}
\includegraphics[width=0.49\textwidth]{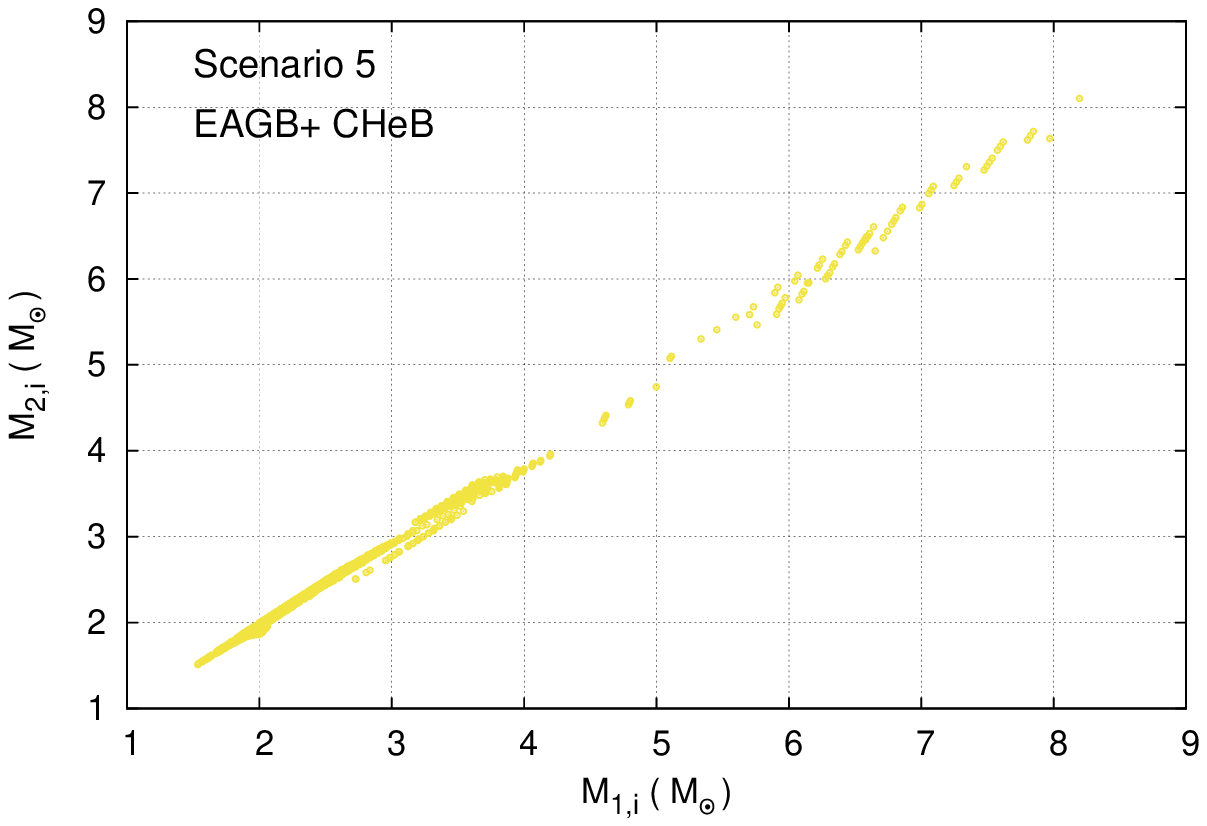}
\includegraphics[width=0.49\textwidth]{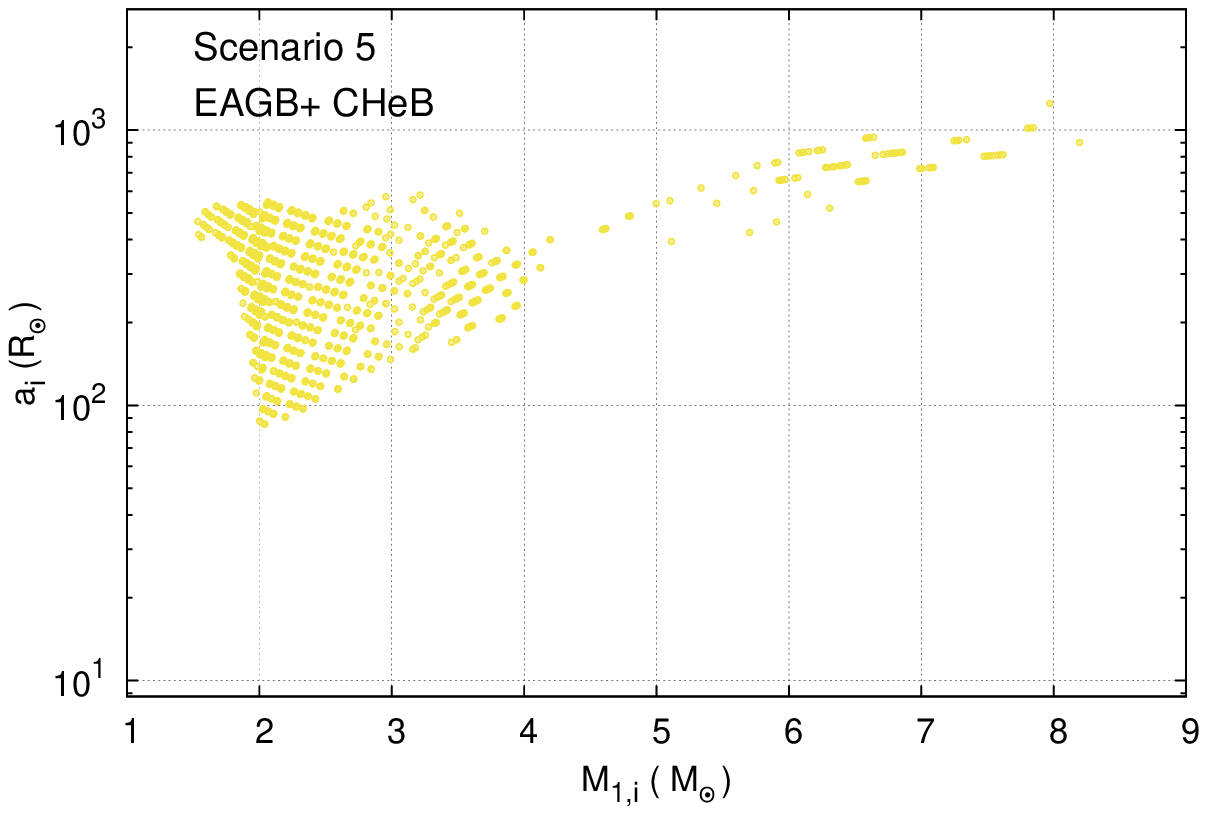}
\includegraphics[width=0.49\textwidth]{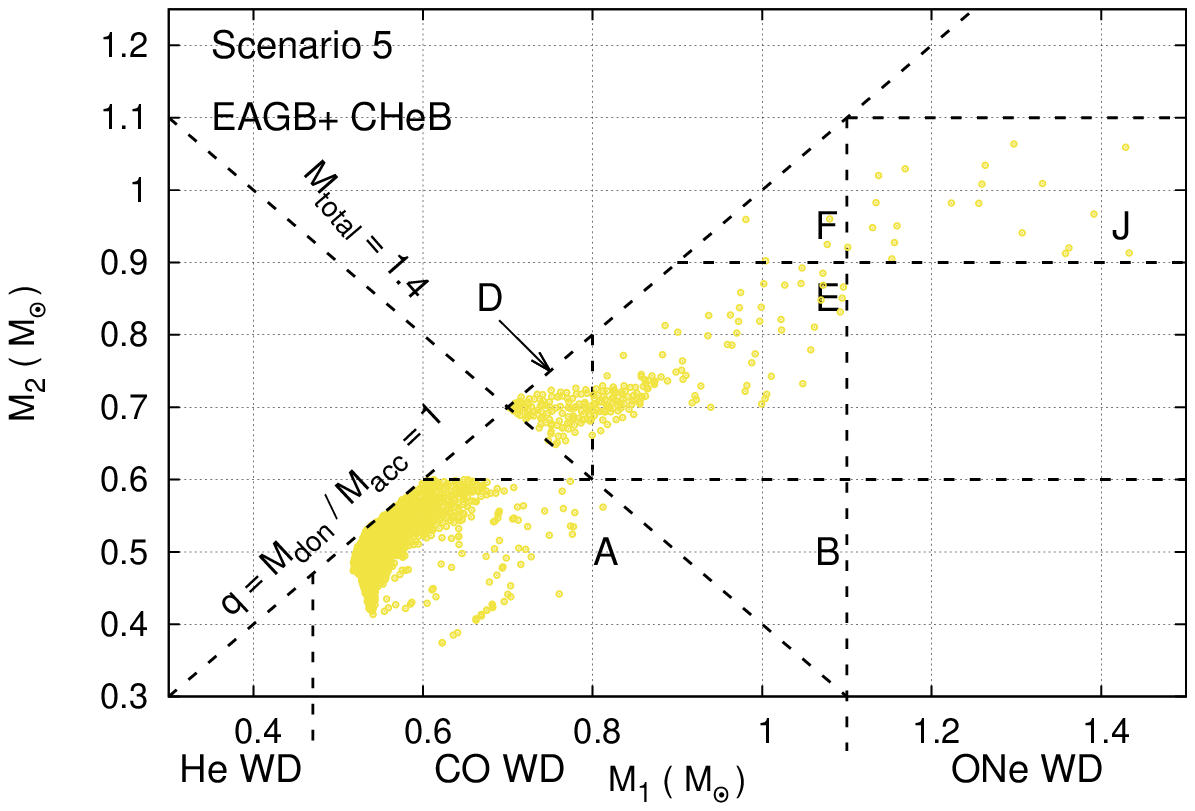}
\caption{As in Fig.~\ref{fig:scen1}, but for scenario 5.}
\label{fig:scen5}\end{figure}

\begin{figure}   
\includegraphics[width=0.49\textwidth]{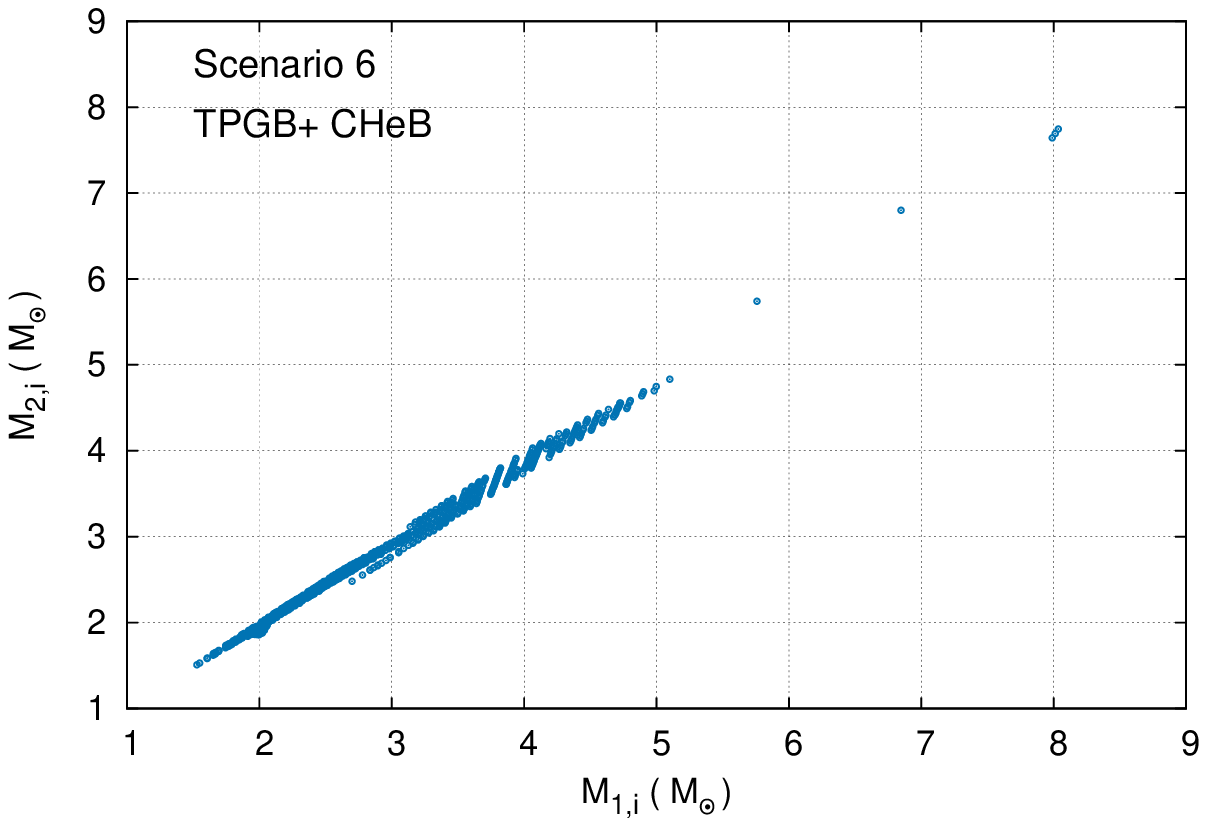}
\includegraphics[width=0.49\textwidth]{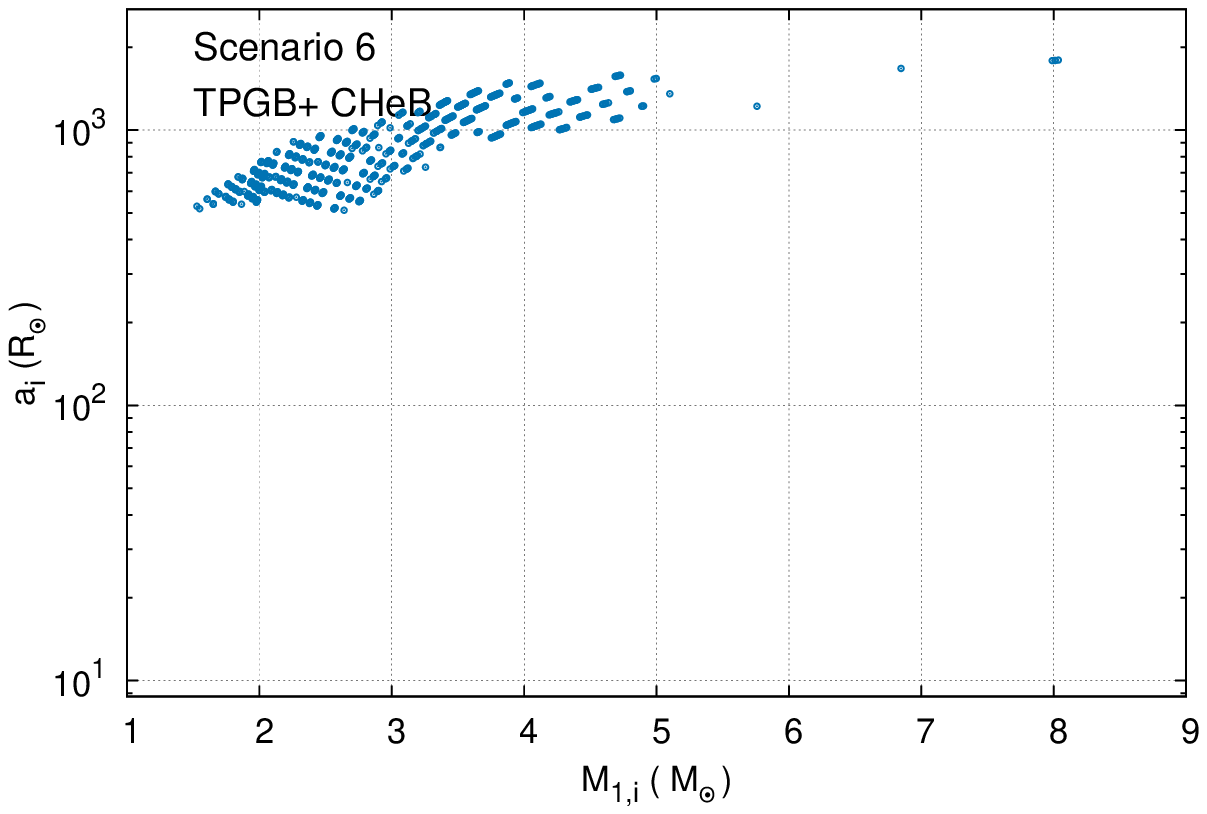}
\includegraphics[width=0.49\textwidth]{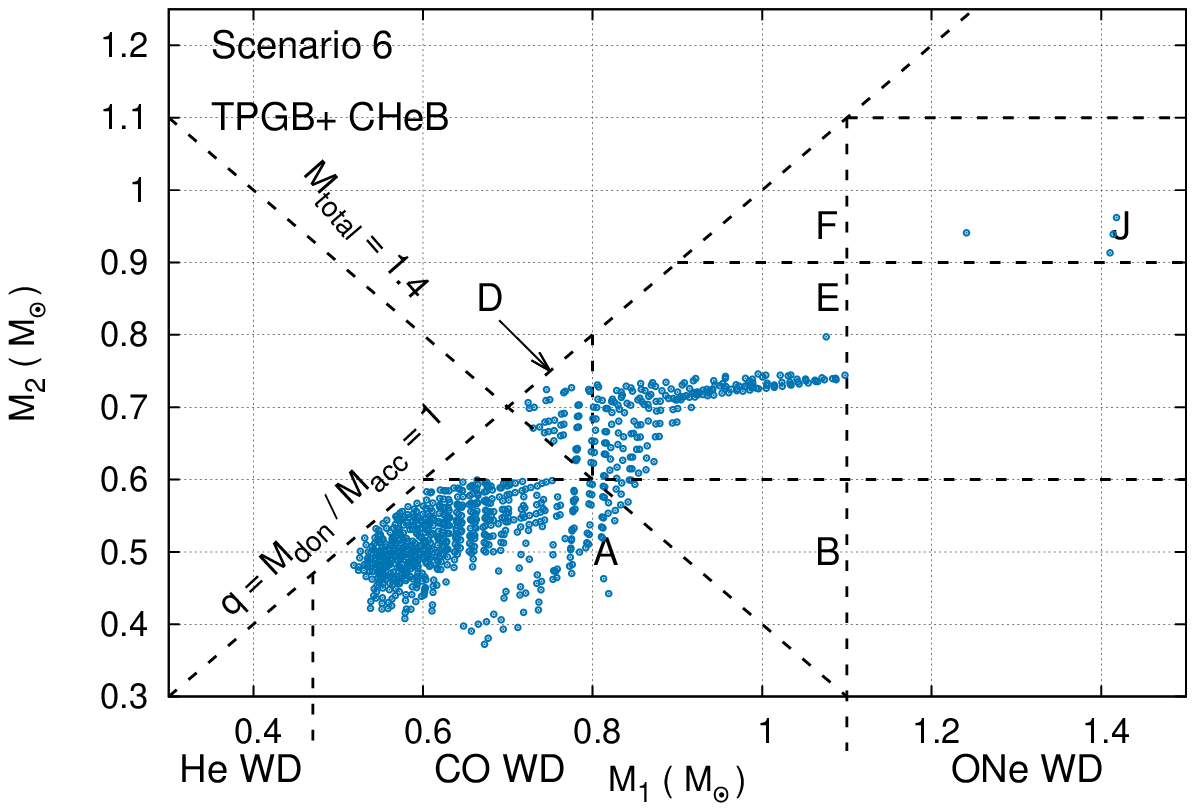}
\caption{As in Fig.~\ref{fig:scen1}, but for scenario 6.}
\label{fig:scen6}
\end{figure}

\begin{figure} 
\includegraphics[width=0.49\textwidth]{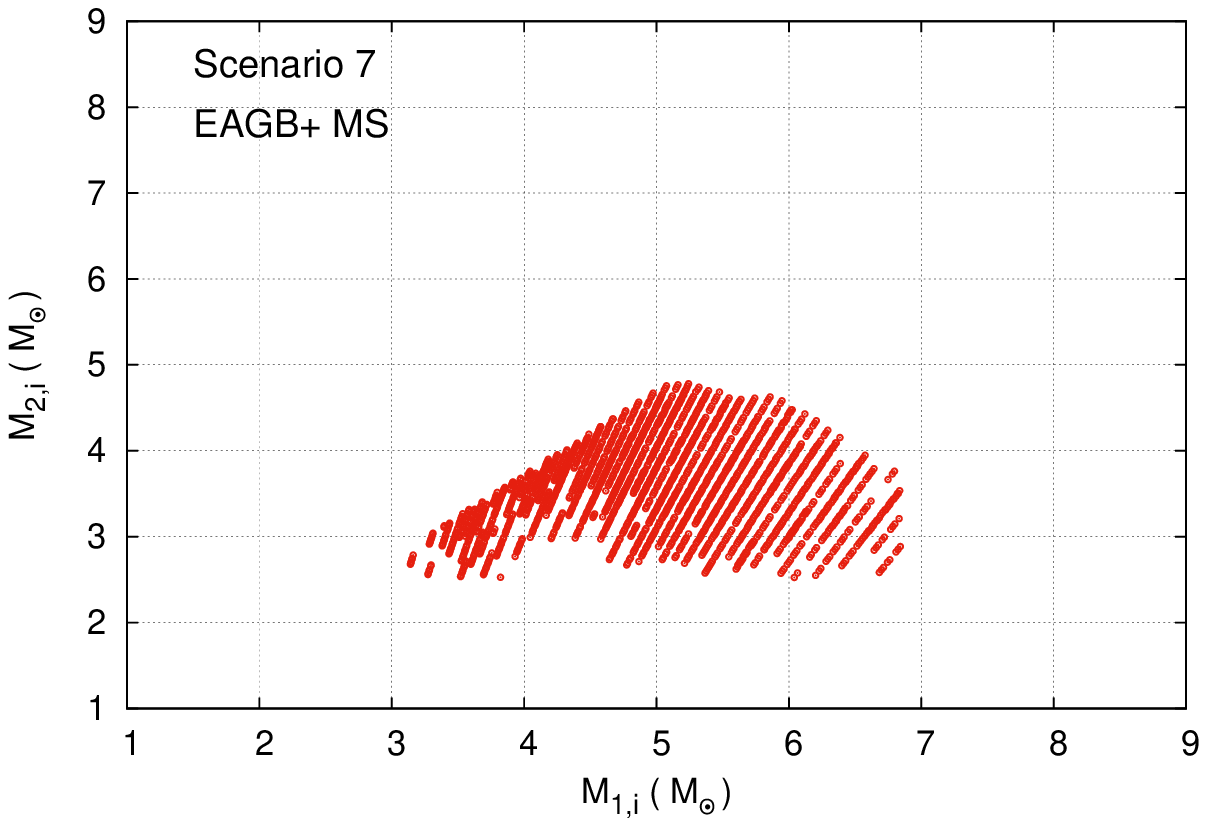}
\includegraphics[width=0.49\textwidth]{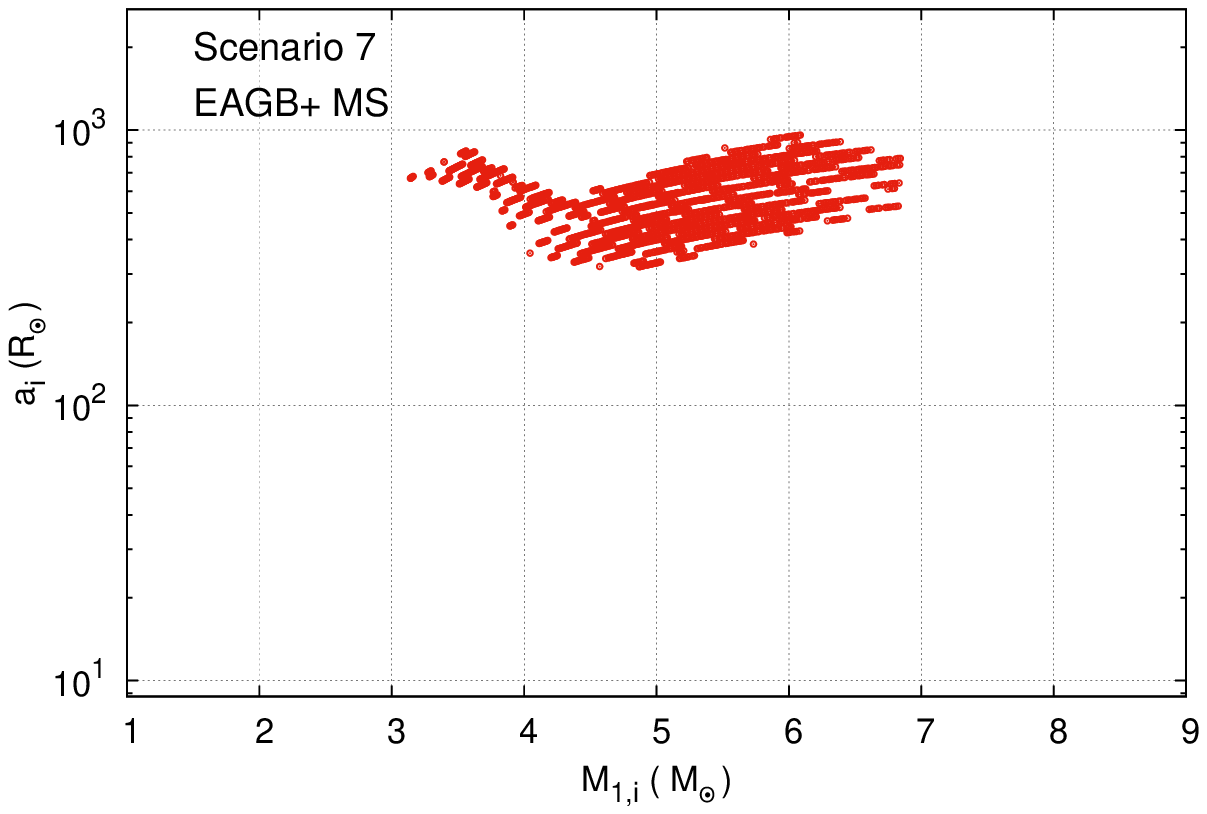}
\includegraphics[width=0.49\textwidth]{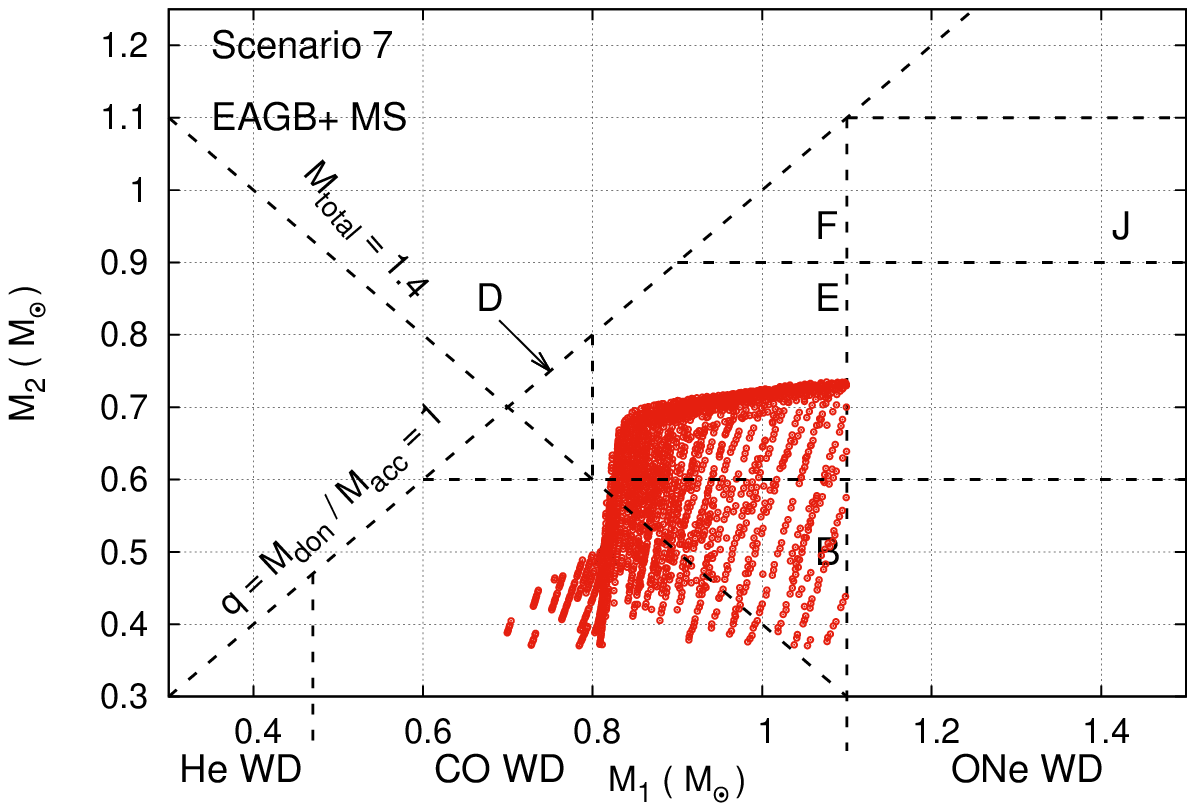}
\caption{As in Fig.~\ref{fig:scen1}, but for scenario 7.}
\label{fig:scen7}\end{figure}

\begin{figure}  
\includegraphics[width=0.49\textwidth]{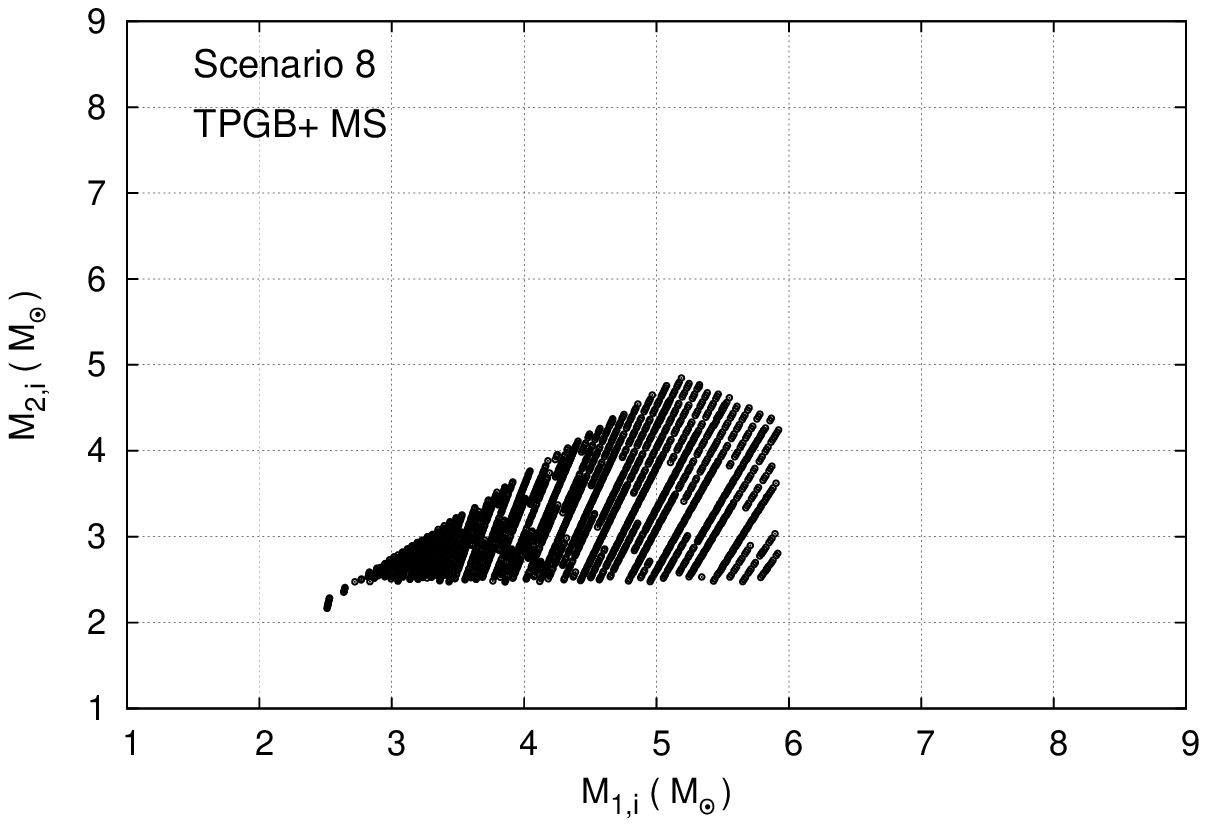}
\includegraphics[width=0.49\textwidth]{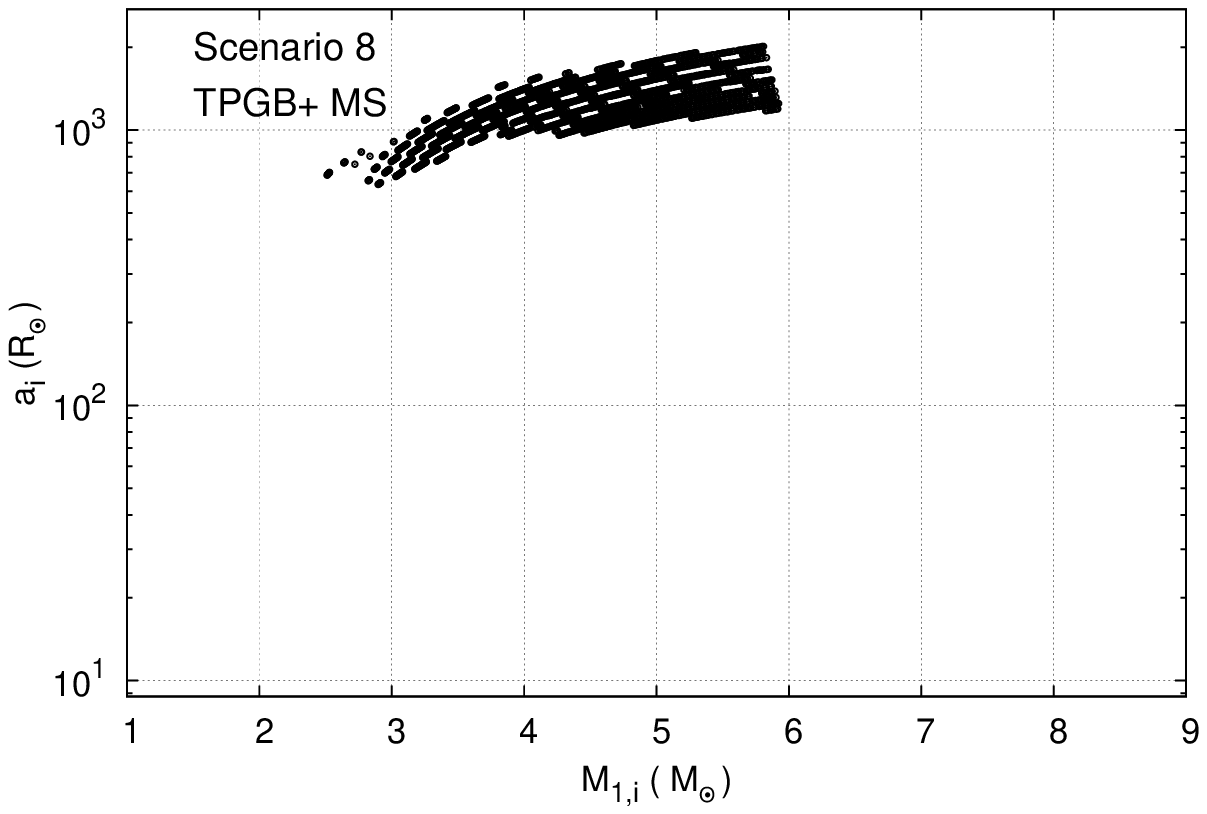}
\includegraphics[width=0.49\textwidth]{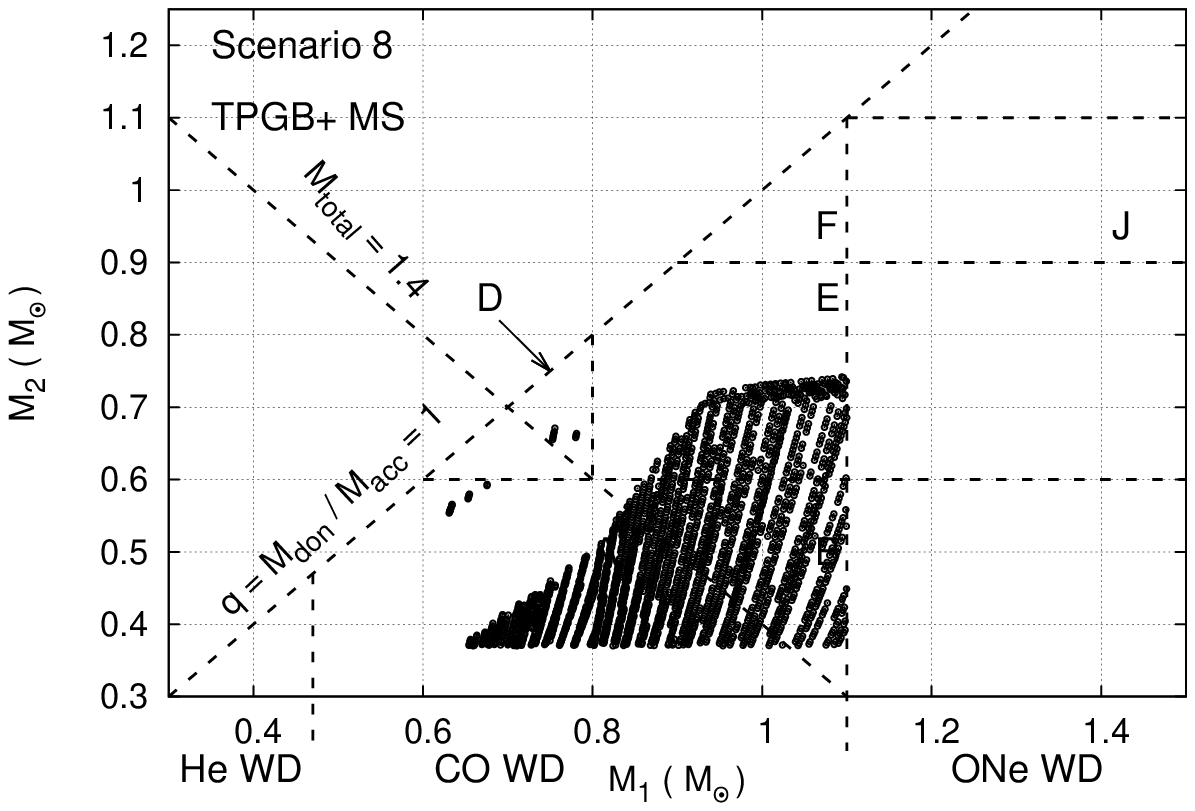}
\caption{As in Fig.~\ref{fig:scen1}, but for scenario 8.}
\label{fig:scen8}
\end{figure}


\begin{table}																
\caption{}																
\tabcolsep 1.0 mm 																
{\scriptsize																
\begin{tabular}{l|cc|cc|ccc}																
\hline																
\multicolumn{8}{c}	{Scenario 1: Formation of a CO WD + CO WD pair} \\															\hline															
	T (Myr)	&	Star1	&	Star2	&	R1/RL1	&	R2/RL2	&	M1	&	M2	&	A	\\
\hline																
	0.0	&	MS	&	MS	&	0.36	&	0.34	&	4.42	&	2.06	&	15.3	\\
	139.8	&	HG	&	MS	&	0.83	&	0.35	&	4.42	&	2.06	&	15.5	\\
	139.9	&	\{HG	&	MS	&	1.00	&	0.35	&	4.42	&	2.06	&	15.6	\\
	140.5	&	\{GB	&	MS	&	1.00	&	0.09	&	0.85	&	4.22	&	55.5	\\
	140.9	&	HeMS	&	MS	&	0.01	&	0.06	&	0.72	&	4.35	&	72.9	\\
	189.4	&	COHe	&	MS	&	0.01	&	0.08	&	0.72	&	4.35	&	74.0	\\
	195.1	&	COWD	&	MS	&	0.001	&	0.08	&	0.72	&	4.35	&	74.0	\\
	276.4	&	COWD	&	HG	&	0.001	&	0.14	&	0.72	&	4.35	&	74.0	\\
	277.1	&	COWD	&	HG\}	&	0.001	&	1.00	&	0.72	&	4.35	&	62.3	\\
\multicolumn{8}{c}	{\textbf{CE}}\\															
	277.1	&	COWD	&	HeMS	&	0.03	&	0.39	&	0.72	&	0.71	&	1.06	\\
	327.4	&	COWD	&	COHe	&	0.03	&	0.40	&	0.72	&	0.71	&	1.02	\\
	333.2	&	COWD	&	COHe\}	&	0.03	&	1.00	&	0.72	&	0.71	&	1.01	\\
	333.3	&	COWD	&	COHe	&	0.03	&	0.99	&	0.74	&	0.69	&	1.01	\\
	333.3	&	COWD	&	COWD	&	0.03	&	0.03	&	0.74	&	0.69	&	1.01	\\
	567.2	&	COWD	&	COWD	&	0.91	&	1.00	&	0.74	&	0.69	&	0.031	\\
	567.2	&	COWD	&	COWD	&	0.91	&	1.00	&	0.74	&	0.69	&	0.031	\\
	567.2	&	COWD	&	COWD\}	&	0.91	&	1.00	&	0.74	&	0.69	&	0.031	\\
\multicolumn{8}{c}	{Merger}\\															
\hline	
\end{tabular}																
}																
\label{tab:scen_1}																
\end{table}	

\begin{table}																
\caption{ }																
\tabcolsep 1.0 mm 																
{\scriptsize																
\begin{tabular}{l|cc|cc|ccc}																
\hline															
\multicolumn{8}{c}	{Scenario 2.1: Formation of a CO WD + He WD pair} \\							\hline																
	T (Myr)	&	Star1	&	Star2	&	R1/RL1	&	R2/RL2	&	M1	&	M2	&	A	\\
\hline																
	0	&	MS	&	MS	&	0.29	&	0.27	&	3.79	&	1.34	&	16.7	\\
	204.9	&	HG	&	MS	&	0.65	&	0.26	&	3.79	&	1.34	&	17.3	\\
	205.2	&	\{HG	&	MS	&	1.00	&	0.27	&	3.79	&	1.34	&	17.1	\\
	206.2	&	\{GB	&	MS	&	1.00	&	0.06	&	0.59	&	2.71	&	65.0	\\
	206.6	&	HeMS	&	MS	&	0.01	&	0.06	&	0.59	&	2.71	&	65.1	\\
	291.5	&	COHe	&	MS	&	0.01	&	0.06	&	0.59	&	2.71	&	65.7	\\
	301.5	&	COWD	&	MS	&	0.00	&	0.06	&	0.59	&	2.71	&	65.7	\\
	687.2	&	COWD	&	HG	&	0.00	&	0.13	&	0.59	&	2.71	&	65.7	\\
	690.3	&	COWD	&	GB	&	0.00	&	0.35	&	0.59	&	2.71	&	65.8	\\
	693.3	&	COWD	&	GB\}	&	0.00	&	1.00	&	0.59	&	2.71	&	48.7	\\
\multicolumn{8}{c}	{\textbf{CE}}\\															
	693.3	&	COWD	&	HeWD	&	0.03	&	0.03	&	0.59	&	0.40	&	0.91	\\
	1090	&	COWD	&	HeWD	&	0.06	&	0.08	&	0.59	&	0.40	&	0.57	\\
	1165	&	COWD	&	HeWD\}	&	0.67	&	1.00	&	0.59	&	0.40	&	0.05	\\
\multicolumn{8}{c}	{Merger}\\[3mm]															
\hline																
\multicolumn{8}{c}	{Scenario 2.2: Formation of a CO WD + CO WD pair} \\															\hline																T (Myr)	&	Star1	&	Star2	&	R1/RL1	&	R2/RL2	&	M1	&	M2	&	A	\\
\hline																
	0	&	MS	&	MS	&	0.20	&	0.18	&	4.38	&	2.34	&	29	\\
	143.4	&	HG	&	MS	&	0.45	&	0.20	&	4.38	&	2.34	&	29	\\
	143.7	&	\{HG	&	MS	&	1.00	&	0.20	&	4.38	&	2.34	&	29	\\
	144.4	&	\{GB	&	MS	&	1.00	&	0.03	&	0.72	&	4.83	&	155	\\
	144.8	&	HeMS	&	MS	&	0.00	&	0.03	&	0.72	&	4.83	&	156	\\
	193.3	&	COHe	&	MS	&	0.00	&	0.04	&	0.71	&	4.83	&	159	\\
	199.1	&	COWD	&	MS	&	0.00	&	0.04	&	0.71	&	4.83	&	159	\\
	246.3	&	COWD	&	HG	&	0.00	&	0.07	&	0.71	&	4.83	&	159	\\
	246.7	&	COWD	&	GB	&	0.00	&	0.50	&	0.71	&	4.83	&	160	\\
	246.9	&	COWD	&	GB\}	&	0.01	&	1.00	&	0.71	&	4.83	&	107	\\
\multicolumn{8}{c}	{\textbf{CE}}\\															
	246.9	&	COWD	&	HeMS	&	0.02	&	0.26	&	0.71	&	0.82	&	1.74	\\
	281.4	&	COWD	&	COHe	&	0.02	&	0.25	&	0.71	&	0.82	&	1.76	\\
	285.2	&	COWD	&	COHe\}	&	0.02	&	1.00	&	0.71	&	0.82	&	1.75	\\
	285.3	&	COWD	&	COWD	&	0.02	&	0.02	&	0.80	&	0.73	&	1.75	\\
	1972	&	COWD	&	COWD\}	&	0.88	&	1.00	&	0.80	&	0.73	&	0.03	\\
\multicolumn{8}{c}	{Merger}\\															
\end{tabular}																
}																
\label{tab:scen_2}																
\end{table}																

\begin{table}																
\caption{																
}																
\tabcolsep 1.0 mm 																
{\scriptsize																
\begin{tabular}{l|cc|cc|ccc}																
\hline																
\multicolumn{8}{c}	{Scenario 3.1: Formation of a CO WD + CO WD pair} \\															
\hline																
	T (Myr)	&	Star1	&	Star2	&	R1/RL1	&	R2/RL2	&	M1	&	M2	&	A	\\
\hline																
	0	&	MS	&	MS	&	0.09	&	0.08	&	5.00	&	2.62	&	70	\\
	103.8	&	HG	&	MS	&	0.20	&	0.09	&	5.00	&	2.62	&	70	\\
	104.2	&	\{HG	&	MS	&	1.00	&	0.09	&	5.00	&	2.62	&	70	\\
	104.3	&	\{GB	&	MS	&	1.00	&	0.05	&	1.70	&	4.38	&	124	\\
	104.8	&	HeMS	&	MS	&	0.00	&	0.02	&	0.87	&	5.21	&	337	\\
	134.7	&	COHe	&	MS	&	0.00	&	0.02	&	0.86	&	5.21	&	344	\\
	138.2	&	\{COHe	&	MS	&	1.00	&	0.02	&	0.86	&	5.21	&	344	\\
	138.3	&	COWD	&	MS	&	0.00	&	0.02	&	0.86	&	5.21	&	345	\\
	189.5	&	COWD	&	HG	&	0.00	&	0.03	&	0.86	&	5.21	&	345	\\
	189.9	&	COWD	&	GB	&	0.00	&	0.28	&	0.86	&	5.21	&	345	\\
	190.1	&	COWD	&	CHeB	&	0.00	&	0.96	&	0.86	&	5.21	&	242	\\
	203.7	&	COWD	&	EAGB	&	0.00	&	0.71	&	0.86	&	5.14	&	246	\\
	203.7	&	COWD	&	EAGB\}	&	0.00	&	1.00	&	0.86	&	5.14	&	180	\\
\multicolumn{8}{c}	{\textbf{CE1}}\\															
	203.7	&	COWD	&	COHe	&	0.01	&	0.23	&	0.86	&	1.25	&	5.04	\\
	203.9	&	COWD	&	COHe	&	0.01	&	1.00	&	0.86	&	1.25	&	5.03	\\
\multicolumn{8}{c}	{\textbf{CE2}}\\															
	203.9	&	COWD	&	COWD	&	0.02	&	0.02	&	0.86	&	0.75	&	1.48	\\
	956.8	&	COWD	&	COWD\}	&	0.84	&	1.00	&	0.86	&	0.75	&	0.03	\\
\multicolumn{8}{c}	{Merger}\\[3mm]															
\hline																
\multicolumn{8}{c}	{Scenario 3.2:  Formation of a CO WD + CO WD pair} \\															
\hline																
	T (Myr)	&	Star1	&	Star2	&	R1/RL1	&	R2/RL2	&	M1	&	M2	&	A	\\
\hline																
	0	&	MS	&	MS	&	0.18	&	0.18	&	2.83	&	1.52	&	24	\\
	440.7	&	HG	&	MS	&	0.42	&	0.19	&	2.83	&	1.52	&	25	\\
	442.9	&	\{HG	&	MS	&	1.00	&	0.19	&	2.83	&	1.52	&	24	\\
	443.4	&	\{GB	&	MS	&	1.00	&	0.10	&	1.04	&	2.51	&	40	\\
	447.8	&	HeMS	&	MS	&	0.00	&	0.03	&	0.43	&	3.12	&	153	\\
	716.2	&	COHe	&	MS	&	0.00	&	0.05	&	0.42	&	3.12	&	154	\\
	743.1	&	COWD	&	MS	&	0.00	&	0.06	&	0.42	&	3.12	&	154	\\
	757.3	&	COWD	&	HG	&	0.00	&	0.06	&	0.42	&	3.12	&	154	\\
	759.3	&	COWD	&	GB	&	0.00	&	0.19	&	0.42	&	3.12	&	154	\\
	761.4	&	COWD	&	CHeB	&	0.00	&	0.78	&	0.42	&	3.12	&	115	\\
	841.0	&	COWD	&	EAGB	&	0.00	&	0.51	&	0.42	&	3.09	&	116	\\
	841.8	&	COWD	&	EAGB\}	&	0.00	&	1.00	&	0.42	&	3.09	&	77	\\
\multicolumn{8}{c}	{\textbf{CE}}\\															
	841.8	&	COWD	&	COHe	&	0.03	&	0.30	&	0.42	&	0.69	&	1.65	\\
	843.7	&	COWD	&	COWD	&	0.03	&	0.02	&	0.42	&	0.69	&	1.65	\\
	4429	&	\{COWD	&	COWD	&	1.00	&	0.60	&	0.42	&	0.69	&	0.05	\\
\multicolumn{8}{c}	{Merger}\\															
\end{tabular}																
}																
\label{tab:scen_3}																
\end{table}																
\begin{table}																
\caption{																
}																
\tabcolsep 1.0 mm 																
{\scriptsize																
\begin{tabular}{l|cc|cc|ccc}																
\hline																
\multicolumn{8}{c}	{Scenario 4.1:  Formation of a CO WD + CO WD pair} \\															
\hline																
	T (Myr)	&	Star1	&	Star2	&	R1/RL1	&	R2/RL2	&	M1	&	M2	&	A	\\
\hline																
	0	&	MS	&	MS	&	0.13	&	0.13	&	3.47	&	2.48	&	40	\\
	257.6	&	HG	&	MS	&	0.31	&	0.16	&	3.47	&	2.48	&	40	\\
	258.8	&	\{HG	&	MS	&	1.00	&	0.16	&	3.47	&	2.48	&	40	\\
	259.0	&	\{GB	&	MS	&	1.00	&	0.10	&	1.66	&	4.05	&	59	\\
	267.6	&	HeMS	&	MS	&	0.00	&	0.02	&	0.57	&	5.14	&	313	\\
	337.6	&	HeMS	&	HG	&	0.00	&	0.04	&	0.57	&	5.14	&	315	\\
	338.0	&	HeMS	&	GB	&	0.00	&	0.28	&	0.57	&	5.14	&	315	\\
	338.2	&	HeMS	&	GB\}	&	0.00	&	1.00	&	0.57	&	5.14	&	190	\\
\multicolumn{8}{c}	{\textbf{CE}}\\															
	338.2	&	HeMS	&	HeMS	&	0.18	&	0.18	&	0.57	&	0.90	&	2.5	\\
	351.4	&	COHe	&	HeMS	&	0.15	&	0.21	&	0.57	&	0.89	&	2.5	\\
	362.8	&	COWD	&	HeMS	&	0.02	&	0.21	&	0.57	&	0.89	&	2.5	\\
	366.1	&	COWD	&	COHe	&	0.02	&	0.18	&	0.57	&	0.88	&	2.5	\\
	369.1	&	COWD	&	COHe\}	&	0.02	&	1.00	&	0.57	&	0.88	&	2.5	\\
	369.2	&	COWD	&	COWD	&	0.01	&	0.01	&	0.71	&	0.74	&	2.3	\\
	6100	&	COWD\}	&	COWD	&	1.00	&	0.95	&	0.71	&	0.74	&	0.03	\\
\multicolumn{8}{c}	{Merger}\\[3mm]															
\hline																
\multicolumn{8}{c}	{Scenario 4.2:  Formation of a CO WD + ONe WD pair} \\															
\hline																
	T (Myr)	&	Star1	&	Star2	&	R1/RL1	&	R2/RL2	&	M1	&	M2	&	A	\\
\hline																
	0	&	MS	&	MS	&	0.12	&	0.11	&	5.29	&	4.15	&	59	\\
	91.2	&	HG	&	MS	&	0.27	&	0.16	&	5.29	&	4.15	&	59	\\
	91.4	&	\{HG	&	MS	&	1.00	&	0.16	&	5.29	&	4.15	&	59	\\
	91.5	&	\{GB	&	MS	&	1.00	&	0.06	&	2.06	&	7.38	&	123	\\
	92.0	&	HeMS	&	MS	&	0.00	&	0.02	&	0.93	&	8.50	&	449	\\
	115.6	&	HeMS	&	HG	&	0.00	&	0.03	&	0.92	&	8.45	&	462	\\
	115.6	&	HeMS	&	GB	&	0.00	&	0.53	&	0.92	&	8.44	&	463	\\
	115.7	&	HeMS	&	GB\}	&	0.00	&	1.00	&	0.92	&	8.44	&	341	\\
\multicolumn{8}{c}	{\textbf{CE1}}\\															
	116.7	&	COHe	&	HeMS	&	0.11	&	0.14	&	0.92	&	1.74	&	5.5	\\
	119.4	&	\{COHe	&	HeMS	&	1.00	&	0.15	&	0.92	&	1.72	&	5.6	\\
	119.5	&	COHe	&	HeMS	&	0.80	&	0.12	&	0.78	&	1.86	&	6.7	\\
	119.5	&	COWD	&	HeMS	&	0.01	&	0.12	&	0.78	&	1.86	&	6.7	\\
	121.5	&	COWD	&	COHe	&	0.01	&	0.10	&	0.78	&	1.82	&	6.9	\\
	122.0	&	COWD	&	COHe\}	&	0.01	&	1.00	&	0.78	&	1.82	&	7.0	\\
\multicolumn{8}{c}	{\textbf{CE2}}\\															
	122.0	&	COWD	&	ONeWD	&	0.02	&	0.01	&	0.78	&	1.04	&	1.44	\\
	592	&	\{COWD	&	ONeWD	&	1.00	&	0.63	&	0.78	&	1.04	&	0.03	\\
\multicolumn{8}{c}	{Merger}\\															
\end{tabular}																
}																
\label{tab:scen_4}																
\end{table}	
															
\begin{table}																
\caption{																
}																
\tabcolsep 1.0 mm 																
{\scriptsize																
\begin{tabular}{l|cc|cc|ccc}																
\hline																
\multicolumn{8}{c}	{Scenario 5: Formation of a CO WD + CO WD pair } \\															
\hline																
	T (Myr)	&	Star1	&	Star2	&	R1/RL1	&	R2/RL2	&	M1	&	M2	&	A	\\
\hline																
	0	&	MS	&	MS	&	0.03	&	0.03	&	2.30	&	2.29	&	177	\\
	785	&	HG	&	MS	&	0.06	&	0.07	&	2.30	&	2.29	&	177	\\
	791	&	GB	&	MS	&	0.12	&	0.07	&	2.30	&	2.29	&	177	\\
	793	&	GB	&	HG	&	0.14	&	0.06	&	2.30	&	2.29	&	177	\\
	799	&	GB	&	GB	&	0.39	&	0.12	&	2.30	&	2.29	&	175	\\
	800	&	CHeB	&	GB	&	0.48	&	0.12	&	2.30	&	2.29	&	174	\\
	808	&	CHeB	&	CHeB	&	0.18	&	0.49	&	2.29	&	2.29	&	172	\\
	1011	&	EAGB	&	CHeB	&	0.33	&	0.29	&	2.27	&	2.27	&	177	\\
	1014	&	\{EAGB	&	CHeB	&	1.00	&	0.32	&	2.27	&	2.27	&	168	\\
\multicolumn{8}{c}	{\textbf{CE}}\\															
	1016	&	COWD	&	HeMS	&	0.03	&	0.28	&	0.57	&	0.56	&	1.24	\\
	1105	&	COWD	&	COHe	&	0.03	&	0.27	&	0.57	&	0.55	&	1.22	\\
	1117	&	COWD	&	COWD	&	0.03	&	0.03	&	0.57	&	0.55	&	1.22	\\
	2121	&	COWD	&	COWD\}	&	0.98	&	1.00	&	0.57	&	0.55	&	0.04	\\
\multicolumn{8}{c}	{Merger}\\
\end{tabular}																
}																
\label{tab:scen_5}																
\end{table}		
														
\begin{table}																
\caption{																
}																
\tabcolsep 1.0 mm 																
{\scriptsize																
\begin{tabular}{l|cc|cc|ccc}																
\hline																	
\multicolumn{8}{c}	{Scenario 6: Formation of a CO WD + CO WD pair} \\															
\hline																
	T (Myr)	&	Star1	&	Star2	&	R1/RL1	&	R2/RL2	&	M1	&	M2	&	A	\\
\hline																
	0	&	MS	&	MS	&	0.00	&	0.00	&	3.60	&	3.38	&	1351	\\
	234.3	&	HG	&	MS	&	0.01	&	0.01	&	3.60	&	3.38	&	1351	\\
	235.5	&	GB	&	MS	&	0.04	&	0.01	&	3.60	&	3.38	&	1351	\\
	236.6	&	CHeB	&	MS	&	0.12	&	0.01	&	3.60	&	3.38	&	1352	\\
	275.4	&	CHeB	&	HG	&	0.06	&	0.01	&	3.57	&	3.38	&	1365	\\
	276.9	&	CHeB	&	GB	&	0.06	&	0.04	&	3.57	&	3.38	&	1366	\\
	278.3	&	CHeB	&	CHeB	&	0.07	&	0.11	&	3.57	&	3.38	&	1367	\\
	283.8	&	EAGB	&	CHeB	&	0.08	&	0.05	&	3.57	&	3.37	&	1375	\\
	285.8	&	TPAGB	&	CHeB	&	0.66	&	0.05	&	3.53	&	3.38	&	1359	\\
	286.3	&	\{TPAGB	&	CHeB	&	1.00	&	0.05	&	3.34	&	3.41	&	1362	\\
\multicolumn{8}{c}	{\textbf{CE}}\\															
	286.3	&	COWD	&	HeMS	&	0.02	&	0.36	&	0.81	&	0.55	&	1.02	\\
	387.5	&	COWD	&	COHe	&	0.03	&	0.39	&	0.81	&	0.55	&	0.91	\\
	400.0	&	COWD	&	COWD	&	0.03	&	0.04	&	0.81	&	0.55	&	0.89	\\
	567.6	&	COWD	&	COWD\}	&	0.63	&	1.00	&	0.81	&	0.55	&	0.04	\\
\multicolumn{8}{c}	{Merger}\\	
\end{tabular}																
}																
\label{tab:scen_6}																
\end{table}		
														
\begin{table}																
\caption{																
}																
\tabcolsep 1.0 mm 																
{\scriptsize																
\begin{tabular}{l|cc|cc|ccc}																
\hline																
\multicolumn{8}{c}	{Scenario 7: Formation of  a CO WD + CO WD pair} \\															
\hline																
	T (Myr)	&	Star1	&	Star2	&	R1/RL1	&	R2/RL2	&	M1	&	M2	&	A	\\
\hline																
	0	&	MS	&	MS	&	0.02	&	0.01	&	4.86	&	4.08	&	449	\\
	111.3	&	HG	&	MS	&	0.03	&	0.02	&	4.86	&	4.08	&	449	\\
	111.7	&	GB	&	MS	&	0.25	&	0.02	&	4.86	&	4.08	&	449	\\
	112.0	&	CHeB	&	MS	&	0.64	&	0.02	&	4.86	&	4.08	&	437	\\
	128.9	&	EAGB	&	MS	&	0.46	&	0.03	&	4.80	&	4.08	&	444	\\
	129.5	&  \{EAGB	&	MS	&	1.00	&	0.03	&	4.79	&	4.08	&	426	\\
\multicolumn{8}{c}	{\textbf{CE1}}\\															
	129.5	&	COHe	&	MS	&	0.42	&	0.20	&	1.15	&	4.08	&	41	\\
	129.5	&	COHe	&	MS	&	1.00	&	0.20	&	1.15	&	4.08	&	41	\\
	129.6	&	COWD	&	MS	&	0.00	&	0.12	&	0.86	&	4.37	&	64	\\
	171.3	&	COWD	&	HG	&	0.00	&	0.17	&	0.86	&	4.37	&	64	\\
	171.9	&	COWD	&	HG\}	&	0.00	&	1.00	&	0.86	&	4.37	&	60	\\
\multicolumn{8}{c}	{\textbf{CE2}}\\															
	171.9	&	COWD	&	HeMS	&	0.02	&	0.37	&	0.86	&	0.72	&	1.18	\\
	221.5	&	COWD	&	COHe	&	0.02	&	0.38	&	0.86	&	0.71	&	1.14	\\
	227.2	&	COWD	&	COHe\}	&	0.02	&	1.00	&	0.86	&	0.71	&	1.13	\\
	227.3	&	COWD	&	COWD	&	0.02	&	0.03	&	0.88	&	0.69	&	1.15	\\
	518.6	&	COWD	&	COWD\}	&	0.73	&	1.00	&	0.88	&	0.69	&	0.03	\\
\multicolumn{8}{c}	{Merger}\\															
\end{tabular}																
}																
\label{tab:scen_7}																
\end{table}																
\begin{table}																
\caption{																
}																
\tabcolsep 1.0 mm 																
{\scriptsize																
\begin{tabular}{l|cc|cc|ccc}																
\hline																
\multicolumn{8}{c}	{Scenario 8.1: Formation of  a CO WD + CO WD pair} \\															
\hline																
	T (Myr)	&	Star1	&	Star2	&	R1/RL1	&	R2/RL2	&	M1	&	M2	&	A	\\
\hline																
	0	&	MS	&	MS	&	0.01	&	0.01	&	3.97	&	3.30	&	1198	\\
	182.4	&	HG	&	MS	&	0.01	&	0.01	&	3.97	&	3.30	&	1198	\\
	183.3	&	GB	&	MS	&	0.06	&	0.01	&	3.97	&	3.30	&	1198	\\
	184.0	&	CHeB	&	MS	&	0.16	&	0.01	&	3.97	&	3.30	&	1198	\\
	217.2	&	EAGB	&	MS	&	0.11	&	0.01	&	3.93	&	3.30	&	1216	\\
	218.7	&	TPAGB	&	MS	&	0.76	&	0.01	&	3.89	&	3.31	&	1177	\\
	219.0	&	\{TPAGB	&	MS	&	1.00	&	0.01	&	3.80	&	3.32	&	1158	\\
\multicolumn{8}{c}	{\textbf{CE1}}\\															
	219.0	&	COWD	&	MS	&	0.00	&	0.07	&	0.85	&	3.32	&	104	\\
	291.7	&	COWD	&	HG	&	0.00	&	0.09	&	0.85	&	3.32	&	104	\\
	293.3	&	COWD	&	GB	&	0.00	&	0.36	&	0.85	&	3.32	&	104	\\
	294.6	&	COWD	&	GB\}	&	0.00	&	1.00	&	0.85	&	3.32	&	80	\\
\multicolumn{8}{c}	{\textbf{CE2}}\\															
	294.6	&	COWD	&	HeMS	&	0.01	&	0.19	&	0.85	&	0.51	&	1.804	\\
	436.4	&	COWD	&	COHe	&	0.01	&	0.19	&	0.85	&	0.50	&	1.791	\\
	452.4	&	COWD	&	COWD	&	0.01	&	0.02	&	0.85	&	0.50	&	1.786	\\
	3288	&	COWD	&	COWD\}	&	0.54	&	1.00	&	0.85	&	0.50	&	0.042	\\
\multicolumn{8}{c}	{Merger}\\[3mm]															
\hline																
\multicolumn{8}{c}	{Scenario 8.2: Formation of  a CO WD + CO WD pair} \\															
\hline																
	T (Myr)	&	Star1	&	Star2	&	R1/RL1	&	R2/RL2	&	M1	&	M2	&	A	\\
\hline																
	0	&	MS	&	MS	&	0.01	&	0.01	&	3.26	&	2.94	&	1093.7	\\
	303.5	&	HG	&	MS	&	0.01	&	0.01	&	3.26	&	2.94	&	1093.7	\\
	305.2	&	GB	&	MS	&	0.04	&	0.01	&	3.26	&	2.94	&	1093.8	\\
	306.9	&	CHeB	&	MS	&	0.13	&	0.01	&	3.26	&	2.94	&	1094.3	\\
	374.7	&	EAGB	&	MS	&	0.08	&	0.01	&	3.23	&	2.94	&	1108.5	\\
	377.3	&	TPAGB	&	MS	&	0.64	&	0.01	&	3.20	&	2.95	&	1094.0	\\
	377.9	&	\{TPAGB	&	MS	&	1.00	&	0.01	&	3.04	&	2.97	&	1097.3	\\
\multicolumn{8}{c}	{\textbf{CE1}}\\															
	377.9	&	COWD	&	MS	&	0.00	&	0.08	&	0.75	&	2.97	&	120.0	\\
	396.6	&	COWD	&	HG	&	0.00	&	0.08	&	0.75	&	2.97	&	120.0	\\
	398.9	&	COWD	&	GB	&	0.00	&	0.24	&	0.75	&	2.97	&	120.0	\\
	401.5	&	COWD	&	CHeB	&	0.00	&	0.98	&	0.75	&	2.97	&	93.5	\\
	495.7	&	COWD	&	EAGB	&	0.00	&	0.64	&	0.76	&	2.95	&	94.5	\\
	496.6	&	COWD	&	EAGB\}	&	0.00	&	1.00	&	0.76	&	2.94	&	76.5	\\
\multicolumn{8}{c}	{\textbf{CE2}}\\															
	496.6	&	COWD	&	COHe	&	0.01	&	0.19	&	0.76	&	0.67	&	2.8	\\
	498.8	&	COWD	&	COWD	&	0.01	&	0.01	&	0.76	&	0.67	&	2.8	\\
	13225	&	COWD	&	COWD\}	&	0.86	&	1.00	&	0.76	&	0.67	&	0.03	\\
\multicolumn{8}{c}	{Merger}\\															
\end{tabular}																
}																
\label{tab:scen_8}																
\end{table}																

\bsp
\label{lastpage}
\end{document}